\documentclass[12pt]{article}
\usepackage{latexsym}
\usepackage{amsthm}
\usepackage{amsmath}
\usepackage[dvips]{graphicx}
\usepackage{wrapfig}
\usepackage{amssymb}
\usepackage{color}
\usepackage{cancel}
\usepackage{multirow}
\usepackage{dsfont}
\usepackage{pst-3d}                  % PSTricks 3D framed boxes
\usepackage{pst-grad}               % PSTricks gradient mode
\usepackage{pst-node}              % PSTricks nodes
\usepackage{pst-slpe}                % Improved PSTricks gradients
\usepackage{pst-plot}                % 
\usepackage{pst-coil}                % 
\usepackage{pst-math}              % 
\usepackage{pstricks-add}         %
\usepackage{arydshln}
\usepackage{rotating}
\usepackage{geometry}

\begin{document}
\baselineskip 0.6cm

\def\simgt{\mathrel{\lower2.5pt\vbox{\lineskip=0pt\baselineskip=0pt
           \hbox{$>$}\hbox{$\sim$}}}}
\def\simlt{\mathrel{\lower2.5pt\vbox{\lineskip=0pt\baselineskip=0pt
           \hbox{$<$}\hbox{$\sim$}}}}

\begin{titlepage}

\begin{flushright}
\end{flushright}

\vskip 2.0cm

\begin{center}

{\LARGE \bf TASI 2011: Four Lectures on TeV \\  [0.5em]
Scale Extra Dimensions }

\vskip 1.0cm

{\large Eduardo Pont\'on}

\vskip 0.4cm

{\it Department of Physics \\ 538 W 120th street \\
Columbia University, New York, NY 10027}

\abstract{Compact spatial dimension at the TeV scale remain an
intriguing possibility that is currently being tested at the LHC. We
give an introductory review of extra-dimensional models and ideas,
from a phenomenological perspective, but emphasizing the appropriate
theoretical tools.  We emphasize the power and limitations of such
constructions, and give a self-contained account of the methods
necessary to understand the associated physics.  We also review a
number of examples that illustrate how extra-dimensional ideas can
shed light on open questions in the Standard Model.  An introduction
to holography is provided.  These are the notes of my TASI 2011
Lectures on Extra Dimensions.}

\end{center}
\end{titlepage}

\def\simgt{\mathrel{\lower2.5pt\vbox{\lineskip=0pt\baselineskip=0pt
           \hbox{$>$}\hbox{$\sim$}}}}
\def\simlt{\mathrel{\lower2.5pt\vbox{\lineskip=0pt\baselineskip=0pt
           \hbox{$<$}\hbox{$\sim$}}}}

\renewcommand{\l}{\langle}
\renewcommand{\r}{\rangle}
\newcommand{\be}{\begin{eqnarray}}
\newcommand{\ee}{\end{eqnarray}}

\newcommand{\dd}[2]{\frac{\partial #1}{\partial #2}}
\newcommand{\NN}{\mathcal{N}}
\newcommand{\LL}{\mathcal{L}}
\newcommand{\MM}{\mathcal{M}}
\newcommand{\ZZ}{\mathcal{Z}}
\newcommand{\WW}{\mathcal{W}}

\newcommand{\FO}{{\rm FO}}
\newcommand{\FI}{{\rm FI}}
\newcommand{\TFO}{T_{\rm FO}}
\newcommand{\TFOp}{T_{\textrm{FO}'}}
\newcommand{\YFO}{Y_{\rm FO}}
\newcommand{\zFO}{z_{\rm FO}}
\newcommand{\FOp}{\textrm{FO}'}
\newcommand{\vev}{\textit{vev}}
\newcommand{\vevs}{\textit{vevs}}

\newcommand{\sv}{\langle \sigma v \rangle}

\newcommand{\MPl}{M_{\rm Pl}}

\tableofcontents

%\newpage
\setcounter{section}{-1}
%%%%%%%%%%%%%%%%%%%%%%%%%%%%%%%%%%%%%%%
\section{Introduction}
\label{intro}
%%%%%%%%%%%%%%%%%%%%%%%%%%%%%%%%%%%%%%%

It is illustrative to consider the first \textit{modern} attempt at
entertaining seriously the possibility that our universe could have
more than $3+1$ dimensions.  In 1921~\cite{Kaluza:1921tu}, not long
after the development of General Relativity (GR), the German
mathematician and physicist Theodor Kaluza extended GR to $4+1$
dimensions.\footnote{Strictly speaking, some of the main ideas had
already been introduced in 1914 by the Finish physicist Gunnar
Nordstr\"om~\cite{Nordstrom5D}.  He observed that extending the
Maxwell theory to 5D, written in terms of a 5-vector $A_M$, contained
the 4-vector $A_\mu$ --to be identified with the electromagnetic
vector potential-- plus a 4D scalar obeying the field equations for
his own proposed \textit{scalar theory of gravity}.  I emphasize here
the Kaluza-Klein theory both because it is based on general
relativity, now known to be the correct description of 4D gravity, as
well as because it allows us to neatly introduce concepts such as that
of a \textit{radion field}.} He noticed that ``splitting" one of the
spatial dimensions from the rest
\begin{wrapfigure}[15]{r}{2in}
\centering
\includegraphics[width=1.5in]{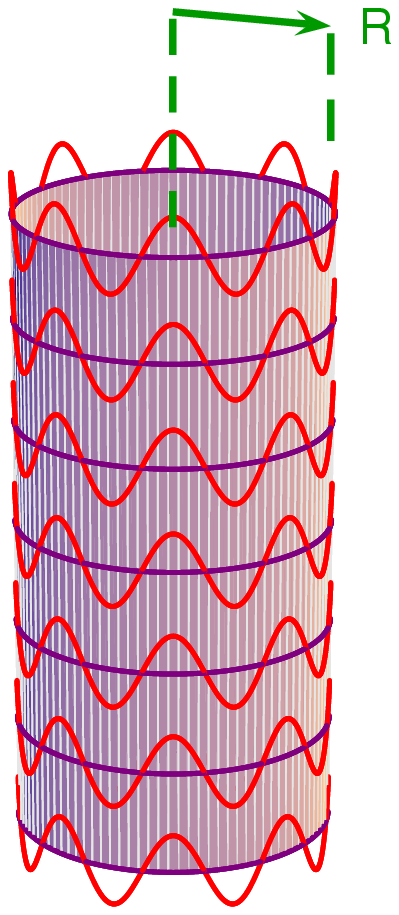}
\end{wrapfigure}
made it apparent that the structure of electromagnetism was contained
within GR. With the introduction of ``curled up", or \textit{compact},
dimensions by the Swedish physicist Oskar Klein in
1926~\cite{Klein:1926tv}, the possibility of electromagnetism simply
being part of $4+1$ dimensional general relativity, with one of the
spatial dimensions compactified, seemed to offer a beautiful path
towards the unification of the known forces at the time.  This is
known as the Kaluza-Klein theory.  The idea is fairly simple, even if
profound, and is illustrated in the figure to the right.  The vertical
dimension stands for the 3+1 (infinitely large) dimensions we are
familiar with.  The fifth dimension, on the other hand, is
\textit{finite}, being \textit{compactified on a circle of radius
$R$}.  Note that our world corresponds to the ``surface of the
cylinder": there is nothing physical inside or outside of it.  When
fields propagate on such a manifold, the fifth component of the
momentum is quantized in units of $1/R$, the \textit{compactification
scale}.  Due to the associated gradients, exciting such field
configurations requires energies of that order, and in fact appear to
4D observers as \textit{massive 4-dimensional fields excitations}.  At
energies low compared to the compactification scale, the relevant
field configurations are constant in the 5th dimension, and are known
as \textit{0-modes}.  The 5D metric tensor of the Kaluza-Klein (KK)
theory takes the schematic form
\be
G_{MN} &=&
\left(
\begin{array}{ccc:c}
 \hspace{4mm}  &   & \hspace{4mm} & \\ [0.3em]
  & \bar{g}_{\mu\nu}  & & ~\bar{A}_\mu  \\ [0.3em]
  &   &  &  \\  [0.3em]
\hdashline
\rule{0mm}{5mm}
 & \bar{A}_\mu^T & & ~\bar{g}_{55} 
\end{array}
\right)
\quad\quad \underset{\textrm{Lorentz group}}{\stackrel{\textrm{Under
4D}}{\rule{0mm}{3mm}\Longrightarrow}} \quad\quad
\begin{array}{l}
\textrm{``spin-2"}~\bar{g}_{\mu\nu}   \\  [0.3em]
+ \textrm{``spin-1"}~\bar{A}_{\mu}   \\ [0.3em]
+ \textrm{``spin-0"}~\bar{g}_{55}  
\end{array}~,
\label{5DMetricKK}
\nonumber
\ee
where we show the expected spin components under the 4D Lorentz group.
It turns out to be more convenient to parametrize the 0-modes as
\be
ds^2 &=& \phi^{-1/3} \left(g_{\mu\nu} - \phi A_\mu A_\nu \right)
dx^\mu dx^\nu
- 2 \phi^{2/3} A_\mu dx^\mu dy - \phi^{2/3} dy^2~,
\ee
where $y\equiv x^5$, and all fields are taken to be $y$-independent.
Replacing this ansatz in the 5D Einstein-Hilbert action gives
\be
\label{5dreduction}
-\frac{1}{2} M_5^3 \int \! d^5 x \, \sqrt{G} \, {\cal R}_5[G] &=&
-\frac{1}{2} M_P^2 \int d^4 x \sqrt{g} \left[~
{\cal R}_4[g] + \frac{1}{4} \, \phi \, F_{\mu\nu} F^{\mu\nu} -
\frac{1}{6} \frac{\partial_\mu \phi \, \partial^\mu
\phi}{\phi^2}~\right]~,
\ee
where, on the r.h.s.~all the contractions are done with
$g^{\mu\nu}(x)$, and $M_P^2 = (2 \pi R) M^3_5$ is to be identified
with the 4D reduced Planck mass.  The appearance of the $F_{\mu\nu}
F^{\mu\nu}$ structure can be understood by realizing that under an
infinitesimal general coordinate transformation $\delta G_{MN} =
\xi_{M;N} + \xi_{N;M}$, and when the infinitesimal parameter $\xi(x)$
is $y$-independent, one has
\be
\delta A_\mu = (\partial_\mu \xi^\rho) A_\rho + \xi^\rho
(\partial_\rho A_\mu ) + \partial_\mu \xi_5~.
\ee
The first term accounts for the transformation of the 4-vector index,
the second for the argument of the field $A_\mu(x)$, and the last one
is recognized as a $U(1)$ gauge transformation, which ensures that
$A_\mu$ must appear as part of $F_{\mu\nu} = \partial_\mu A_\nu -
\partial_\nu A_\mu$.  After the field redefinition $\phi \to
e^{-\sqrt{6} \phi/M_P}$, the scalar kinetic term becomes canonical,
and $\phi$ acquires the couplings of a \textit{dilaton field}.  Thus,
at low energies the Kaluza-Klein theory describes 4D gravity $+$ a
$U(1)$ gauge theory $+$ a real scalar field with dilaton couplings.
The latter can also be thought as parameterizing small oscillations in
the size of the extra dimension, and is often called a \textit{radion
field}.

From a modern perspective we may draw some lessons:
\begin{itemize}

\item The principle of general coordinate invariance in 4D is
well-tested experimentally, in the form of General Relativity (the
natural framework to describe massless spin-$2$ particles, in quantum
mechanical language).  In our applications, the assumption that the
full spacetime is also described by GR gives a set of well-defined
rules to write down extra-dimensional models.

\item The KK theory provides an intriguing way to achieve the
unification of apparently disparate forces, by obtaining a $U(1)$
field from a fifth dimension.  In fact, non-abelian fields can arise
from spacetimes of higher dimensionality, when appropriately
compactified.  Unfortunately, describing what we know today about the
Standard Model (SM) cannot be done within the Kaluza-Klein
framework~\cite{Witten:1981me}.  The problem arises from the fact that
the SM fermions are \textit{chiral} under the SM gauge group, while
Kaluza-Klein theories give rise to a \textit{vector-like} fermion
spectrum.

\item In modern (early $21^{\rm st}$ century) proposals, the previous
difficulty is overcome by considering compactifications with some sort
of (mild) singularities.  To illustrate what we mean, consider the
case of two extra-dimensions.  A simple way to compactify them is the
\textit{torus compactification}, where adjacent sides of a rectangular
region are identified, as shown in Fig.~\ref{fig:Torus}.
\begin{figure}[t]
\begin{center}
\includegraphics[width=1 \textwidth]{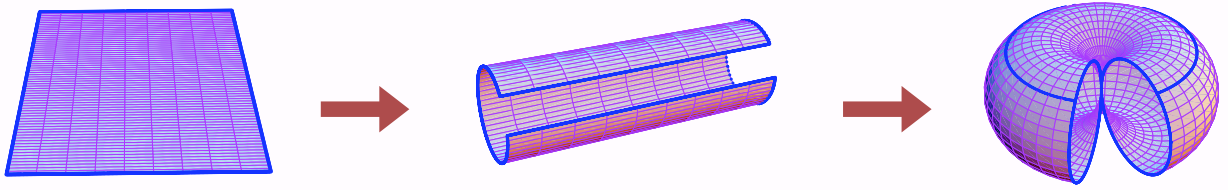}
\caption{Torus compactification: the identification of opposite sides
leads to a smooth manifold.  Fermions propagating on the torus lead to
a vector-like spectrum at low energies.}
\label{fig:Torus}
\end{center}
\end{figure}
This is a straightforward generalization of the 5D Kaluza-Klein theory
to 6D, and gives rise necessarily to a vector-like theory at low
energies, that is not suitable for embedding the SM.

A simple modification is achieved with the \textit{Chiral
Square}~\cite{Dobrescu:2004zi} compactification, where \textit{adjacent},
rather than \textit{opposite} sides are identified, as illustrated in
Fig.~\ref{fig:ChiralSquare}.
\begin{figure}[t]
\begin{center}
\includegraphics[width=0.75 \textwidth]{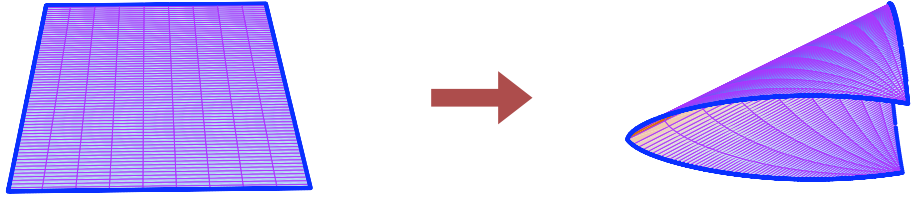}
\caption{The \textit{Chiral Square}: adjacent sides are identified to
form a closed 2D manifold with three conical singularities.  Fermions
propagating on the chiral square necessarily have a chiral 0-mode.}
\label{fig:ChiralSquare}
\end{center}
\end{figure}
In this case, the compact space has three \textit{conical
singularities}.  The boundary conditions that codify the chiral square
compactification are
\be
\begin{array}{rclcrcl}
\displaystyle
\Phi(y,0) &=& e^{in\pi/2} \Phi(0,y)~,
  & \hspace{5mm} &
\displaystyle
\left. \partial_5 \Phi \right|_{(x^4,x^5)=(y,0)} &=& - e^{in\pi/2}
\left. \partial_4 \Phi \right|_{(x^4,x^5)=(0,y)}~, 
\\ [0.5em]
\displaystyle
\Phi(y,\pi R) &=& e^{in\pi/2} \Phi(\pi R,y)~,
  &  &
\displaystyle
\left. \partial_5 \Phi \right|_{(x^4,x^5)=(y,\pi R)} &=& -
e^{in\pi/2} \left. \partial_4 \Phi \right|_{(x^4,x^5)=(\pi R,y)}~, 
\end{array}
\nonumber
\ee
which say that, up to a possible phase $e^{in\pi/2}$, the field and
its derivatives are continuous across the glued boundaries.  The
integer $n=0,1,2,3$ labels the consistent possibilities.  As it turns
out, fermions propagating on the chiral square are necessarily chiral
at low energies, hence the name of the compactification, which has
been used to describe possible physics beyond the SM. In the body of
these lectures, we will consider an even simpler possibility: the
\textit{interval compactification} in 5 dimensions.~\footnote{However,
6D spaces have intriguing properties not present in 5D, that we will not 
have space to explore.}

\item Extra-dimensional models beg questions such as: what fixes the
size of the extra dimensions?  (Or, in more technical language, what
sets the mass for the radion field?)  What about their ``shape"?  Or
for that matter, what selects how many dimensions are
compact/``infinite"?

\end{itemize}

Compared to the old Kaluza-Klein proposal, our aim here is at the same
time more and less ambitious.  It is less ambitious in that we do not
attempt to achieve a unification of forces through the geometric
structure.  We will simply assume the presence of higher-dimensional
fields of various spins, apart from the higher-dimensional metric.
The ``chiral compactifications" will typically get rid of the $U(1)$
Kaluza-Klein gauge field, but not of the radion mode.  In addition, we
will be content with assuming the number of compact dimensions
(together with the familiar $3+1$ non-compact dimensions), and also
the type of compactification.  Our guide will be phenomenological.  At
the same time we are more ambitious in that these extra-dimensional
models are built to be consistent with all aspects of the SM. Also, at
least in principle, many of these extensions can be tested or ruled
out at the TeV scale.

A very important question concerns the size of the compact extra
dimensions, if they exist.  Within string theory it may be natural for
this size to be of the order of the Planck or string scales.  If so,
such dimensions will not be probed directly, although even very small
extra dimensions can have observable indirect effects.  Another
possibility is that the compactification scale may be related in some
deep way to the weak scale, perhaps even playing a role in the
breaking of the electroweak symmetry (EWSB).  This type of new physics
might very well be testable during the coming decade.

\begin{figure}[t]
\begin{center}
\includegraphics[width=0.5 \textwidth]{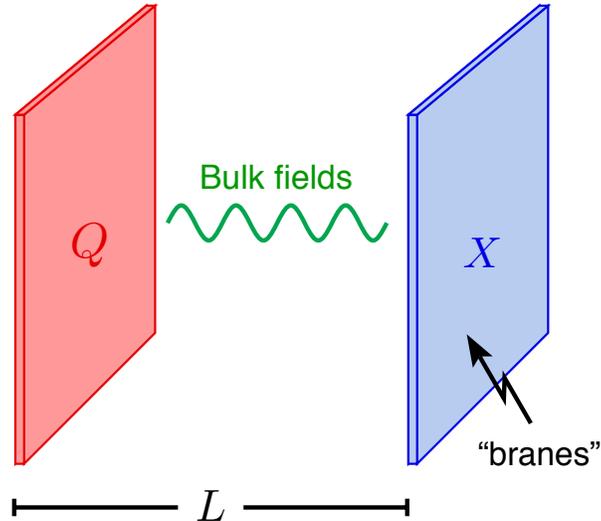}
\caption{General extra-dimensional setup, with defects of reduced
dimensionality.  These may be thought to have a thickness much smaller
than the typical size of the compact space, denoted by $L$.  Some
fields, denoted here by $Q$ and $X$, may be restricted to propagate on
such ``branes", while other may propagate in the bulk of the space.}
\label{fig:Branes}
\end{center}
\end{figure}

As already mentioned, realistic extra-dimensional scenarios possess
certain singularities, or defects, often also called \textit{branes}
(loosely borrowing stringy terminology).  Such defects can support
localized fields of various spins, while other fields may propagate in
the bulk of the space.  The situation is illustrated in
Fig.~\ref{fig:Branes}.
The possibility of localizing fields at different places of the extra
dimensional real estate marks an important difference with
4-dimensional model building.  The use of locality in the extra
dimensions has been successfully used in a number of applications.
This also means that specifying the localization (or not) of the
various fields in the theory becomes part of the \textit{definition}
of the model, beyond the specification of the field content and
relevant symmetries.

The above brings an additional issue: the cosmological constant.
Presumably, the gravitational background on which various fields
propagate is a solution to the 5D Einstein equations, possibly sourced
by bulk fields.  One can also write a 5D cosmological constant term,
\be
S_5 = \int \! d^5 x \,
\sqrt{g} \left\{ -\frac{1}{2} M^{3}_{5} \left( \mathcal{R}_{5} +
\Lambda_5 \right) + \textrm{other fields} \right\}~.
\label{ActionCC}
\ee
The only constraint we have on $\Lambda_5$ is that the
\textit{effective} 4D cosmological constant should be very small:
\be
M^2_P \Lambda_4 &\sim& \left( \textrm{contribution from
gravit.~background} \right)
+ \left( \textrm{contribution from } \Lambda_5 \right)
\nonumber \\ 
& & \mbox{} + \left( \textrm{contribution from other sources,
e.g.~Casimir energies} \right)
~~~\lesssim~ {\cal O}(10^{-3}~{\rm eV})^4~.
\label{4DCC}
\nonumber
\ee
One may envision two extremes here:
\begin{enumerate}

\item The various contributions have a \textit{natural} size of order
$M^4_5$, but they cancel out almost precisely.  This cancellation is
just the well-known cosmological constant problem, which so far seems
to necessitate an extraordinary fine-tuning (even if the final
understanding is based on ``anthropic considerations").

\item Perhaps, for some unspecified reason, $\Lambda_5$ is very
suppressed so that the extra dimensions are essentially flat.  Note
that the ``contribution from other sources" can be expected to contain
EWSB effects, or the effects from QCD chiral symmetry breaking, as
well as the quantum contribution from the compactification, itself of
order $1/R^4$ (often called Casimir energy).  Thus, fine-tuning is
still necessary, beyond the suppression of $\Lambda_5$.  Nevertheless,
flat extra dimensions illustrate an interesting benchmark from the
point of view of phenomenology.

\end{enumerate}

Naturally, one can also consider scenarios where $\Lambda_5$ takes an
intermediate value (perhaps related to the weak scale?).  In any case,
from our point of view, $\Lambda_5$ is a free parameter, and it is
important to be aware that it can have a significant impact on the
phenomenology.  This is the basis for the classification between
\textit{flat} versus \textit{warped} extra dimensions.

\begin{figure}[ht]
\begin{center}
\includegraphics[width=0.75 \textwidth]{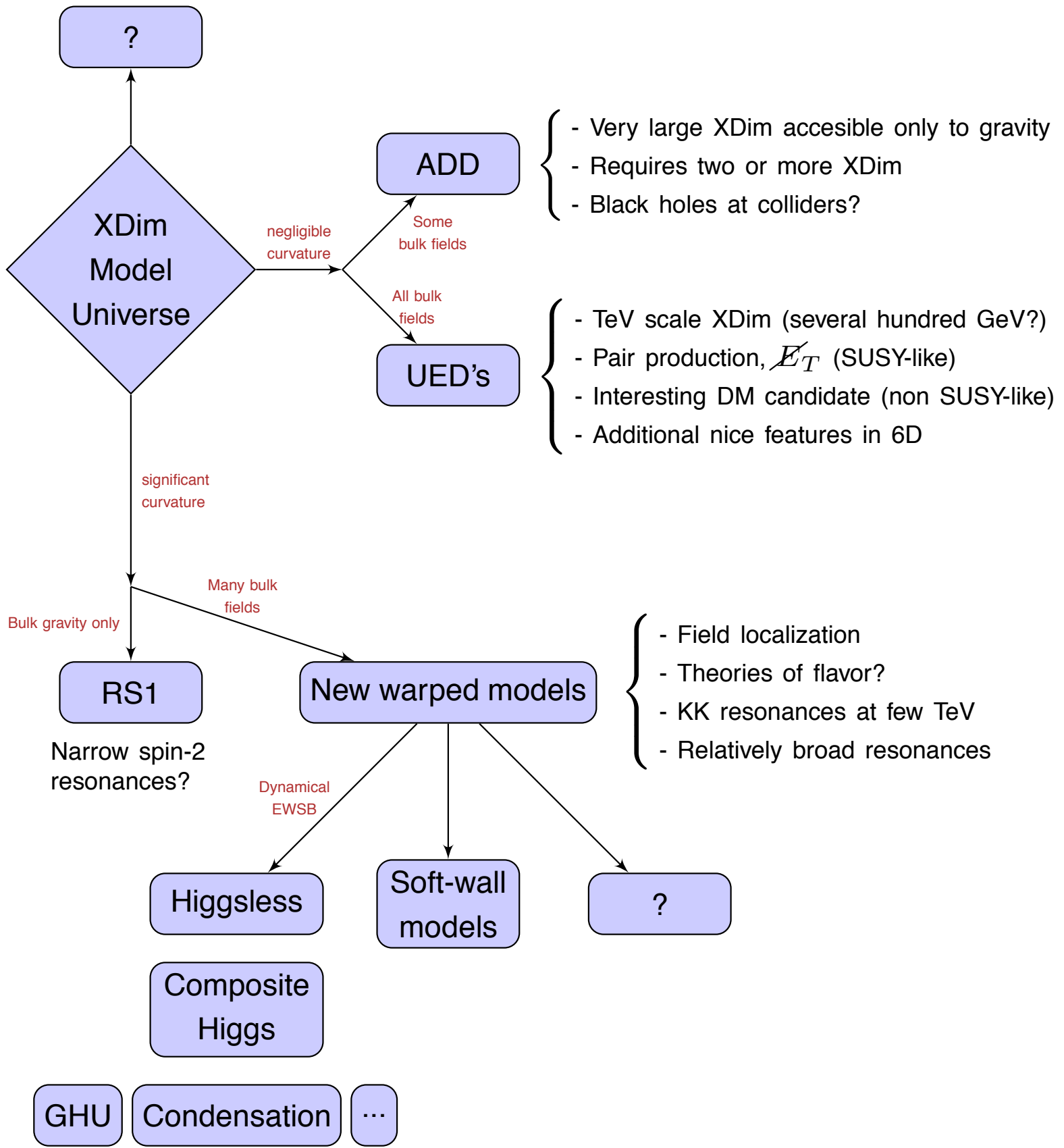}
\caption{A guide to \textit{representative} extra-dimensional
scenarios (not intended to be complete!).}
\label{fig:Guide}
\end{center}
\end{figure}

In these lectures, we explore various aspects of extra-dimensional
model building.  In the first lecture, we discuss a number of general
features that should be confronted when considering field theories in
more than four dimensions.  The student may want to have a cursory
look at this short lecture, and come back to it later on.  In the
second lecture, we develop the basic language required to discuss
extra-dimensional theories and their phenomenology.  We restrict to
the case of one compact extra spatial dimension.  This simplest case
already contains most of the main features and it pays to get fully
familiar with it.  In addition, many models of physical interest fall
in this category.  We derive our results in a general background,
specializing to particular cases only when necessary.  In the third
lecture, we review a number of such models, without intending to be
exhaustive.  Nevertheless, the student should be able to get a good
exposure to the possibilities that arise in extra-dimensional
frameworks.  In Fig.~\ref{fig:Guide} we show a schematic guide to a
few popular extra-dimensional scenarios and some of their prominent
features.  We will only touch on a subset of these, in particular only
those with extra dimensions at the TeV scale.  Finally, in Lecture 4
we give an elementary introduction to holography.

There have already been a number of excellent TASI lectures on Extra
Dimensions from a phenomenological point of view.  The student
interested in the subject is encouraged to look at all of these since
they cover different aspects and present different points of view:

\begin{itemize}

\item
%\cite{Csaki:2004ay}
%\bibitem{Csaki:2004ay} 
  C.~Cs\'aki,
  ``TASI lectures on extra dimensions and branes,''
%  In *Shifman, M. (ed.) et al.: From fields to strings, vol. 2* 967-1060
  hep-ph/0404096.
  %%CITATION = HEP-PH/0404096;%%

\item
%\cite{Sundrum:2005jf}
%\bibitem{Sundrum:2005jf} 
  R.~Sundrum,
  ``Tasi 2004 lectures: To the fifth dimension and back,''
  hep-th/0508134.
  %%CITATION = HEP-TH/0508134;%%

\item
%\cite{Csaki:2005vy}
%\bibitem{Csaki:2005vy} 
  C.~Cs\'aki, J.~Hubisz and P.~Meade,
  ``TASI lectures on electroweak symmetry breaking from extra
dimensions,''
  hep-ph/0510275.
  %%CITATION = HEP-PH/0510275;%%

\item
%\cite{Kribs:2006mq}
%\bibitem{Kribs:2006mq} 
  G.~D.~Kribs,
  ``TASI 2004 lectures on the phenomenology of extra dimensions,''
  hep-ph/0605325.
  %%CITATION = HEP-PH/0605325;%%

\item
%\cite{Dobrescu:2008zz}
%\bibitem{Dobrescu:2008zz} 
  B.~A.~Dobrescu,
  ``Particle physics in extra dimensions,''
  FERMILAB-CONF-08-703-T.
  %%CITATION = FERMILAB-CONF-08-703-T;%%
  
\item
%\cite{Cheng:2010pt}
%\bibitem{Cheng:2010pt} 
  H.~-C.~Cheng,
  ``2009 TASI Lecture -- Introduction to Extra Dimensions,''
  arXiv:1003.1162 %[hep-ph].
  %%CITATION = ARXIV:1003.1162;%%

\item
%\cite{Gherghetta:2010cj}
%\bibitem{Gherghetta:2010cj} 
  T.~Gherghetta,
  ``TASI Lectures on a Holographic View of Beyond the Standard Model
Physics,''
  arXiv:1008.2570 %[hep-ph].
  %%CITATION = ARXIV:1008.2570;%%

\end{itemize}

\newpage

\newgeometry{left=-3cm,right=-3cm,top=-0.5cm,bottom=0cm}
\restoregeometry

\newpage
%\renewcommand{\thesection}{}
%%%%%%%%%%%%%%%%%%%%%%%%%%%%%%%%%%%%%%%
\section{Lecture 1: Quantum Field Theory in More than 4D}
\label{QFT}
%%%%%%%%%%%%%%%%%%%%%%%%%%%%%%%%%%%%%%%

Recall that our expectation is that any Quantum Field Theory (QFT)
should be regarded as an effective low-energy description.  Even in a
``renormalizable" theory such as the Standard Model, higher-dimension
operators are most likely present, although they may be suppressed by
a very high scale.  In our discussion of extra-dimensional physics it
will be useful to have in mind the classic example of an effective
theory, the Fermi theory of $\beta$-decay:
\be
{\cal L}_{\rm Fermi} &=& G_F \, \bar{\psi}_1 \gamma^\mu \psi_2 \,
\bar{\psi}_3 \gamma_\mu \psi_4~,
\label{Fermi}
\ee
where $G_F \sim 1/v^2$.  The mass scale associated with the Fermi
constant implies that Eq.~(\ref{Fermi}) could have been valid at most
up to a cutoff of order $4\pi v$.  As it happens, the Fermi theory
breaks down at the somewhat lower scale associated with the weak gauge
bosons, due to the perturbative nature of the UV completion (i.e.~the
electroweak theory).  Generic higher-dimensional theories are very
similar to the Fermi theory in this respect.  In other words,
extra-dimensional models contain an intrinsic UV cutoff, that forces
us to treat them with the methods and spirit appropriate to Effective
Field Theories (EFT's): $i)$ processes with characteristic energies
parametrically lower than the cutoff can be reliably described, $ii)$
there are UV dominated observables displaying a power-law sensitivity
to the unspecified UV completion, $iii)$ IR-dominated observables
become particularly interesting.

Thus, one should keep in mind that these models may have limitations
in terms of calculability.  The basic reason for the above arises from
\textit{dimensional analysis}.  Consider a generic field theory in
$D$-dimensions that describes interacting fields of various spins,
\be
{\cal L}_D &\sim& \int \! d^D x \left\{
-\frac{1}{4g_D^2} F_{MN} F^{MN} + i \overline{\Psi} \cancel{D} \Psi +
| D_M \Phi |^2 + \cdots
\right\}~,
\label{LD}
\ee
where $D_M = \partial_M + i A_M$ implies the mass dimension $[A_M] =
1$, i.e.~$[F_{MN}] = 2$ and $g^2_D = 4-D$.  In particular, the gauge
coupling has inverse mass dimension for $D>4$, which introduces an
intrinsic scale into the theory.  Similarly, since fermions and
scalars have $[\Psi] = \frac{1}{2}(D-1)$ and $[\Phi] =
\frac{1}{2}(D-2)$, respectively, if we add a Yukawa interaction of the
type ${\cal L}_{\rm Yuk} = y_D \Phi \overline{\Psi} \Psi$, it follows
that $[y_D] = D - \frac{1}{2}(D-2) - (D-1) = 2 - \frac{1}{2} D$ has
also inverse mass dimensions for $D>4$.

%%%%%%%%%%%%%%%%%%%%%
\begin{quote}

\textbf{\underline{Exercise}:}
What is the mass dimension of a scalar quartic coupling?

\end{quote}
%%%%%%%%%%%%%%%%%%%%%

%%%%%%%%%%%%%%%%%%%%%
\begin{quote}

\textbf{\underline{Exercise}:} 
We have mentioned the possibility of localizing fields on 4D
subspaces, e.g.
\be
{\cal L}_5 &\supset& \delta(y-y_0) \left\{ \frac{1}{2} \left(
\partial_\mu \phi \right)^2 + i \bar{\psi} \cancel{\partial} \psi +
\cdots \right\}~,
\nonumber
\ee
[one can also consider $\partial_5$ derivatives, although these
require some care in the interpretation].  Work out the mass dimension
for
\begin{enumerate}

\item[a)] a Yukawa coupling between a bulk fermion $\Psi$ and a
localized scalar $\phi$.

\item[b)] a Yukawa coupling between a localized fermion $\psi$ and a
bulk scalar $\Phi$.

\item[c)] Explore various combinations of bulk/localized scalar
self-interactions.

\end{enumerate}

\end{quote}
%%%%%%%%%%%%%%%%%%%%%

The point of these observations is that most interesting (from our
low-energy 4D point of view) interactions become non-renormalizable
when considering $D>4$.  To better appreciate the meaning of this,
consider the 1-loop self-renormalization of a bulk Yukawa interaction:

\vspace*{-0.5cm}
\be
\put(-35,-35){
\resizebox{3.5cm}{!}{\includegraphics{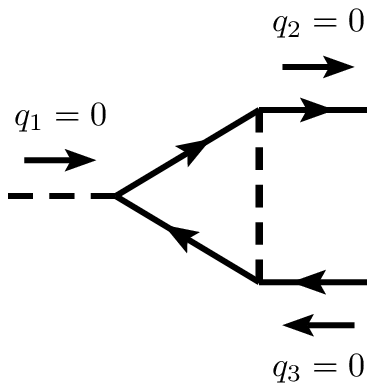}}
}
\hspace*{2cm}
&\sim& (i y_D)^3 \int \! \frac{d^{D}p}{(2\pi)^D} \,
\frac{i}{\cancel{p} - M_\Psi}  \, \frac{i}{\cancel{p} - M_\Psi}  \,
\frac{i}{p^2 - M_\Phi^2}
\nonumber \\ [-1.5em]
&\sim& \frac{y_D^3}{(2\pi)^D} \, \Omega_D \int^\Lambda \! p^{D-2}
\frac{1}{2} dp^2 \, \frac{1}{\cancel{p}^2 \, p^2}
\nonumber \\  [0.5em]
&\sim& \frac{\Omega_D}{2(2\pi)^D} \, y_D^3 \frac{2}{D-4}
\Lambda^{D-4}~,
\label{YukawaRen}
\ee
where $\Omega_D \equiv \int \!  d \Omega_D = 2\pi^{D/2} / \Gamma(D/2)$
is the $D$-dimensional solid angle.  A few comments are in order:
\begin{enumerate}

\item[i)] We have assumed that $D>4$, so that the integral is
dominated by the region near the cutoff $\Lambda$, as indicated in the
second line.  We also take the fermion and scalar masses to be small
compared to $\Lambda$.

\item[ii)] One obtains a D-dimensional loop factor
\be
\frac{1}{l_D} &\equiv& \frac{\Omega_D}{2(2\pi)^D} =
\frac{1}{(4\pi)^{D/2} \Gamma(D/2)} ~=~
\left\{
\begin{array}{ccc}
\frac{1}{16\pi^2}  &  \textrm{for}  &  D = 4  \\ [0.5em]
\frac{1}{24\pi^3}  &  \textrm{for}  &  D = 5
\end{array}~,
\right.
\label{DLoopFactor}
\ee
where we exhibit the 4D case to make contact with a familiar result,
as well as the 5D case that will be of use in the following.

\item[iii)] It may be illuminating to consider the regulator
dependence of the above result, which was obtained by imposing a
hard-momentum cutoff.  Let us consider again the momentum integral,
taking for concreteness $D=5$ and $M_\Psi = 0$.  In Euclidean space,
we have
\be
\int^\Lambda_0 \! p^{3} dp^2 \, \frac{1}{\cancel{p}^2}  \,
\frac{1}{p^2 + M_\Phi^2} &=& 2\Lambda - 2 M_\Phi \tan^{-1} \left(
\frac{\Lambda}{M_\Phi} \right)
\nonumber \\
&=& 2\Lambda - \pi M_\Phi \left[ 1 + {\cal O}\left( \frac{M_\Phi}{\Lambda}
\right) \right]~,
\nonumber
\ee
where we show also the subleading term in the second line.  If we were
to regulate the momentum integral with a \textit{Pauli-Villars}
(anticommuting) scalar, we would then get
\be
\int^\infty_0 \! p^{3} dp^2 \, \frac{1}{\cancel{p}^2}  \, \left[
\frac{1}{p^2 + M_\Phi^2} - \frac{1}{p^2 + \Lambda_{\rm PV}^2} \right]
&=& \pi \left( \Lambda_{\rm PV} - M_\Phi \right)~.
\nonumber
\ee
The fact that, for odd $D$, the integrand is not an analytic function
of $p^2$ (depends on $|p|$ through the measure) can lead to the
appearance of factors of $\pi$ on top of those in the loop factor,
Eq.~(\ref{DLoopFactor}).  Nevertheless, here we see that the
connection between the momentum cutoff and the Pauli-Villars mass is
given by $2\Lambda \leftrightarrow \pi \Lambda_{\rm PV}$, which shows
that they differ by an order one factor.  For the most part, we will
use a momentum cutoff to estimate UV-dominated contributions, but
should remember that there is some uncertainty as to where to
``expect" the new physics to come in.

\item[iv)] Since $[y_D] = 2 - \frac{1}{2} D$, we will often write it
in terms of the cutoff scale, $y_D \equiv c_y\Lambda^{2 - \frac{1}{2}
D}$, for some dimensionless $c_y$.  Note that the dimensions of
Eq.~(\ref{YukawaRen}) come from $y_D^3 \Lambda^{D-4} = c_y^3
\Lambda^{2 - \frac{1}{2} D}$, which coincide with those of $y_D$, as
required by dimensional analysis (remember that the amputated diagram
corresponds to a Yukawa interaction in the quantum effective action).

\item[v)] We have assumed above that we are working in a flat and
infinite $D$-dimensional space.  We will comment further on this later
on, but suffice it to say here that if the cutoff length is small
compared to the ``size" of the extra-dimensional space, as well as the
curvature radius (if in curved spacetime), it should be intuitively
clear that the above estimates continue to hold.  Only the IR
(i.e.~truly calculable) contributions depend on such details.

\end{enumerate}

\noindent
The above considerations are useful because
\begin{quote}
\begin{itemize}

\item They exhibit the \textit{power-law} dependence on the UV cutoff,
generally present in $D>4$ (for $D = 4$, we would recover
$\frac{1}{D-4} \Lambda^{D-4} \to \ln \Lambda$).  This signals that
certain quantities are, strictly speaking, incalculable.

\item They also allow to define a reasonable criterion for how large
the cutoff of a higher-dimensional theory can be, in terms of the
(dimensionful) couplings defining it.

\end{itemize}
\end{quote}

\noindent As we will see, the above does not mean that we cannot use
extra-dimensional models to understand physics questions or even make
predictions.  It only means that we should regard them as an effective
description valid over a finite energy regime, with uncertainties
pertaining to the actual (unknown) UV completion.  We therefore turn
to making the second bullet above more precise.

\medskip
\noindent
\textbf{\underline{Naive Dimensional Analysis} (NDA)}
\medskip

Recall from 4D experience that, after renormalization, the effects of
higher-dimension operators on physical observables scale like powers
of $E/\Lambda$, where $E$ characterizes the energy scale of the
process under consideration and $\Lambda$ characterizes the mass scale
of the coefficient of the higher-dimension operator.  Similarly, in
$D$-dimensions, cutoff-dependent effects, as estimated in
Eq.~(\ref{YukawaRen}) are absorbed, through the process of
renormalization, into the experimentally measured values of the
couplings of the theory (a Yukawa coupling in our example).  As in
``non-renormalizable" 4D theories, an infinite tower of interactions
is induced in this manner.  Nevertheless, at energies low compared to
the cutoff, only a finite number of them is relevant, at least if we
are content with a finite precision.  Only when the characteristic
energy approaches the cutoff, does the full set of higher-dimension
operators become important, and therefore the EFT looses predictive
power.  This is not surprising, as the details of the UV completion
should start to become crucial.  The above can also be related to the
breakdown of the loop expansion, i.e.~when diagrams at all loop orders
contribute equally (and are also comparable to the ``tree-level"
effects).  This implies a breakdown of the perturbative expansion, and
amounts to the onset of ``strong dynamics" (or more generally, to the
appearance of new degrees of freedom at or below $\Lambda$).  There
are therefore two aspects to our discussion:
\begin{enumerate}

\item Operators of arbitrary dimension can contribute equally at
tree-level.

\item The quantum (i.e.~loop) contributions associated with a given
interaction can all contribute equally (and at the same order as
tree-level).

\end{enumerate}

Let us start by considering loop effects at $E\sim\Lambda$, again
considering a Yukawa interaction for definiteness.  Note that now we
regard the coupling $y_D$ as the \textit{renormalized} Yukawa
interaction, so that we are talking about a result that is
\textit{finite} (i.e.~cutoff-independent, other than through the value
of $y_D$).  Nevertheless, when setting the external $E \sim \Lambda
\gg M_\Psi, M_\Phi$, our estimate proceeds as in
Eq.~(\ref{YukawaRen}), except that now the $\Lambda$-dependence is
understood as coming from the external momenta.  Thus, we find that
the renormalized three-point function evaluated at external momenta of
order $\Lambda$ takes the form
\be
y_D \left\{ 1 + {\cal O}(1) \times \frac{1}{l_D} y_D^2 \Lambda^{D-4}
+ \textrm{higher-order loops} \right\}
& \sim& y_D \left\{ 1 + \frac{c_y^2}{l_D} + \cdots \right\}~.
\ee
This suggests that the loop expansion breaks down when $c_y^2 \sim
l_D$.  In other words, the cutoff in the above sense is estimated as
\be
\Lambda &\sim& \left[ \frac{c_y^{\rm NDA}}{ y_D} \right]^{\textstyle
\frac{2}{4-D}} ~\equiv~ \left[ \frac{l_D^{1/2}}{ y_D}
\right]^{\textstyle \frac{2}{4-D}}~,
\ee
where we defined the NDA value $c_y^{\rm NDA} = l_D^{1/2}$ based on
the above considerations.
%%%%%%%%%%%%%%%%%%%%%
\begin{quote}

\textbf{\underline{Exercise}:}
Estimate by power counting, as above, some two-loop contributions to
the 3-point correlator above, and show that they are also as large as
the tree-level and one-loop contributions, for $c_y = c_y^{\rm NDA}$.

\end{quote}
%%%%%%%%%%%%%%%%%%%%%

\noindent 
Thus, the previous ``NDA rule" provides a reasonable criterion for the
largest the cutoff $\Lambda$ can be.  We note that this criterion also
addresses the first bullet above regarding the interplay of several
higher-dimension operators: if we estimate their coefficients by NDA
(say, by quick 1-loop estimates), then they also give similar
tree-level contributions to a given observable when $E \sim \Lambda$.
This will be illustrated in the following
%%%%%%%%%%%%%%%%%%%%%
\begin{quote}

\textbf{\underline{Exercise}:}
consider the operator $y_D^{(2)} (\Box \Phi) \overline{\Psi} \Psi$,
where all fields propagate in the bulk.
\begin{enumerate}

\item[a)] What is the mass dimension of $y_D^{(2)}$?

\item[b)] Writing $y_D^{(2)}$ in units of $\Lambda$, where $\Lambda$
is defined by $y_D$ above, estimate the NDA value of $y_D^{(2)}$ from
a 1-loop computation.

\item[c)] Show that the \textit{tree-level} contribution of
$y_D^{(2)}$ to the $\langle \Phi \Psi \overline{\Psi} \rangle$
3-point function is as large as $y_D$, for $E\sim\Lambda$.

\end{enumerate}
\end{quote}
%%%%%%%%%%%%%%%%%%%%%

Thus, the NDA couplings give an estimate for the ``boundary" between
perturbativity and non-perturbativity. Before giving an efficient
prescription for obtaining the NDA estimates, it will be useful to
look at two more examples.

\medskip
\noindent
\textbf{\underline{Gauge Self-energy and Gauge Coupling}}
\medskip

As was already noted, the gauge couplings become dimensionful for $D
\neq 4$. Consider the 1-loop contribution of a fermion to the gauge
self-energy:
\be
\put(-35,-30){
\resizebox{3.5cm}{!}{\includegraphics{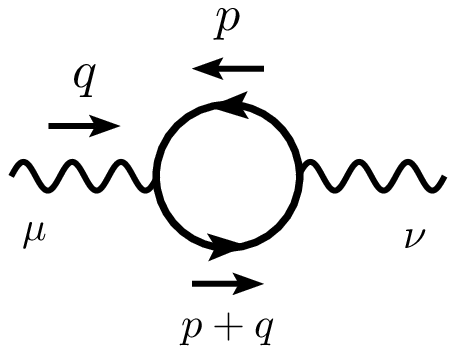}}
}
\hspace*{2cm}
&\sim& (i g_D)^2 \int \! \frac{d^{D}p}{(2\pi)^D} \, {\rm Tr} \left(
\Gamma_\mu \frac{i}{\cancel{p} - M_\Psi}  \, \Gamma_\nu
\frac{i}{(\cancel{p} + \cancel{q}) - M_\Psi} \right)~.
\nonumber
\ee
Superficially, the integral diverges like $\Lambda^{D-2}$. However,
gauge invariance requires that it be proportional to $q^2
\eta_{\mu\nu} - q_\mu q_\nu$. As is well-known, to obtain such a
result it is necessary to use a gauge-invariant regulator, such as
Pauli-Villars regularization.\footnote{Although dimensional
regularization is very convenient in practical computations, by
defining the power-law divergences to zero, it obscures the physical
points that interest us here.} Nevertheless, for power-counting
arguments it is sufficient to extract the above transverse projector,
and estimate the remaining integral with a straightforward momentum
cutoff:
\vspace*{-0.3cm}
\be
\put(-35,-30){
\resizebox{3.5cm}{!}{\includegraphics{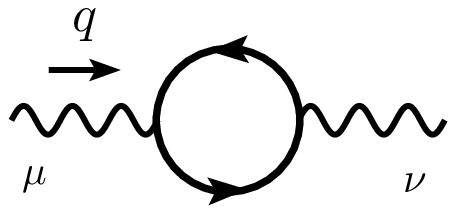}}
}
\hspace*{2cm}
&\sim& (q^2 \eta_{\mu\nu} - q_\mu q_\nu) \, \frac{g_D^2}{l_D} \,
\int^\Lambda \! p^{D-2} dp^2 \, \frac{1}{p^4}
\nonumber \\ [-1.0em]
&\sim& (q^2 \eta_{\mu\nu} - q_\mu q_\nu) \, \frac{g_D^2}{l_D} \,
\frac{1}{D-4} \Lambda^{D-4}~,
\label{GaugeRen}
\ee
where we omitted the factors from the trace (which count the
fermionic degrees of freedom). Adding this to the tree-level
contribution, which is simply $q^2 \eta_{\mu\nu} - q_\mu q_\nu$, we
get the correction factor
\be
1 + {\cal O}(1) \times \frac{1}{l_D} \, g_D^2 \Lambda^{D-4} + \cdots
&\sim& 1 + \frac{c_g^2}{l_D} + \cdots~,
\ee
where we wrote $g_D = c_g / \Lambda^{\frac{1}{2} D -2}$. Thus, the
NDA value is $c_g^{\rm NDA} = l_D^{1/2}$. 
%%%%%%%%%%%%%%%%%%%%%
\begin{quote}

\textbf{\underline{Exercise}:}
For non-abelian groups there are several states in a given
irreducible representation, leading to a counting factor ${\rm
Tr}(T^A T^B) = d_r \delta^{AB}$ [in the standard physics
normalization, $d_F = \frac{1}{2}$ in the fundamental rep.] By
considering the self-energy due to the gauge self-interactions, show
that the correction goes like $N_c c_g^2 / l_D$, where $N_c$ is the
``number of colors". Hence a somewhat better estimate is $c_g^{\rm
NDA} = (l_D/N_c)^{1/2}$. 

\end{quote}
%%%%%%%%%%%%%%%%%%%%%

As we will see, the connection between the (dimensionful) 5D gauge
coupling and the (dimensionless) 4D gauge coupling takes the form
\be
g_4^2 &=& \frac{g_5^2}{L} ~=~ \frac{c_g^2}{\Lambda L}~,
\hspace{2cm}
\textrm{(tree-level matching)}
\ee
where $L$ is the size of the compactified fifth dimension. Since for
the SM gauge couplings [e.g.~$g_s$ for $SU(3)_C$], $g^2_4 \sim {\cal
O}(1)$, we find based on NDA~\footnote{One may want to write $\Lambda
L \sim l_5/(\pi N_c)$ anticipating numerical factors in the direction
suggested by the Pauli-Villars regularization.
\label{PV}}
\be
\Lambda L &\sim& \frac{l_5}{N_c}~,
\ee
where $\Lambda$ is interpreted as the cutoff associated to the 5D
$SU(3)_C$ gauge interactions.
\begin{quote}

\textbf{\underline{Important lesson}:}
the cutoff $\Lambda$ cannot be too far above $1/L$, of order the
compactification scale (in flat space).

\end{quote}
%

%%%%%%%%%%%%%%%%%%%%%
\begin{quote}

\textbf{\underline{Exercise}:}
\begin{enumerate}

\item[a)] Using that, in flat space, $L \sim \pi R$, where $1/R \equiv
m_c$ characterizes the scale of the new 5D physics (beyond the SM),
estimate $LR \sim$ ``number of KK-modes below the cutoff".

\item[b)] Repeat the exercise for arbitrary $D$.

\end{enumerate}
\end{quote}
%%%%%%%%%%%%%%%%%%%%%

\medskip
\noindent
\textbf{\underline{Localized Interactions}}
\medskip

We have already mentioned that realistic extra-dimensional models
contain special subspaces (defects) on which fields, or interactions
that may involve both bulk and localized fields, are localized. The
presence of the defect implies that momentum is not conserved in
certain directions (due to the breaking of translational invariance
by the defect). For instance, consider the operator in 5D:
\be
{\cal L}_5 &\supset& \delta(y-y_0) \, \tilde{y}_5 \Phi
\overline{\Psi} \Psi~,
\ee
where $[\tilde{y}_5] = -3/2$. The presence of the $\delta$-function
implies that when going to momentum space, the $dy$ integration is
trivial and cannot be used to recover the representation of the
momentum-conserving $\delta$-function along the fifth dimension.
Thus, some of the $dp_5$ integrals coming from the $\int \! d^5p
\ldots$ representation of the propagators cannot be ``eliminated" and
remain as part of the loop integration. For example, taking vanishing
external momenta,
%\vspace*{-0.5cm}
%
\be
\put(-35,-28){
\resizebox{3.5cm}{!}{\includegraphics{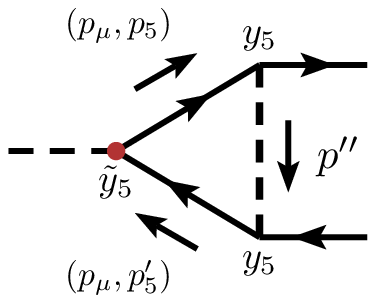}}
}
\hspace*{2cm}
&\sim& (i \tilde{y}_5) (i y_5)^2 \int \! \frac{d^{5}p}{(2\pi)^5} \,
\frac{dp'_5}{(2\pi)} \, \frac{i}{\cancel{p}^{\, \prime} - M_\Psi}  \,
\frac{i}{\cancel{p} - M_\Psi}  \, \frac{i}{p^{\prime\prime 2} -
M_\Phi^2}~.
\nonumber
\label{LocYukawaRen}
\ee
%

%%%%%%%%%%%%%%%%%%%%%
\begin{quote}

\textbf{\underline{Exercise}:}
if $q_1$, $q_2$ and $q_3$ denote the external momenta, what relation
should their fifth components satisfy, using only momentum
conservation at the two $y_5$ (bulk) vertices?

Based on the above, which operator is being renormalized by the above
diagram? Double-check your conclusion by computing the mass dimension
of the above diagram, and comparing against those of $y_5$ and
$\tilde{y}_5$.
\end{quote}
%%%%%%%%%%%%%%%%%%%%%

\noindent
We can now estimate the diagram by power counting:
\be
\frac{1}{2\pi} \, \frac{1}{l_5} \, \tilde{y}_5 y_5^2 \Lambda^2
&\sim&
y_5 \times \frac{1}{2\pi} \, \frac{1}{l_5} \,
\frac{\tilde{c}_y}{\Lambda^{3/2}} \, \frac{c_y}{\Lambda^{1/2}} \,
\Lambda^2~.
\ee
Plugging in $c_y = c_y^{\rm NDA} = \sqrt{l_5}$, and applying the NDA
rules to this diagram, we conclude that
\be
\frac{1}{2\pi} \, \frac{\tilde{c}^{\rm NDA}_y}{\sqrt{l_5}} \sim 1 
\hspace{1cm}
\Longrightarrow
\hspace{1cm}
\tilde{c}^{\rm NDA}_y = 4 \sqrt{6} \pi^{5/2} ~\sim~ 1.1 \times l_4~,
\ee
where $l_4 = 16\pi^2$. The fact that a 4D loop factor is recovered is
not a coincidence.

\medskip
\noindent
\textbf{\underline{NDA Made Trivial}}
\medskip

As we have seen, the game is to use standard dimensional analysis by
expressing all quantities in units of $\Lambda$, as well as keeping
track of the explicit $\pi$-factors from the phase space integrations
(the "naive" part), and maybe also an even more naive estimate to
count the number of states.

Recall that the loop expansion (with an additional
$\frac{1}{16\pi^2}$ loop factor per loop in 4D) follows from keeping
track of $\hbar$, which appears as an \textit{overall} factor in
front of the action in the path-integral formulation of QFT. The same
argument suggests that we write~\cite{Chacko:1999hg}
\be
{\cal L}_D &=& \frac{N}{l_D} {\cal L}_{\rm Bulk}(A_M, \Phi, \Psi) +
\frac{N}{l_4}  \delta^{(D-4)}(y-y_0) \, {\cal L}_{\rm Brane}(\Phi,
\phi, \ldots)~, 
\label{LNDA}
\ee
where \textit{all} the couplings in ${\cal L}_{\rm Bulk}$ and ${\cal
L}_{\rm Brane}$ are expressed in units of $\Lambda$ with
\textit{coefficients of order one} (except for well-understood
combinatoric factors when several identical fields are involved). For
instance,
\be
{\cal L}_D &=& \frac{N}{l_D} 
\left\{
- \frac{1}{4} \Lambda^{D-4} F_{MN} F^{MN} + i \overline{\Psi}
\cancel{D} \Psi + | D_M \Phi |^2 + \Lambda^{2-\frac{1}{2}D} \Phi
\overline{\Psi} \Psi + \cdots
\right\}
\nonumber \\ [0.3em]
&& \mbox{} +  \frac{N}{l_4}  \delta^{(D-4)}(y-y_0) \, 
\left\{  
| D_\mu \phi |^2 + \cdots + \Lambda^{4-D} \phi \overline{\Psi} \Psi +
\Lambda^{6-\frac{3}{2}D} \Phi \overline{\Psi} \Psi 
 + \cdots 
\right\}~.
\label{LDNDA}
\ee
If there are self-interactions for a real scalar, we would write
e.g.~$\frac{1}{n!} \Lambda^k \phi^n$, where $k$ depends on whether
this is a bulk or localized interaction/field. 

The idea behind Eq.~(\ref{LNDA}), borrowed from the semi-classical
expansion, is that the explicit factors in front of ${\cal L}_{\rm
Bulk}$ and ${\cal L}_{\rm Brane}$ will precisely cancel the loop
factors appearing at each order in the loop expansion, thus realizing
the ``strong coupling" regime discussed above.  Of course, it is
possible that some of the operators have smaller coefficients than
above.  The aim here is simply to characterize the maximum size of
those coefficients.~\footnote{If we include in Eq.~(\ref{LNDA}) an
overall factor of $1/\epsilon$, for $\epsilon <1$, we can characterize
a situation where each additional loop brings a suppression of
$\epsilon$ in processes with $E \sim \Lambda$.} Regarding the factor
of $N$, the cancellation need not work perfectly since the dependence
on the specific field content of the theory is more complicated than
can be parameterized above.  Nevertheless, the factors of $N$ give us
a rough idea of how the multiplicity of states affects the results.

Note that we chose to write a 4D defect simply described by a
$\delta$-function, and that we write a 4D loop factor in front (the
generalization to defects of other dimensionalities is
straightforward).  The 4D loop factor is easily understood for
processes that involve only localized fields and interactions (works
identically to $1/l_D$ in front of ${\cal L}_{\rm Bulk}$).  As we will
see, it also reproduces our less trivial result involving both bulk
and localized interactions (and works more generally).

In order to use the NDA parametrization given in Eq.~(\ref{LNDA}), we
first normalize the fields canonically:
\be
\Psi &\to& \left( \frac{l_D}{N} \right)^{1/2} \Psi~,
\nonumber \\
\Phi &\to& \left( \frac{l_D}{N} \right)^{1/2} \Phi~, 
\\
\phi &\to& \left( \frac{l_4}{N} \right)^{1/2} \phi~,
\nonumber
\ee
Although one can also normalize canonically the gauge field, it is
more useful here to keep conventions with $D_M = \partial_M + i A_M$,
so that $[A_M] = 1$ and the gauge coupling can be read from the
coefficient of the gauge kinetic term, as in Eq.~(\ref{LD}). Then the
Lagrangian in Eq.~(\ref{LDNDA}) becomes
\be
{\cal L}_D &=& 
- \frac{1}{4} \, \frac{N}{l_D}  \Lambda^{D-4} F_{MN} F^{MN} + i
\overline{\Psi} \cancel{D} \Psi + | D_M \Phi |^2 + \left(
\frac{l_D}{N} \right)^{1/2}  \Lambda^{2-\frac{D}{2}} \Phi
\overline{\Psi} \Psi + \cdots
\nonumber \\ [0.3em]
&& \mbox{} + \delta^{(D-4)}(y-y_0) \, 
\left\{
| D_\mu \phi |^2 + \left( \frac{l_4}{N} \right)^{1/2}  \frac{l_D}{N}
\, \Lambda^{4-D} \phi \overline{\Psi} \Psi + \frac{l_D}{l_4} \left(
\frac{l_D}{N} \right)^{1/2} \Lambda^{6-\frac{3D}{2}} \Phi
\overline{\Psi} \Psi 
 + \cdots 
 \right\}~,
\nonumber
\ee
from which we can immediately read
\be
\begin{array}{lcl}
\displaystyle 
g_D^2 = \frac{l_D/N}{\Lambda^{D-4}}~,  & \hspace{1cm} &  
\displaystyle 
y_D = \left( \frac{l_D}{N} \right)^{1/2} \, \Lambda^{2-\frac{1}{2}
D}~,    \\ [1.5em]
\displaystyle 
y'_D = \left( \frac{l_4}{N} \right)^{1/2} \frac{l_D}{N} \,
\Lambda^{4-D}~,  & \hspace{1cm} &
\displaystyle 
\tilde{y}_D = \frac{l_D}{l_4} \left( \frac{l_D}{N} \right)^{1/2} \,
\Lambda^{6-\frac{3}{2}D}~.
\end{array}
\ee
%
%%%%%%%%%%%%%%%%%%%%%
\begin{quote}

\textbf{\underline{Exercise}:}
\begin{enumerate}

\item[a)] Compare to the previous results and convince yourself that
this works.

\item[b)] Choose a 1-loop example that renormalizes $y'_D$, and check
the above NDA estimate.

\end{enumerate}
\end{quote}
%%%%%%%%%%%%%%%%%%%%%

\medskip
\noindent
\textbf{\underline{What have we achieved?}}
\medskip

\noindent
We have emphasized that higher-dimensional theories have an intrinsic
cutoff $\Lambda$, and that some observables are very sensitive to the
physics at and above $\Lambda$, and are therefore incalculable within
the extra-dimensional model.  One should simply include in the theory
the corresponding local operators, with coefficients \textit{to be
determined by experiment}.  In addition, the simple considerations
described above allow us to
\begin{itemize}

\item Estimate the largest that the cutoff can be (in a given model).

\item Estimate the maximum (and perhaps, expected) size of the
coefficients of the tower of operators consistent with the assumed
symmetries.

\end{itemize}
The above situation is not intrinsically different from other 4D
effective theories, but it is important to keep in mind the
limitations imposed by the existence of a cutoff relatively close to
the scale of the new extra-dimensional physics. In many models of
interest, the SM interactions are promoted to higher-dimensional bulk
interactions. To estimate the cutoff, we can then focus on the
strongest of these, namely the $SU(3)_C$ and top Yukawa interactions.
In flat 5D space, we then expect that the number of KK-modes below
the cutoff, $N_{\rm KK} \lesssim 24\pi/N_c \sim 8\pi < {\cal O}(30)$,
where we used the estimate suggested by a Pauli-Villars
regularization (see footnote~\ref{PV}). We will see later that this
is reduced in the presence of 5D curvature (even in flat space, other
considerations based on unitarity can put more stringent bounds,
see~\cite{Chivukula:2003kq}). 

\medskip
\noindent
\textbf{\underline{Extra Dimensions and the Hierarchy Problem}}
\medskip

The previous discussion also highlights a possible application to
particle physics.  With a weakly coupled SM(-like) Higgs boson, one
faces a potential naturalness problem, associated with the quadratic
sensitivity of the Higgs mass parameter to ultra-short distance
physics.  This suggests that there should exist a physical cutoff at
or near the TeV scale, such that the Higgs quadratic divergences can
be under theoretical control, and hence the EW scale can be understood
without fine-tuned cancellations within the (unknown) high-energy
contributions.

Having extra-dimensions at the TeV scale, provides a strong rationale
for why a cutoff should exist near that scale.  Although this
observation by itself does not fully address the Higgs hierarchy
problem, it does provide a possible starting point for a potentially
deeper, and perhaps more satisfactory, understanding of the
naturalness issue.  Possible additional features one would like to
have in this context are:
\begin{enumerate}

\item A mechanism that selects the compactification scale to be near
the TeV scale. If one regards other higher scales (such as the Planck
scale) as fundamental, one would like to ``generate" the
compactification scale dynamically.

\item A dynamical connection between the compactification
scale/extra-dimensional physics, and the breaking of the EW symmetry. 

\end{enumerate}

Note, however, that even in the absence of such theoretical
``goodies", if an extra-dimensional structure at the TeV scale was
discovered experimentally, we would be forced to accept the existence
of a cutoff not far above. In this sense, the big Planck/weak scale
hierarchy problem of the 4D SM would become less urgent to fully
solve: the priority would be to understand the potential ``little
hierarchy" between the weak scale and the cutoff associated with the
extra-dimensional physics.

In the above sense, generic extra-dimensional models (not necessarily
addressing points 1.~and 2.~above) can be regarded as motivated by the
hierarchy problem, provided the compactification scale is near the
weak scale, similar to the SUSY solution to the hierarchy problem when
the superpartner masses are around a TeV (although, of course, there
are significant differences between the two approaches).  This
provides our motivation for focusing on TeV scale extra dimensions in
these lectures.

\medskip
\noindent
\textbf{\underline{Additional comments}}
\medskip

Our NDA discussion was framed in the language of uncompactified
extra-dimensional field theories, even though we know that any
extra-dimension would have to be compactified.  Technically, the
distinction is contained in the replacement (in flat 5D, for
illustration)
\be
\int \! \frac{dp_5}{2\pi}  &\longrightarrow& \frac{1}{2\pi R} \sum_n~,
\ee
where $\Delta p_5 = 1/R$. Since the compact extra dimension is
different from the four non-compact ones, strictly speaking we cannot
assume spherical symmetry as we did when we computed $\Omega_D$, the
$D$-dimensional solid angle. Nonetheless, our interest was mainly in
the UV contributions (which is what NDA is useful for). Thus, to the
extent that $\Lambda R \gg 1$, the discrete sums can be approximated
by integrals, and we conclude that, up to ${\cal O}(1)$ factors, our
discussion carries over to the compact case without change.

We further assumed that the background was exactly flat. Non-zero
curvature is an intrinsic feature of many extra-dimensional models.
Nevertheless, again the fact that we were interested in ultra-short
distance physics allows us to extend our discussion to warped
backgrounds. All we need is that the curvature $k$ be small compared
to the cutoff $\Lambda$. In this case it is intuitively clear (and
can be checked in explicit examples) that curvature effects play a
negligible role in the NDA discussion. Note that the hierarchy $k \ll
\Lambda$ is typically a tacit assumption, or else we would have
little control over the gravitational background. 

The most important difference between flat and warped backgrounds is
the relation between the compactification scale, $m_{\rm KK}$, and
the ``size" of the extra dimension, e.g.~as measured with a meter
stick (if we could do that):
\be
\begin{array}{lclcl}
\textrm{ Flat space:}          & \hspace{5mm} & 
\displaystyle 
m_{\rm KK} = \frac{1}{R} ~=~\frac{\pi}{L}~,   \\ [0.8em]
\begin{array}{l}
\textrm{Warped space:} \\
\textrm{(e.g.~Randall-Sundrum)}
\end{array}   & \hspace{5mm} & 
\displaystyle 
m_{\rm KK} \sim k \,e^{-kL}  & & 
\displaystyle 
\textrm{if}~kL \gg 1~.
\end{array}
\nonumber
\ee
In both cases, as argued above, one has that $\Lambda L \sim l_5 /
(\pi N_c)$, but the number of KK modes below $\Lambda$ read
\be
\begin{array}{lcl}
\textrm{Flat space:}          & \hspace{5mm} & 
\displaystyle 
N_{\rm KK} = \Lambda R ~=~\frac{l_5}{\pi^2 N_c}~,   \\ [1em]
\textrm{Warped space:}
& \hspace{5mm} & 
\displaystyle 
N_{\rm KK} = \frac{\Lambda}{k} ~=~ \frac{1}{kL} \, \frac{l_5}{\pi
N_c}~.
\end{array}
\nonumber
\ee
The fact that in the warped case the number of KK modes below the
cutoff is given by $\Lambda/k$ will become clear when we have
developed the necessary technology. The point here is that the
suppression factor from $1/kL$ can be significant, and therefore one
expects far fewer KK modes before reaching the strong-coupling regime
in warped scenarios than in flat space models.

\newpage 

%%%%%%%%%%%%%%%%%%%%%%%%%%%%%%%%%%%%%%%
\section{Lecture 2: The Tools of the Trade}
\label{Tools}
%%%%%%%%%%%%%%%%%%%%%%%%%%%%%%%%%%%%%%%

In this lecture we will collect a number of basic results that will
allow us to discuss specific extra-dimensional models.  It will be
worthwhile doing it in some generality, that can be specialized
according to the cases of interest.

We have already mentioned that the compactification of the extra
dimensions must be accompanied by certain ``singularities" or
``defects".  This is essential to obtain models that are chiral at low
energies (e.g.~the LH and RH electrons have different quantum
numbers).  A useful construction is called a ``Field theory orbifold".
However, in 5D, a more straightforward and somewhat more general
prescription is to compactify on an \textit{interval}, i.e.
\be
ds^2 &=& g_{MN} \, dx^M dx^N
\nonumber \\ [0.3em]
&=& e^{-2A(y)} \eta_{\mu\nu} dx^\mu dx^\nu - dy^2~,
\label{metric}
\ee
where $y \in [0,L]$, and $A(y)$ is arbitrary at this point (although
we can always rescale the $x^\mu$ to require $A(0) = 0$, which we
shall do). Thus, one assumes that the extra dimension is a simply
connected compact space. Our ansatz for the line element is the most
general one exhibiting 4D Lorentz invariance. The ``flat 4D sections"
correspond to the vanishing of the \textit{effective} 4D cosmological
constant. The $y$-coordinate is chosen here as the \textit{proper
distance along the extra dimension}. If $A(y) = $const then the 5D
spacetime is flat; otherwise we say it is \textit{warped}.

%%%%%%%%%%%%%%%%%%%%%%%%%%%%%%%%%%%%%%%
\subsection{Boundary Conditions}
\label{BCs}
%%%%%%%%%%%%%%%%%%%%%%%%%%%%%%%%%%%%%%%

An important point to realize is that when the spacetime has a
boundary at a finite distance, the theory is not fully
\textit{defined} until one specifies the boundary conditions (b.c.'s).
In infinite spacetime, the issue of boundary conditions is often
implicit because there is a natural choice: the fields should vanish
at infinity ``sufficiently fast". This corresponds to the physical
notion that we are dealing with \textit{local} physics, and the same
prescription will be used for the 4 non-compact dimensions in our
case. However, there can be several boundary conditions at $y = 0$
and $y = L$ that are physically acceptable. 
Nevertheless, not quite anything is allowed: the boundary conditions
should reflect the fact that \textit{nothing escapes the interval,
just because nothing is supposed to exist outside}. 

A general prescription for finding the b.c.'s (in the above sense)
was described in detail in the ``TASI Lectures on EWSB and Extra
Dimensions" by Cs\'aki, Hubisz and Meade~\cite{Csaki:2005vy}. We will
briefly review the idea, and refer the student to those lectures for
further details.

For simplicity, consider a real scalar field:
\be
S &=& \int \! d^5x \sqrt{g} \, \left\{ \frac{1}{2} g^{MN} \partial_M
\Phi \partial_N \Phi - V(\Phi) \right\}~,
\ee
where it is assumed that the metric background [i.e.~the function
$A(y)$] is specified. Under a variation $\delta \Phi$, one finds
\be
\delta S &=& \delta S_V + \delta S_S~,
\ee
where
\be
\delta S_V &=& - \int \! d^5x \sqrt{g} \, \left\{ \frac{1}{\sqrt{g}}
\partial_M \left[ \sqrt{g} \, g^{MN} \partial_N \Phi \right] +
\frac{\partial V}{\partial \Phi} \right\} \delta \Phi~,
\ee
is a \textit{volume term}, and
\be
\delta S_S &=& \int \! d^4x \sqrt{g} \, g^{5N} \partial_N \Phi \cdot
\left.\delta \Phi \rule{0mm}{3.5mm} \right|_{y=0}^{y=L}~,
\ee
is a \textit{surface term} arising from the integration by parts. We
assumed that $\Phi \to 0$ as $x^\mu \to \pm \infty$. 
We recognize $\delta S = 0$ as the variational principle that allow
us to derive the equations of motion (EOM) for $\Phi$, provided the
b.c.'s are such that the surface term vanishes.

The idea then is to impose, for arbitrary $\delta \Phi$,
\be
\delta S_V = 0
\hspace{1cm} \& \hspace{1cm}
\delta S_S = 0~.
\label{VariationalPrinciple}
\ee
The first condition leads to the EOM
\be
\frac{1}{\sqrt{g}} \partial_M \left[ \sqrt{g} \, g^{MN} \partial_N
\Phi \right] + V'(\Phi) &=& 0~,
\ee
which in our background reads
\be
e^{2A} \partial_\mu \partial^\mu \Phi - e^{4A} \partial_y \left[
e^{-4A} \partial_y \Phi \right] + V' &=& 0~.
\label{ScalarEOM}
\ee
Note that here $\partial_\mu \partial^\mu$ is understood to be
contracted with the Minkowski metric.  The second condition in
Eq.~(\ref{VariationalPrinciple}) defines the allowed set of b.c.'s (in
our diagonal background):
\be
- \left. \partial_y \Phi \, \delta \Phi \rule{0mm}{3.5mm} \right|_{y
= L} +
\left. \partial_y \Phi \, \delta \Phi \rule{0mm}{3.5mm} \right|_{y =
0} &=& 0~.
\ee
One way to satisfy this condition is to impose periodicity: $\Phi(L)
= \Phi(0)$ and $\Phi'(L) = \Phi'(0)$. However, this defines the
compactification on a circle, which is a smooth manifold and leads to
a \textit{vector-like} low-energy theory, as we will soon see when we
describe bulk fermions. We therefore discard this option as
uninteresting for our phenomenological applications.

Other simple possibilities are to impose Neumann or Dirichlet
boundary conditions:
\be
\begin{array}{lcrcl}
\textrm{Neumann (N):}          & \hspace{5mm} & 
\displaystyle 
\left. \partial_y \Phi \rule{0mm}{3.5mm} \right| &=& 0~,   \\ [1em]
\textrm{Dirichlet (D):}
& \hspace{5mm} & 
\displaystyle 
\left. \Phi \rule{0mm}{3.5mm} \right| &=& 0~,
\end{array}
\nonumber
\ee
in various combinations, that we denote by
\be
\begin{array}{lcl}
\displaystyle 
(+,+)          & = & 
\displaystyle 
(\textrm{N at } y = 0, \textrm{N at } y = L)~,   \\ [1em]
\displaystyle 
(+,-) 
& = & 
\displaystyle 
(\textrm{N at } y = 0, \textrm{D at } y = L)~,   \\ [1em]
\displaystyle 
(-,+) 
& = & 
\displaystyle 
(\textrm{D at } y = 0, \textrm{N at } y = L)~,   \\ [1em]
\displaystyle 
(-,-) 
& = & 
\displaystyle 
(\textrm{D at } y = 0, \textrm{D at } y = L)~.
\end{array}
\nonumber
\ee
The upshot, from a technical point of view, is that 
\begin{itemize}

\item We can freely integrate by parts and discard the surface terms
in all our manipulations.

\item The EOM with the allowed b.c.'s defines a \textit{self-adjoint}
eigenvalue problem:

\begin{enumerate}

\item[a)] Real eigenvalues

\item[b)] Orthogonal eigenfunctions

\end{enumerate}

\end{itemize}

\noindent
These properties turn out to be extremely convenient in the
manipulations and physical interpretation that follow.  From a physics
perspective, the above properties allow us to define
\textit{conserved} charges, i.e.~no flow outside $[0,L]$.  In this
sense, there is really nothing outside the interval!

Nonetheless, N or D boundary conditions are not the most general
ones. Sometimes one finds \textit{mixed} b.c.'s, e.g.
\be
\left. \partial_y \Phi \rule{0mm}{3.5mm} \right|_{y=0} - \left. m
\Phi \rule{0mm}{3.5mm} \right|_{y=0} &=& 0~.
\ee
Typically, this indicates that there is a \textit{source} on the
boundary. For instance, adding a boundary term to the action
\be
\Delta S_{\rm boundary} &=& - \int \! d^4 x \left. \sqrt{\bar{g}} \,\, 
\frac{1}{2} \, m \Phi^2 \right|_{y=0}
\ee
leads to a contribution to $\delta S_S$ that results in the mixed N/D
boundary condition above.  Note that we use the induced metric,
denoted by $\bar{g}_{\mu\nu}$, to write the boundary terms.  In
Section~\ref{Holography:Scalars} we provide expressions for the case
with the most general quadratic boundary terms.  What matters is that,
as long as the b.c.'s can be derived from the variational procedure
described above, we are guaranteed to have the mathematical properties
that lead to sensible physics.

%%%%%%%%%%%%%%%%%%%%%%%%%%%%%%%%%%%%%%%
\subsection{Fermions in 5D}
\label{Fermions}
%%%%%%%%%%%%%%%%%%%%%%%%%%%%%%%%%%%%%%%

The description of 5D fermions proceeds as follows.  First, we need
\textit{five} anticommuting Dirac $\Gamma$-matrices, for which we can
take
\be
\Gamma^A &=& (\gamma^\alpha, -i \gamma_5)~,
\hspace{2cm}
\alpha = 0,1,2,3
\ee
where, in the Weyl representation,
\be
\gamma^\mu &=&
\left(
\begin{array}{cc}
              0                & \sigma^\mu   \\
\bar{\sigma}^\mu  &       0       
\end{array}
\right)~,
\hspace{2cm}
\begin{array}{rcl}
 \sigma^\mu & = & (\mathds{1}_{2\times 2}, \vec{\sigma})~,  \\ [0.5em]
 \bar{\sigma}^\mu & = & (\mathds{1}_{2\times 2}, -\vec{\sigma})~,
\end{array}
\nonumber \\ [0.5em]
\gamma_5 &=&
\left(
\begin{array}{cc}
  - \mathds{1}_{2\times 2}     &                   0
\\
               0                                &
\mathds{1}_{2\times 2}       
\end{array}
\right)~,
\nonumber
\ee
which obey $\{ \Gamma^A, \Gamma^B \} = 2 \eta^{AB}$. Here
$\vec{\sigma}$ stands for the three Pauli matrices, while $\gamma_5$
is the 4D chirality operator. It is also useful to define the
projectors on 4D chirality, $P_{L,R} = \frac{1}{2} (1 \mp \gamma_5)$. 

One also needs the f\"unfbein, $e_M^{\,\,\,\,\,A}$ defined by
\be
g_{MN} = e_M^{\,\,\,\,\,A} e_N^{\,\,\,\,\,B} \eta_{AB}~,
\ee
and the inverse f\"unfbein $e^M_{\,\,\,\,\,A}$ obeying
$e^M_{\,\,\,\,\,A} e_M^{\,\,\,\,\,B} = \delta^B_A$. In our
background, the latter is
\be
e^\mu_{\,\,\, \alpha} &=& e^{+ A(y)} \delta^\mu_\alpha~,
\hspace{1.5cm}
e^y_{\,\,\,5} ~=~ + 1~, 
\ee
with all other components vanishing. 

In general backgrounds one also needs a spin connection to define the
covariant derivative w.r.t.~both general coordinate and local Lorentz
transformations. In our case, the covariant derivative, $D_M =
\partial_\mu + \frac{1}{8} \, \omega_{MAB} [\Gamma^A, \Gamma^B]$,
takes the form~\footnote{For completeness, recall that
$\omega_M^{\,\,\,\,AB} = e_N^{\,\,\,\,A} (\partial_M e^{NB} + e^{SB}
\Gamma^N_{SM})$, where the Christoffel symbols are $\Gamma^{K}_{MN} =
\frac{1}{2} g^{KL} \left\{ \partial_N g_{LM} + \partial_M g_{LN} -
\partial_L g_{MN} \right\}$. Here the only non-vanishing components
are $\omega_\mu^{\,\,\,a5} = - \omega_\mu^{\,\,\,5a} =
\partial_5(e^{-A}) \delta^a_\mu$.}
\be
D_\mu &=& \partial_\mu - \frac{i}{2} \, e^{-A} A' \gamma_\mu
\gamma_5~,
\hspace{1.5cm}
D_5 ~=~ \partial_5~.
\ee

The (explicitly hermitian) action for a free 5D fermion reads
\be
S_{\Psi} &=&
\int\! d^5x \sqrt{g} \left\{ \frac{i}{2}
\overline{\Psi} e^M_{\,\,\,A} \Gamma^{A} D_{M} \Psi - \frac{i}{2}
(D_{M}\Psi)^\dagger \Gamma^0 e^M_{\,\,\,A} \Gamma^{A} \Psi - M
\overline{\Psi} \Psi \right\}~.
\label{FermionAction}
\ee
The determination of the allowed b.c.'s proceeds via the variational
principle described in the scalar case. Here the most direct
procedure is to replace first the metric background, noting that the
spin connection cancels out in the action and we might as well
replace $D_M \to \partial_M$:
\be
S_{\Psi} &=&
\int\! d^5x \, e^{-3A} \left\{
i \, \overline{\Psi} \gamma^{\mu} \partial_{\mu} \Psi + \frac{1}{2}
e^{-A}
\left[
\overline{\Psi} \gamma_5 \partial_{5} \Psi - (\partial_5
\overline{\Psi}) \gamma_{5} \Psi \right] - 
e^{-A} M \overline{\Psi} \Psi \right\}~,
\label{FermionAction2}
\ee
where we integrated by parts along $x^\mu$ (but not $y$) and dropped
the corresponding terms, assuming that the field vanishes at
infinity. Now we can proceed with the variation w.r.t.~$\Psi$. For
instance, under $\delta \overline{\Psi}$:
\be
\delta S^V_{\Psi} &=&
\int\! d^5x \, e^{-3A} \delta \overline{\Psi} \left\{
i \, \gamma^{\mu} \partial_{\mu} \Psi + e^{-A} \gamma_5 \partial_{5}
\Psi -  \frac{1}{2} A'  e^{-A} \gamma_{5} \Psi - 
e^{-A} M \Psi \right\}~,
\ee
while the integration by parts along $x^5 = y$ generates the surface
term
\be
\delta S^S_{\Psi} &=& -
\int\! d^4x \left.  e^{-4A} \delta \overline{\Psi} \gamma_{5} \Psi
\right|_{y=0}^{y=L}~.
\ee
Our prescription of requiring $\delta S^V_{\Psi} = \delta S^S_{\Psi}
= 0$ under any $\delta \overline{\Psi}$ leads to the EOM
\be
\left\{ i \, e^{A}  \gamma^{\mu} \partial_{\mu} + \left( \partial_{5}
-  \frac{1}{2} A'  \right) \gamma_{5} - 
M \right\} \Psi &=& 0~,
\label{EOMFermion}
\ee
and to the requirement
\be
- \left. \delta \overline{\Psi} \gamma_{5} \Psi \right|_{y=L} + 
\left. \delta \overline{\Psi} \gamma_{5} \Psi \right|_{y=0} &=& 0~.
\ee
It is useful to express this in terms of $\Psi_{L,R} \equiv P_{L,R}
\Psi$:
\be
\left. \delta \overline{\Psi}_L \Psi_R - \delta \overline{\Psi}_R
\Psi_L
\right|_{y=0}^{y = L} &=& 0~.
\label{GeneralBCFermion}
\ee
We can now be more explicit about the relation between chirality and
compactification:
\begin{enumerate}

\item[i)] If we were to impose periodic b.c.'s (compactification on a
circle)
\be
\left. \Psi_L \rule{0mm}{3.5mm} \right|_{y=L} &=& \left. \Psi_L
\rule{0mm}{3.5mm} \right|_{y=0}
\hspace{1cm} \& \hspace{1cm}
\left. \Psi_R \rule{0mm}{3.5mm} \right|_{y=L} ~=~ \left. \Psi_R
\rule{0mm}{3.5mm} \right|_{y=0}~,
\ee
there would be nothing that distinguishes between the two
chiralities, L and R.

\item[ii)] If, on the other hand, we treat the two boundaries in our
general condition (\ref{GeneralBCFermion}) separately (interval
compactification), we see that we can use
\be
\left. \Psi_L \rule{0mm}{3.5mm} \right| &=& 0
\hspace{1cm} \textrm{\underline{or}} \hspace{1cm}
\left. \Psi_R \rule{0mm}{3.5mm} \right| ~=~ 0~.
\ee
Either one of these (with the four possible combinations at the two
boundaries) is sufficient to ensure $\delta S^S_{\Psi} = 0$. In fact,
we do not have the right to impose Dirichlet b.c.'s conditions on
\textit{both} chiralities at a given boundary!

Indeed, recalling the EOM (\ref{EOMFermion}), which can be split into
L and R as
\be
i \, e^{A}  \gamma^{\mu} \partial_{\mu} \Psi_L + \left[ \left(
\partial_{5} -  \frac{1}{2} A'  \right) - 
M \right] \Psi_R  &=& 0~,
\\ [0.5em]
i \, e^{A}  \gamma^{\mu} \partial_{\mu} \Psi_R + \left[ - \left(
\partial_{5} -  \frac{1}{2} A'  \right) - 
M \right] \Psi_L  &=& 0~,
\ee
we see that, if $\left. \Psi_L \rule{0mm}{3.5mm} \right| = 0$, then
the first equation implies
\be
\left. \partial_5 \Psi_R  \rule{0mm}{3.5mm} \right| &=& 
\left. \left( \frac{1}{2} A'  + M \right) \Psi_R \right|~.
\ee
If we were to also impose $\left. \Psi_R \rule{0mm}{3.5mm} \right| =
0$, we would automatically have $\left. \partial_5 \Psi_R
\rule{0mm}{3.5mm} \right| = 0$, and then --from the second equation--
we would find $\left. \partial_5 \Psi_L \rule{0mm}{3.5mm} \right| =
0$. But then the only allowed solution of the system of two
first-order differential equations (in $y$) is $\Psi \equiv 0$.

Thus, we settle on two interesting possibilities:
\be
\begin{array}{llclcrcl}
\displaystyle 
& (-)         & \equiv & 
\displaystyle 
\left. \Psi_L \rule{0mm}{3.5mm} \right| = 0~,   
& \textrm{hence} &
\left. \partial_5 \Psi_R  \rule{0mm}{3.5mm} \right| &=& 
\left. \left( \frac{1}{2} A'  + M \right) \Psi_R \right|~,
\\ [0.5em]
\textrm{\underline{or}}
\\ [0.5em]
\displaystyle 
& (+)         & \equiv & 
\displaystyle 
\left. \Psi_R \rule{0mm}{3.5mm} \right| = 0~,   
& \textrm{hence} &
\left. \partial_5 \Psi_L  \rule{0mm}{3.5mm} \right| &=& 
\left. \left( \frac{1}{2} A'  - M \right) \Psi_L \right|~.
\end{array}
\nonumber
\ee
[Note the sign flip on the mass term]. This is at the basis of the
advertised result: \textit{compactification on an interval
necessarily leads to boundary conditions that distinguish L from R},
which will allow us to easily embed the SM structure. 

Similar to the scalar case, specifying the boundary conditions on
\textit{both} boundaries  leads to four possibilities that we label
as 
\be
\begin{array}{lcclccl}
\displaystyle 
(+,+) & &
(+,-) & &
(-,+) & &
(-,-)
\end{array}~.
\nonumber
\ee
Also, as in the scalar case, one can generalize these boundary
conditions by including localized terms in the action (via a
$\delta$-function, i.e.~a boundary term), which contribute directly
to $\delta S^S_{\Psi}$.

\end{enumerate}
%

%%%%%%%%%%%%%%%%%%%%%%%%%%%%%%%%%%%%%%%
\subsection{5D Gauge Fields}
\label{Gauge}
%%%%%%%%%%%%%%%%%%%%%%%%%%%%%%%%%%%%%%%

The case of gauge fields, which will be the last case we will review
here, involves a new ingredient, the need for \textit{gauge fixing}:
\be
S_{A} &=& \int \! d^5x \sqrt{g} \left\{  -\frac{1}{4}
  g^{MN} g^{KL} F_{MK} F_{NL}  \right\} + S_{\rm GF}~,
\label{GaugeAction}
\ee
where the gauge-fixing term $S_{\rm GF}$ will be specified shortly.
In this section we will be interested in the quadratic part of the
action, so that there is no distinction between the abelian and
non-abelian cases, and we can take $F_{MN} = \partial_M A_N -
\partial_N A_M$.  To motivate the choice of $S_{\rm GF}$ let us expand
the gauge kinetic term in our background, Eq.~(\ref{metric}):
\be
%& & \int \! d^5x \, e^{-4A} \left\{
%-\frac{1}{4} \, e^{4A}  F_{\mu\nu} F^{\mu\nu} + \frac{1}{2} \,
%e^{2A} F_{\mu5} F^{\mu}_5
%\right\}
%\nonumber \\ [0.5em]
%&=&
& &
\int \! d^5x \, \left\{
-\frac{1}{4} F_{\mu\nu} F^{\mu\nu} + \frac{1}{2} \, e^{-2A}
\partial_{\mu} A_5 \partial^{\mu} A_5
- \partial_{5} \! \left[ e^{-2A} A_5 \right] \partial_{\mu} A^\mu
+ \frac{1}{2} \, e^{-2A} \partial_{5} A_\mu \partial_{5} A^\mu
\right\}~,
\ee
where all the contractions are now done with the Minkowski metric. We
also integrated by parts in both $x^\mu$ and $y$ and dropped the
terms at infinity, as usual. However, we must keep track of the
generated surface term in the $y$ direction:
\be
S_S &=&
\int \! d^4x \left. e^{-2A} A_5 \partial_{\mu} A^\mu
\right|_{y=0}^{y=L}~.
\ee
We see that there is a bulk term mixing the $A_\mu$ and $A_5$
components, which also results in mixing in the EOM. The gauge-fixing
action is chosen so as to cancel this term~\cite{Randall:2001gb},
specifically:
\be
S_{\rm GF} &=& \int \! d^5x -\frac{1}{2\xi} \left\{
\partial_\mu A^\mu - \xi \, \partial_5 \left[ e^{-2A} A_5 \right]
\right \}^2~,
\label{SS}
\ee
where $\xi$ is an arbitrary \textit{gauge-fixing parameter}. The
total gauge action then becomes
\be
S_A &=& \int \! d^5x \, \left\{
-\frac{1}{4} F_{\mu\nu} F^{\mu\nu} -\frac{1}{2\xi} \left(
\partial_\mu A^\mu \right)^2 +
\frac{1}{2} \, e^{-2A}  \partial_{5} A_\mu \partial_{5} A^\mu
\right.
\nonumber \\ [0.5em]
&& \hspace{1.5cm} \left. \mbox{} + \frac{1}{2} \, e^{-2A}
\partial_{\mu} A_5 \partial^{\mu} A_5
- \frac{1}{2} \, \xi \left( \partial_5 \left[ e^{-2A} A_5 \right]
\right)^2
\right\}~,
\ee
together with the surface term, Eq.~(\ref{SS}). We may now apply the
variational procedure.

\medskip
\underline{Under $\delta A_\mu$}:
\be
\delta S^V_A &=& \int \! d^5x \, \delta A_\mu \left\{
\left[ \eta^{\mu\nu} \Box - \left( 1 - \frac{1}{\xi} \right)
\partial^\mu \partial^\nu \right]
-\eta^{\mu\nu} \partial_5 \, e^{-2A} \partial_5
\right\} A_\nu~,
\nonumber \\ [0.5em]
\delta S^S_A &=& \int \! d^4x \, e^{-2A} \delta A_\mu \left. \left\{
\partial_5 A^\mu - \partial^\mu \! A_5
\right\} 
\rule{0mm}{3.5mm} \right|_{y=0}^{y=L}~,
\nonumber
\ee
where $\Box = \eta^{\mu\nu} \partial_\mu \partial_\nu$, and we took
into account the variation of Eq.~(\ref{SS}).

\medskip
\underline{Under $\delta A_5$}:
\be
\delta S^V_A &=& \int \! d^5x \, e^{-2A} \delta A_5 \left\{
-\Box + \xi \, \partial^2_5 e^{-2A}
\right\} A_5~,
\nonumber \\ [0.5em]
\delta S^S_A &=& - \int \! d^4x \, e^{-2A} \delta A_5 \left. \left\{
\xi \, \partial_5 \left[ e^{-2A} A_5 \right] + \partial_\mu A^\mu
\right\} 
\rule{0mm}{3.5mm} \right|_{y=0}^{y=L}~.
\nonumber
\ee
Therefore, the EOM are:
\be
0 &=& \left[ \eta^{\mu\nu} \Box - \left( 1 - \frac{1}{\xi} \right)
\partial^\mu \partial^\nu \right] A_\nu
- \partial_5 \! \left[ e^{-2A} \partial_5 A^\mu \right]~,
\\ [0.5em]
0 &=& \Box A_5 - \xi \, \partial^2_5 \! \left[ e^{-2A} A_5 \right]~.
\ee
Focusing on \textit{interval boundary conditions}, we also identify
two interesting possibilities:
\be
\begin{array}{llclcrcl}
\displaystyle 
& (+)         & \equiv & 
\displaystyle 
\left. A_5 \rule{0mm}{3.5mm} \right| = 0~,   
& \textrm{hence} &
\left. \partial_5 A_\mu \rule{0mm}{3.5mm} \right| &=& 
0~,
\\ [0.5em]
\textrm{\underline{or}}
\\ [0.5em]
\displaystyle 
& (-)         & \equiv & 
\displaystyle 
\left. A_\mu \rule{0mm}{3.5mm} \right| = 0~,   
& \textrm{hence} &
\left. \partial_5 \! \left[ e^{-2A} A_5 \right] \rule{0mm}{3.5mm}
\right| &=& 
0~.
\end{array}
\nonumber
\ee
Similarly to the case of fermions, we see that the b.c.'s for $A_\mu$
and $A_5$ are correlated. Again, there are four cases that we label by
\be
\begin{array}{lcclccl}
\displaystyle 
(+,+) & &
(+,-) & &
(-,+) & &
(-,-)
\end{array}~,
\nonumber
\ee
where the first (second) entry refers to the b.c.~at $y=0$ ($y=L$).

Additional boundary terms in the action translate into a
modification of the boundary conditions above, and are read with the
same methods.  One interesting case corresponds to localized
gauge-kinetic terms, and we refer the interested reader
to~\cite{Georgi:2000ks,Ponton:2001hq,Carena:2002me} for further
details (see also Section~\ref{Holography:AdS} in the fourth lecture).
We should also comment that the Faddeev-Popov procedure, when applied
to our gauge-fixing choice, leads to ghost fields.  However, we will
not need them in these lectures.

%%%%%%%%%%%%%%%%%%%%%
\begin{quote}

\textbf{\underline{Exercise}:}
\begin{enumerate}

\item[a)] Show in full detail how the $(+)$ and $(-)$ b.c.'s arise
from $\delta S^S_A = 0$. In particular, argue for the correlation
between the b.c.'s of $A_\mu$ and $A_5$.

\item[b)] Are there other types of b.c.'s consistent with $\delta
S^S_A = 0$?

\end{enumerate}
\end{quote}
%%%%%%%%%%%%%%%%%%%%%

%%%%%%%%%%%%%%%%%%%%%
\begin{quote}

\textbf{\underline{Exercise}:}
It is possible that the bulk gauge field is coupled to a ``Higgs
field" that gets a non-zero vev. This can lead to an effective
$y$-dependent mass for the gauge field, $M^2_A(y)$. How is the above
formalism modified?
\end{quote}
%%%%%%%%%%%%%%%%%%%%%

\noindent
As we will see in the next lecture
\begin{itemize}

\item The SM gauge fields arise from $(+,+)$ b.c.'s.

\item Models where the Higgs is interpreted as an extra-dimensional
polarization of higher-dimensional gauge fields use $(-,-)$ b.c.'s.

\item $(+,-)$ and $(-,+)$ b.c.'s show up in models with custodial
symmetries, among others.

\end{itemize}
%

%%%%%%%%%%%%%%%%%%%%%%%%%%%%%%%%%%%%%%%
\subsection{The Metric Background}
\label{metricbackground}
%%%%%%%%%%%%%%%%%%%%%%%%%%%%%%%%%%%%%%%

So far, we have kept the metric background arbitrary, imposing only
that it respects 4D Lorentz invariance. However, the choice of
background is not only necessary to solve explicitly the previous
EOM; it also largely determines the resulting phenomenology.

Presumably, the background is a solution to the higher-dimensional
Einstein equations.\footnote{As long as we assume that the
extra-dimensions share the properties of the dimensions we have
observed, at least at short distances.} There may exist (scalar)
fields sourcing these equations, e.g.
\be
S = \int_{M} \! d^5 x \,
\sqrt{g} \left[ -\frac{1}{2} M^{3}_{5} \, \mathcal{R}_{5} +
\frac{1}{2} \nabla_M \Phi_i
\nabla^M \Phi_i  -V(\Phi_i)\right]
+ \int_{\partial M} \! d^4x \, \sqrt{\bar{g}} \, {\cal
L}_{4}(\Phi_i)~,
\label{GravityScalarAction}
\ee
where $M_{5}$ is the (reduced) 5D Planck mass, $\mathcal{R}_{5}$ is
the 5D Ricci scalar, and the last term allows for operators localized
on the boundary.~\footnote{\label{GibbonsHawking} We also assume the
presence of the Gibbons-Hawking term, $\int_{\partial M} \!  d^4x \,
\sqrt{\bar{g}} \, M^3_5 K$, where K is the trace of the extrinsic
curvature and $\bar{g}_{\mu\nu}$ is the induced metric.  In our
coordinates, we have $K_{\mu\nu} = \frac{1}{2} \partial_y
\bar{g}_{\mu\nu}$, which implies that $K \equiv \bar{g}^{\mu\nu}
K_{\mu\nu }= 4 A'$ in our background.  This term is important when
performing the variation of the action w.r.t.~$A(y)$ in order to
determine the appropriate boundary conditions governing the gravity
sector.}

Solving the EOM that follow from the above action is, in general,
quite hard. However, it turns out that there is a class of potentials
that is amenable to fully analytic solutions to the coupled
metric/scalar system. Before presenting the method, let us make a few
remarks.

Although the class of potentials that allows for closed solutions is
rather special (and not necessarily stable under radiative
corrections), we can mention a number of reasons for why it is
worthwhile to study these cases:
\begin{enumerate}

\item Having explicit solutions for varied (even if special)
potentials allows to develop an intuition for the effects associated
with backreaction.

\item In many cases, the physical properties of interest can be seen
not to depend on the special relations that make the problem
analytically soluble. Thus, the analytic solutions can be taken as a
convenient way to understand the relevant physics.

\item Models based on this method have been recently proposed.

\item ``Traditional" backgrounds, such as flat space or AdS$_5$ are
particular examples of this scheme. Although the machinery we will
present is overkill for these simple cases, it allows for a unified
treatment, and generalizations.

\end{enumerate}
%

%%%%%%%%%%%%%%%%%%%%%%%%%%%%%%%%%%%%%%%
\subsubsection{The ``Superpotential" Method}
\label{superpotential}
%%%%%%%%%%%%%%%%%%%%%%%%%%%%%%%%%%%%%%%

This is a supergravity-inspired approach that was applied to
extra-dimensional scenarios in~\cite{Brandhuber:1999hb} and in a more
general form in~\cite{DeWolfe:1999cp}. We assume a \textit{single}
scalar field in 5D (or, more generally, in codimension one). The
class of potentials of interest here are those that can be written as 
\be
V(\Phi) = \frac{1}{8} \left(\frac{\partial W(\Phi)}{\partial \Phi}
\right)^2 -
\frac{1}{6M^{3}_{5}} W(\Phi)^2~,
\label{PotFromSuperpot}
\ee
for some ``superpotential" $W(\Phi)$. We emphasize that, in spite of
the language, there is no assumption that supersymmetry is involved
here. 

The 5D EOM for the metric/scalar system are, \textit{in the bulk}
\be
4 (A^\prime)^2 - A^{\prime\prime} &=& - \frac{2}{3M_5^3} V~,
\nonumber \\ [0.3em]
(A^\prime)^2 &=& \frac{1}{12 M_5^3} (\Phi^{\prime})^2 -
\frac{1}{6M_5^3} V~,
\label{metricscalareqns}
 \\ [0.3em]
\phi^{\prime\prime} - 4 A^\prime \phi^\prime &=& \frac{\partial
V}{\partial \phi}~,
\nonumber
\ee
where we assumed the line element Eq.~(\ref{metric}), and that the
scalar vev $\langle \Phi \rangle = \phi(y)$ is $x^\mu$ independent
(as required by 4D Lorentz invariance). The solutions with vanishing
effective 4D Cosmological Constant (C.C.) (i.e.~flat 4D sections, as
in our ansatz) lead to a particularly simple solution,\footnote{The
method can be generalized to cases where the 4D geometry is de Sitter
or Anti-de Sitter~\cite{DeWolfe:1999cp}.} when $V$ is given by
Eq.~(\ref{PotFromSuperpot}). Indeed, it is straightforward to check
that \underline{if}
\be
\phi'(y) &=& \frac{1}{2} \frac{\partial W(\phi)}{\partial \phi}~,
\label{phiEOM}
\\ [0.3em]
A'(y)  &=& \frac{1}{6M^{3}_{5}} W(\phi(y))~,
\label{AEOM}
\ee
then Eqs.~(\ref{metricscalareqns}) are fulfilled. The above
represents a huge simplification: given $W(\phi)$, Eq.~(\ref{phiEOM})
is a first-order ODE for $\phi$ that we can always solve by
separation of variables. Having an explicit solution for $\phi(y)$,
Eq.~(\ref{AEOM}) has an explicit function of $y$ on the r.h.s., and
can be integrated immediately to find $A(y)$. Thus, the problem has
been reduced to quadrature.

We have described the \textit{bulk} solution, but still need to
specify the b.c.'s that allow such a solution.  This will require
specifying appropriate sources at the boundary $\partial M$.  Indeed,
Eqs.~(\ref{phiEOM}) and (\ref{AEOM}) imply
\be
\left. \phi' \rule{0mm}{3.5mm} \right| &=& \left. \frac{1}{2}
\frac{\partial W}{\partial \phi} \right|~,
\hspace{1.5cm}
\left. A' \rule{0mm}{3.5mm} \right| ~=~ \left. \frac{1}{6M^{3}_{5}} W
\right|~.
\label{APhiBulkBoundary}
\ee
We want to check that these conditions can be derived from a
variational principle.  Assuming that ${\cal L}_{4}$ in
Eq.~(\ref{GravityScalarAction}) does not involve $y$ derivatives (e.g.
if it is a pure potential term), varying the action with respect to
both $\Phi$ and $A$, and requiring that the surface terms obtained
from the integrations by parts vanish, leads to~\footnote{Here it is
important to include the terms arising from the variation of the
Gibbons-Hawking term (see footnote~\ref{GibbonsHawking}).  This
generates a boundary term proportional to $\delta A'$ that cancels a
similar term from the variation of the Einstein-Hilbert action, and
leaves behind a term proportional to $\delta A$ that results in the
l.h.s. of the second equation in (\ref{BoundaryConditons}).}
\be
\mp \left. \Phi' \rule{0mm}{3.5mm} \right|_{0,L} &=& \left.
\frac{\partial {\cal L}_{4}}{\partial \Phi} \right|_{0,L}~,
\hspace{1cm}
\mp \left. A' \rule{0mm}{3.5mm} \right|_{0,L} ~=~ \left. \frac{1}{3
M^{3}_{5}} \, {\cal L}_{4} \right|_{0,L}~.
\label{BoundaryConditons}
\ee
We see that Eqs.~(\ref{APhiBulkBoundary}) follow from
Eqs.~(\ref{BoundaryConditons}) when
\be
\mp \left. {\cal L}_{4}(\Phi) \rule{0mm}{3.5mm} \right|_{0,L} &=&
\frac{1}{2} \, W (\bar{\phi}_{0,L}) +
\frac{1}{2}W'(\bar{\phi}_{0,L})(\Phi - \bar{\phi}_{0,L}) \mp  \Delta
{\cal L}_{4}(\Phi)~,
\label{L4}
\ee
where $\bar{\phi}_{0,L}$ are constants to be interpreted as the
boundary field values
\be
\Phi(y = 0) &=& \bar{\phi}_0~,
\hspace{1.5cm}
\Phi(y = L) ~=~ \bar{\phi}_L~,
\label{BoundaryFields}
\ee
and $\Delta  {\cal L}_{4}$ is arbitrary, except for the requirement
that ${\cal L}_{4}(\Phi = \bar{\phi}_{0,L}) = 0$. This freedom is
often used to write a \textit{stiff} potential that enforces
Eqs.~(\ref{BoundaryFields}) in a simple way. 

%%%%%%%%%%%%%%%%%%%%%
\begin{quote}

\textbf{\underline{Exercise}:}
Check the above statements in detail.
\end{quote}
%%%%%%%%%%%%%%%%%%%%%

%%%%%%%%%%%%%%%%%%%%%%%%%%%%%%%%%%%%%%%
\subsubsection{Application: Radion Stabilization}
\label{RadionStabilization}
%%%%%%%%%%%%%%%%%%%%%%%%%%%%%%%%%%%%%%%

\begin{figure}[t]
\begin{center}
\includegraphics[width=0.5 \textwidth]{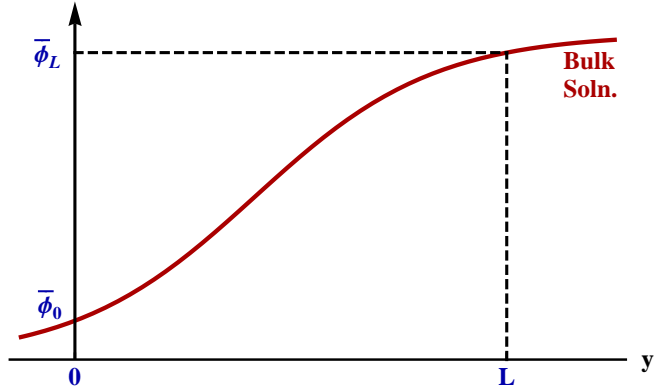}
\caption{Illustration of ``radion stabilization" as arising from the
interplay between the bulk scalar solution and the boundary
conditions. We may assume, by a shift of the $y$ coordinate, that
$\langle \Phi(0) \rangle = \bar{\phi}_0$. The size of the extra
dimension is then determined by $\langle \Phi(L) \rangle =
\bar{\phi}_L$.}
\label{fig:RadionStab}
\end{center}
\end{figure}

Generically, non-trivial scalar profiles lead to \textit{radion
stabilization}, i.e.~to a preferred value for the size of the extra
dimension, $L$.~\footnote{This happens here at tree-level, but it is
possible to imagine stabilization mechanisms that rely on quantum
effects, see e.g.~\cite{Ponton:2001hq,Garriga:2002vf}.} This can be
transparently seen from the above formalism. Assume, for simplicity,
that $\left. \Delta {\cal L}_{4} \right|_{y = 0,L}$ are stiff
potentials that essentially fix the scalar boundary values to
$\bar{\phi}_{0}$ and $\bar{\phi}_{L}$. For instance, 
\be
\left. \Delta {\cal L}_{4}(\Phi) \rule{0mm}{3.5mm} \right|_{0,L}  &=&
-\gamma \left( \Phi^2 - \bar{\phi}^2_{0,L} \right)^{2}~,
\hspace{1.5cm}
\textrm{for } \gamma \to \infty~.
\label{FixedBoundaryVEV}
\ee
The argument is illustrated in Fig.~\ref{fig:RadionStab}: basically,
given the bulk solution and $\bar{\phi}_{0}$ and $\bar{\phi}_{L}$,
only special values of $L$ lead to consistency between the bulk
solution and the boundary conditions. Thus, a particular size for the
extra dimension is selected \textit{dynamically}. 

\noindent
Let us add a few comments:
\begin{itemize}

\item Depending on the bulk solution (i.e.~the superpotential $W$)
and the choice of the parameters $\bar{\phi}_{0}$ and
$\bar{\phi}_{L}$, it may happen that there is no ``intersection" in
Fig.~\ref{fig:RadionStab}, for example, if $|\bar{\phi}_{L} -
\bar{\phi}_{0}|> {\rm max}_y \, \phi(y) - {\rm min}_y \, \phi(y)$,
where $\phi(y)$ is the bulk profile. In such a case, one of the
assumptions must break down, most likely the flat 4D sections ansatz.
There may be solutions for the same theory which are de Sitter or
Anti-de Sitter, for fixed $y$.

\item Somewhat related to the above, if the boundary Lagrangians
${\cal L}_4$ do not have the special form given above (which require
tuning), then what happens in general is that the solutions have
non-vanishing 4D curvature.

\item Assuming that the solution with 4D flat sections is allowed,
what happens if we displace ``by hand" the branes from their
preferred separation?  This is a dynamical, time-dependent situation
that goes beyond our ansatz. If the static solution is stable, then
the system will oscillate about the preferred brane separation. This
motion corresponds to ``radion excitations" and will show up as a
particle with properties rather similar to the Higgs boson! 

The issue of stability can be investigated on a case by case basis by
analyzing small oscillations of the scalar modes in the gravity/bulk
scalar system, to check that there are no negative mass eigenvalues
(tachyonic modes). This also allows to determine the \textit{radion
mass} by the KK-decomposition procedure to be presented below.

\item The stabilization of the radion by non-trivial bulk profiles
was first studied in~\cite{Goldberger:1999uk} in the limit that the
scalar backreaction is small. In this limit the stabilization can be
understood from a 4D effective potential, and the curvature at the
minimum of this potential gives the radion mass. This is just a
different language to study the same physics as above. It allows for
more general potentials than the ``superpotential approach", but the
parameters must be chosen so that the backreaction on the metric is a
small effect.

\item If W is odd in $\Phi$, the resulting backgrounds have a
KK-parity symmetry, as in Universal Extra Dimensional models (UED's)
--to be discussed in the next lecture-- but allowing for non-trivial
warping (see~\cite{Medina:2010mu}). 

\end{itemize}
%

%%%%%%%%%%%%%%%%%%%%%%%%%%%%%%%%%%%%%%%
\subsubsection{Simple Limits}
\label{SimpleLimits}
%%%%%%%%%%%%%%%%%%%%%%%%%%%%%%%%%%%%%%%

Two simple backgrounds may be easily obtained in the framework of the
superpotential approach.

\medskip
\underline{Flat space}:
Taking $W=0$ (one may just remove $\Phi$ since it plays no role), one
has $A'(y) = 0$, hence $A(y) = {\rm constant}$. By rescaling $x^\mu$,
we simply have $ds^2 = \eta_{\mu\nu} dx^\mu dx^\nu - dy^2$, i.e.~flat
space.

\begin{figure}[t]
\begin{center}
\includegraphics[width=0.8 \textwidth]{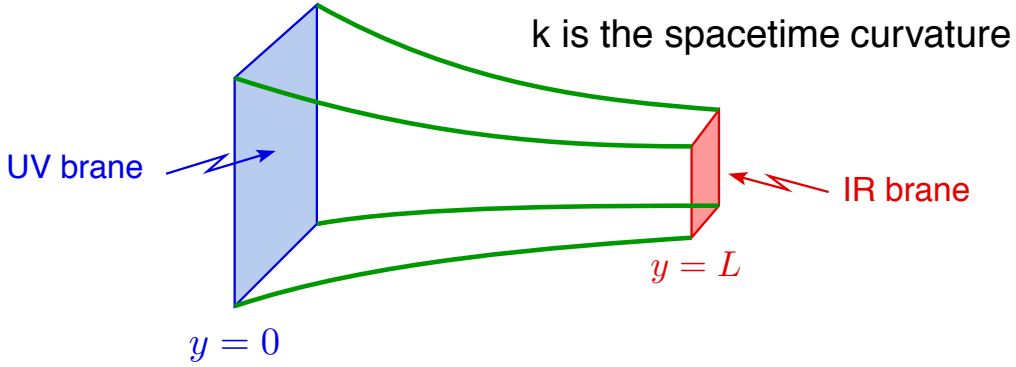}
\caption{A cartoon of the AdS$_5$ setup.  The fifth dimension is
bounded by the UV and IR branes, where the warp factor, $e^{-A(y)}$, is
largest and smallest, respectively.  These are $3+1$ dimensional
infinite flat surfaces.  The green lines indicate regions of constant
coordinate $x^\mu$.}
\label{fig:AdS}
\end{center}
\end{figure}

\medskip
\underline{AdS$_5$}:
Removing again $\Phi$ and taking $W = 6 M_5^3 k$, for some constant
$k$ (with mass dimension one), we have $A'(y) = k$, hence $A(y) = k y
+ {\rm constant}$. By rescaling $x^\mu$, we can again set the
constant to zero, and obtain
\be
ds^2 = e^{-2ky} \, \eta_{\mu\nu} dx^\mu dx^\nu - dy^2~,
\ee
which can be recognized as Anti-de Sitter space with curvature $k$
(see Fig.~\ref{fig:AdS}). In this case, the required boundary
Lagrangians read
\be
\left. {\cal L}_4 \rule{0mm}{3.5mm} \right|_{0,L} &=& \pm 6 M^3_5 k
~\equiv~ T_{0,L}~,
\ee
while the bulk potential is $V = -6 M^3_5 k^2 \equiv \frac{1}{2} M^3_5
\Lambda_5 < 0$.  This reproduces the two fine-tunings found
in~\cite{Randall:1999ee}~\footnote{Note that my $M_5^3$ is their $4
M^3$, and my $\frac{1}{2} M^3_5 \Lambda_5$ is their $\Lambda$.}
\be
T_0 &=& -T_L = - \frac{M_5^3 \Lambda_5}{2k}~.
\ee
In this limit the radion has no potential (i.e.~the size of the extra
dimension is not stabilized), which accounts for one of the
fine-tunings. The second tuning amounts to setting the 4D C.C. to
zero, and thus is tied to the C.C.~problem. Although this second
fine-tuning is always present, the first one disappears once the
radion is stabilized in the more general setup involving the scalar
field. 
%

%%%%%%%%%%%%%%%%%%%%%%%%%%%%%%%%%%%%%%%
\subsection{The Kaluza-Klein Decomposition}
\label{KKDecomp}
%%%%%%%%%%%%%%%%%%%%%%%%%%%%%%%%%%%%%%%

A central concept in extra-dimensional physics is that the theory of
a bulk field propagating in a \textit{compact} extra dimension can be
\textit{rewritten} as a 4D theory involving an infinite number of 4D
fields. For example, for a free scalar
\be
\int \! d^5x \sqrt{g} \, \left\{ \frac{1}{2} \partial_M \Phi
\partial^M \Phi - \frac{1}{2} M^2 \Phi^2 \right\}
&=& 
\sum_n \int \! d^4x \, \frac{1}{2} \left\{ \partial_\mu \phi_n
\partial^\mu \phi_n - m^2_n \phi^2_n \right\}~,
\label{ScalarKK}
\ee
where we assumed 4D flat sections. The $\phi_n(x^\mu)$ are known as
Kaluza-Klein (KK) modes. The information about the background (and
$M$) is fully contained in the KK spectrum, $m_n$.

The above is just a generalization of the Fourier-mode decomposition,
written here as
\be
\Phi(x^\mu,y) &=& \frac{e^{A(y)}}{\sqrt{L}} \sum_n \phi_n(x^\mu)
f_n(y)~,
%\hspace{1.5cm}
%\textrm{(scalar)}
\label{KKDecompScalar}
\ee
where $\{ f_n \}$ is any set of complete functions that allow us to
expand an arbitrary $y$-dependence, with $x^\mu$-dependent
coefficients, $\phi_n(x^\mu)$. We will see soon how to choose this
set in the most convenient way. They will be called ``KK
wavefunctions". Note that in our expansion,
Eq.~(\ref{KKDecompScalar}), we pulled out an explicit factor of
$e^{A(y)}$, which gives the $f_n$'s so defined a rather direct
physical interpretation in terms of \textit{localization properties
of the KK-modes along the extra dimension}. Hence I will call them
``physical wavefunctions" (This will become clearer in subsequent
sections.) In addition, we also wrote a factor of $1/{\sqrt{L}}$ in
Eq.~(\ref{KKDecompScalar}). This allows us to display more
transparently the different dimensions of the fields involved: $\Phi$
has mass dimension $3/2$, while the $\phi_n$ have mass dimension $1$,
as appropriate to their 4D interpretation. Our conventions will
always be that the KK wavefunctions $f_n$ are dimensionless.

To understand how to choose the set $f_n$, recall the scalar EOM
derived in Eq.~(\ref{ScalarEOM}), which reads here
\be
e^{2A} \partial_\mu \partial^\mu \Phi - e^{4A} \partial_y \! \left[
e^{-4A} \partial_y \Phi \right] + M^2 \Phi &=& 0~.
\label{FreeScalarEOM}
\ee
The interpretation of $\phi_n$ as a (free) 4D scalar field of mass
$m_n$ means that
\be
\partial_\mu \partial^\mu \phi_n + m^2_n \phi_n &=& 0~.
\label{4DScalarEOM}
\ee
If we take a special field configuration with a single $n$,
$\Phi(x^\mu,y) = \frac{e^{A(y)}}{\sqrt{L}} \phi_n(x^\mu) f_n(y)$, and
replace in Eq.~(\ref{FreeScalarEOM}) using Eq.~(\ref{4DScalarEOM}),
we get
\be
- e^{2A} m^2_n f_n - e^{3A} \partial_y \! \left[ e^{-4A} \partial_y
\left( e^A f_n \right) \right] + M^2 f_n &=& 0~,
\ee
or more explicitly
\be
f^{\prime\prime}_n - 2 A' f'_n + \left[ A^{\prime\prime} - 3 (A')^2 -
M^2 + e^{2A} m^2_n \right] f_n &=& 0~.
\label{ScalarfnEOM}
\ee
Here is where our discussion of boundary conditions pays off, for if
we multiply by $e^{-2A}f_m$ and subtract the same expression with $m
\leftrightarrow n$, we find
\be
\partial_y \left\{ e^{-2A} \left( f_m f'_n - f_n f'_m \right)
\right\} + \left( m^2_n - m^2_m \right) f_m f_n &=& 0~.
\label{ScalarKKEOM}
\ee
If we integrate $\int_0^L \! dy$, the first term gives rise to
surface terms of exactly the form found when applying the variational
procedure. Since these are required to vanish, we find
\be
 \left( m^2_n - m^2_m \right) \int_0^L \! dy \, f_m f_n &=& 0~,
\ee
i.e.~we have derived an orthogonality relation for $m^2_n \neq
m^2_m$.~\footnote{\label{Wronskian}The same line of reasoning can be
used to show that for two solutions, $h_1$ and $h_2$, of
Eq.~(\ref{ScalarfnEOM}), one has $e^{-2A(y)} \, W(y) = e^{-2A(y')} \,
W(y')$, where $W(y) = h_1(y) h'_2(y) - h_2(y) h'_1(y)$ is the
\textit{Wronskian}.} In fact, the differential equation for $f_n$,
Eq.~(\ref{ScalarKKEOM}), together with the boundary conditions derived
from the variational principle, define a self-adjoint problem that
guarantees not only the orthogonality of the solutions, but also that
the $\{ f_n \}$ form a complete set that we can use to ``Fourier"
expand the $y$-dependence of any field configuration $\Phi(x^\mu, y)$.

Choosing the normalization
\be
\frac{1}{L} \int_0^L \! dy \, f_m f_n &=& \delta_{mn}~,
\label{KKNorm}
\ee
and replacing the KK expansion, Eq.~(\ref{KKDecompScalar}) in the free
bulk scalar action, it is straightforward to obtain the advertised
result, Eq.~(\ref{ScalarKK}).  Of course, this 4D action just embodies
the EOM for the KK modes, Eq.~(\ref{4DScalarEOM}).  The KK masses,
$m^2_n$, are found by solving the ODE for the $f_n$,
Eq.~(\ref{ScalarfnEOM}), and requiring that the desired b.c.'s be
satisfied.

Note the philosophy here: the KK decomposition is defined in terms of
the \textit{free} action, while interactions are to be treated
perturbatively after inserting the previous KK decomposition. For
instance, for a $\frac{1}{3!} \lambda_5 \Phi^3$ interaction, we get
in the equivalent KK theory:
\be
{\cal L}_{\rm int} ~=~ \int_0^L \! dy \, \sqrt{g} \, \frac{1}{3!}
\lambda_5 \Phi^3
&=&
\sum_{m,n,r} \frac{1}{3!} \lambda_{mnr} \phi_m \phi_n \phi_r~,
\ee
where
\be
\lambda_{mnr} &=& \frac{\lambda_5}{L^{3/2}} \, e^{-A(L)} \int_0^L \!
dy \, e^{-[A(y) - A(L)]} f_m(y) f_n(y) f_r(y)~.
\ee
We have written the factors of $e^{\pm A(L)}$ for later physical
interpretation. These overlap integrals carry information about the
relative localization of the KK-modes. 

%%%%%%%%%%%%%%%%%%%%%
\begin{quote}

\textbf{\underline{Exercise}:}
What are the mass dimensions of $\lambda_5$ and $\lambda_{mnr}$?
\end{quote}
%%%%%%%%%%%%%%%%%%%%%

%%%%%%%%%%%%%%%%%%%%%%%%%%%%%%%%%%%%%%%
\subsubsection{KK Decomposition for Fermions}
\label{FermionKKDecomp}
%%%%%%%%%%%%%%%%%%%%%%%%%%%%%%%%%%%%%%%

We now quickly summarize the KK decomposition for a 5D fermion field.
As in the scalar case, the KK wavefunctions are chosen to obey the
classical (free) EOM, which together with the allowed b.c.'s
guarantees that they form a complete, orthogonal set.
We write now 
\be
\Psi_{L,R}(x^\mu,y) &=& \frac{e^{\frac{3}{2} A(y)}}{\sqrt{L}} \sum_n
\psi^n_{L,R}(x^\mu) f^n_{L,R}(y)~,
%\hspace{1.5cm}
%\textrm{(fermion)}
\label{KKDecompFermion}
\ee
where 
\be
\left( \partial_y + M - \frac{1}{2} A' \right) f^n_L &=& m_n e^A
f^n_R~,
\label{LHFermionKKEq}
\\ [0.5em]
\left( \partial_y - M - \frac{1}{2} A' \right) f^n_R &=& - m_n e^A
f^n_L~,
\label{RHFermionKKEq}
\ee
and 
\be
\frac{1}{L} \int_0^L \! dy \, f^m_{L,R} f^n_{L,R} &=& \delta_{mn}~,
\label{FermionKKNorm}
\ee

Note the factor $e^{\frac{3}{2} A(y)}$ in Eq.~(\ref{KKDecompFermion})
that defines the $f^n_{L,R}(y))$ as \textit{physical wavefunctions},
and that we allow for different KK expansions for the LH and RH
chiralities, in accord with the fact that these are distinguished by
the b.c.'s.
The two types of boundary conditions discussed in
Section~\ref{Fermions} read
\be
\begin{array}{llclcrcl}
\displaystyle 
& (-)         & : & 
\displaystyle 
\left. \partial_y f^n_R - \left( M + \frac{1}{2} A' \right) f^n_R
\rule{0mm}{3.5mm} \right| 
&=& 
0
& \textrm{\&} &
\left. f^n_L \rule{0mm}{3.5mm} \right| = 0~,   
\\ [0.5em]
\textrm{\underline{or}}
\\ [0.5em]
\displaystyle 
& (+)         & : & 
\displaystyle 
\left. \partial_y f^n_L + \left( M - \frac{1}{2} A' \right) f^n_L
\rule{0mm}{3.5mm} \right| &=& 
0
& \textrm{\&} &
\left. f^n_R \rule{0mm}{3.5mm} \right| = 0~,
\end{array}
\label{FermionBCs}
\ee
while the KK theory simply reads
\be
S_\Psi &=& 
\sum_n \int \! d^4x \, \bar{\psi}^n \left( i \gamma^\mu \partial_\mu
- m_n \right) \psi^n~.
\label{FermionKK}
\ee
%

%%%%%%%%%%%%%%%%%%%%%
\begin{quote}

\textbf{\underline{Exercise}:}
Check the above results. Derive also decoupled $2^{\rm nd}$ order
ODE's for the massive modes $f^n_L$ and $f^n_R$.
\end{quote}
%%%%%%%%%%%%%%%%%%%%%

\medskip
\noindent
\textbf{\underline{0-modes}}
\medskip

The KK fermion equations, together with the b.c.'s
(\ref{FermionBCs}), have the very interesting property that they
always allow a massless mode. Indeed, setting $m_{n=0} = 0$, the
first order ODE's, Eqs.~(\ref{LHFermionKKEq}) and
(\ref{RHFermionKKEq}) decouple and can be integrated immediately to
give:
\be
f_{L,R}^0(y) &=& N_{L,R} \exp \left(\frac{1}{2} A(y) \mp \int^y_{0}
\! dz \, M (z) \right)~,
\label{zeromodes}
\ee
where the $N_{L,R}$ are fixed by the
normalization~(\ref{FermionKKNorm}), and we allowed for a
$y$-dependent mass, as would arise from a Yukawa coupling to a bulk
scalar that gets a non-trivial vev profile:
\be
- y_5 \Phi \overline{\Psi} \Psi &\to& - y_5 \phi(y) \overline{\Psi}
\Psi ~\equiv~ - M(y) \overline{\Psi} \Psi~.
\ee
Note also that the $(+,+)$ b.c.'s allow for $f^0_L$ but imply $N_R =
0$, while the $(-,-)$ b.c.'s allow for $f^0_R$ but imply $N_L = 0$.
Thus, the 0-mode level is indeed chiral. Presumably, the SM fermions
could be identified with such 0-modes. Note that the above KK-masses
are related to the compactification of the extra dimension, but that
there can be additional contributions to the physical masses arising
from EWSB. Thus, in realistic models, the ``0-modes" may not be
strictly massless. Nevertheless, since typically the masses
associated to EWSB are small compared to the compactification scale,
it is customary to continue referring to these would-be zero-modes
simply as 0-modes.

\begin{figure}[t]
\begin{center}
\includegraphics[width=0.44 \textwidth]{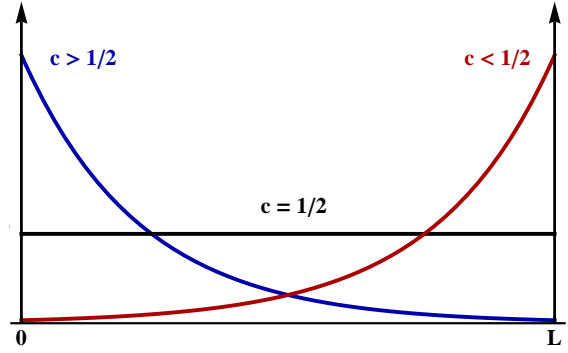}
\caption{Illustration of the localization of the fermion 0-modes,
depending on the $c$-parameter defined by Eq.~(\ref{MproptoA}) .}
\label{fig:FermionZeroModes}
\end{center}
\end{figure}

A special case arises when
\be
M(y) &\approx& \pm c A'(y)
\hspace{1,5cm}
\left[
\begin{array}{rcl}
\multirow{2}{*}{\textrm{for}}  & (+,+) &
\multirow{2}{*}{\textrm{b.c.'s}}  \\
  & (-,-) &   
\end{array}
\right]~,
\label{MproptoA}
\ee
for some dimensionless ``$c$-parameter". In such a case we have
\be
f_{L,R}^0(y) &\approx& f^0_c(y) ~\equiv~ N_{0} \, e^{-(c - \frac{1}{2}) A(y) }~,
\label{zeromodescpar}
\ee
independently of chirality [L for $(+,+)$, R for $(-,-)$]. The
localization properties can then be conveniently described by $c$.
For instance, in the case of AdS$_5$, with $A'= k$, $c$ is just the
Dirac mass $M$ in units of the curvature scale $k$. However, it is
worth pointing out that the relation~(\ref{MproptoA}) can arise in
more general settings. We illustrate the localization of the fermion
0-modes in Fig.~\ref{fig:FermionZeroModes}.
%

%%%%%%%%%%%%%%%%%%%%%%%%%%%%%%%%%%%%%%%
\subsubsection{KK Decomposition for Gauge Fields}
\label{GaugeKKDecomp}
%%%%%%%%%%%%%%%%%%%%%%%%%%%%%%%%%%%%%%%

The 5D gauge fields are written as [here $A_{M}(x^\mu,y)$ has mass
dimension $3/2$, see Eq.~(\ref{GaugeAction})]
\be
A_{\mu,5}(x^\mu,y) &=& \frac{1}{\sqrt{L}} \sum_n A^n_{\mu,5}(x^\mu)
f^n_{A,5}(y)~,
%\hspace{1.5cm}
%\textrm{(gauge)}
\label{KKDecompGauge}
\ee
where
\be
\partial_y \left[ e^{-2 A} \partial_y f^n_A \right] + m^2_n f^n_A &=&
0~,
\label{fAEOM}
\ee
while the appropriate equation for $f^n_5$ is 
\be
\partial^2_y \left( e^{-2 A} f^n_5 \right) + m^2_n f^n_5 &=& 0~.
\label{f5EOM}
\ee
Note that a solution to Eq.~(\ref{fAEOM}) with $m^2_n \neq 0$
immediately gives a solution to Eq.~(\ref{f5EOM}) as
$f^n_5 = \frac{1}{m_n} \partial_y f^n_A$. As we further comment
below, this relation is not accidental.

The KK wavefunctions are normalized as expected:
\be
\frac{1}{L} \int_0^L \! dy \, f^m_{A,5} f^n_{A,5} &=& \delta_{mn}~.
\label{GaugeKKNorm}
\ee
For $(+,+)$ b.c.'s, i.e.~$\left. \partial_y f^n_A \right| = \left.
f^n_5 \right| = 0$ on both branes, one finds a spin-1 zero-mode with
$f^0_A(y) = 1$ and $f^0_5(y) = 0$. Presumably, such a state could be
identified with one of the observed SM gauge bosons. Notice that,
unlike the case of fermion 0-modes (which are controlled by the 5D
Dirac mass), there are no free parameters to control the localization
of the gauge zero-mode. The fact that the 0-mode is flat in our
\textit{physical normalization} is closely related to gauge
invariance, and the necessary universality of the gauge interactions.
If the gauge 0-mode was not flat, the non-trivial localizations of
fermions or scalars would allow us to adjust their gauge couplings in
an arbitrary way. As it happens, the overlap integrals arising from
the gauge vertices $\overline{\Psi} \cancel{A} \Psi$, $\partial_\mu
\Phi^\dagger A^\mu \Phi$ and $\Phi^\dagger A_\mu A^\mu \Phi$, after
replacing the KK-mode decompositions with the gauge 0-mode, always
reduce to the fermion or scalar orthonormality conditions, so that
the corresponding 4D gauge interactions are indeed universal. In
particular, taking $f^0_A(y) = 1$, one finds the \textit{tree-level
matching relation} [see also Eq.~(\ref{LoopLevelMatching}) later on]
\be
g_4^2 &=& \frac{g_5^2}{L}~.
\label{TreeLevelMatching}
\ee

For $(-,-)$ b.c.'s, i.e.~$\left. \partial_y ( e^{-2A} f^n_5) \right|
= \left. f^n_A \right| = 0$ on both branes,  there is no spin-1
zero-mode, but there is a spin-0 zero-mode with $f^0_5(y) \propto
e^{2A(y)}$. In certain constructions, such scalars can be identified
with the SM Higgs field. Notice again that, given the background, the
profile of such a 4D scalar 0-mode is fixed (no adjustable parameters
to control its localization).

\medskip
\noindent
\textbf{\underline{Comment}:} the massive $A^n_5$, with $f^n_5 =
\frac{1}{m_n} \partial_y f^n_A$, provide the \textit{longitudinal}
polarizations of the associated massive gauge fields. What happens is
that the compactification breaks the gauge invariance associated to
the $n$-th KK-level \textit{spontaneously}, and there is a Higgs
mechanism at work at each KK-level. Thus, the physical massive KK
spectrum consists only of massive spin-1 fields (with three physical
polarizations each), and there are no massive physical scalars. Such
massive gauge fields can, and do have non-trivial wavefunction
profiles, and their interactions with fermion or scalar pairs can
depend on the details of those fields. The spontaneous breaking of
the $n$-th KK level gauge invariance leaves no obvious trace of the
underlying 5D gauge invariance, although it is still manifested
in a less trivial manner, e.g.~in the form of sum rules.

%%%%%%%%%%%%%%%%%%%%%
\begin{quote}

\textbf{\underline{Exercise}:}
Establish as many of the the facts stated in this section as possible.
\end{quote}
%%%%%%%%%%%%%%%%%%%%%

%%%%%%%%%%%%%%%%%%%%%%%%%%%%%%%%%%%%%%%
\subsubsection{Special Cases}
\label{SpecialCases}
%%%%%%%%%%%%%%%%%%%%%%%%%%%%%%%%%%%%%%%

Here we collect the KK decompositions for the two simple backgrounds
discussed in Section~\ref{SimpleLimits}. 

\medskip
\noindent
\textbf{\underline{Flat space}:} $A(y) = 0$
\medskip

\noindent
This case is particularly simple as far as KK decompositions: all KK
equations reduce to the Simple Harmonic Oscillator.
\be
f^{\prime\prime}_n + \left( m^2_n - M^2 \right) f_n &=& 0~,
\ee
so that the $f_n$'s are just sines and cosines.

The most widely studied application is to UED's where all fields obey
$(+,+)$ or $(-,-)$ boundary conditions. In addition, the 5D fermion
Dirac masses are assumed to vanish. In such a case:
\be
f^{(+,+)}_n &=& 
\left\{
\begin{array}{ccl}
\displaystyle 
 1                               & \textrm{for} & n = 0  \\ [0.5em]
\displaystyle 
\sqrt{2} \cos m_n y  & \textrm{for} & n \neq 0
\end{array}
\right.~,
\\ [0.5em]
f^{(-,-)}_n &=& \sqrt{2} \sin m_n y~,
\ee
where $m_n = \frac{\pi n}{L} \equiv \frac{n}{R}$ for fermions or
gauge fields, while $m_n^2 = M^2 +  (n/R)^2$ for the Higgs field. 

%%%%%%%%%%%%%%%%%%%%%
\begin{quote}

\textbf{\underline{Exercise}:}
If the 5D fermion Dirac mass does not vanish, the KK wavefunctions
are linear superpositions of sines and cosines, and the KK masses are
not as above. Work out this case in detail.
\end{quote}
%%%%%%%%%%%%%%%%%%%%%

\medskip
\noindent
\textbf{\underline{AdS$_5$ (or RS)}:} $A(y) = k y$
\medskip

\noindent
This case can also be treated analytically, since the wave equations
reduce to Bessel equations.

\medskip
\underline{Scalars}: writing the bulk mass as $M^2 = (\alpha^2 - 4)
k^2$, one finds
\be
f_n &=& N_n \, e^{ky} \left\{  
J_{\alpha} \left( \frac{m_n}{k} \, e^{ky} \right) + b_n \,
Y_{\alpha} \left( \frac{m_n}{k} \, e^{ky} \right)
\right\}~,
\ee
where $b_n$ depends on the type of boundary conditions. It can be
easily worked out in each case.

\medskip
\underline{Fermions}: writing the bulk mass as $M = \pm c k$, where
the $+$ sign applies to $(+,+)$ b.c.'s while the  $-$ sign applies to
$(-,-)$ b.c.'s, one finds
\be
f^n_{L,R} &=& N_n \, e^{ky} \left\{  
J_{c \pm \frac{1}{2}} \left( \frac{m_n}{k} \, e^{ky} \right) + b_n \,
Y_{c \pm \frac{1}{2}} \left( \frac{m_n}{k} \, e^{ky} \right)
\right\}~.
\ee
Due to our sign convention for the Dirac mass in terms of $c$, this
expression applies to both types of boundary conditions (at the
massive KK-level). The constants $N_n$ and $b_n$ are the same for
both chiralities.

\medskip
\underline{Gauge}: (unbroken case)
\be
f^n_A &=& N_n \, e^{ky} \left\{  
J_{1} \left( \frac{m_n}{k} \, e^{ky} \right) + b_n \, Y_{1} \left(
\frac{m_n}{k} \, e^{ky} \right)
\right\}~.
\ee

\medskip
\noindent
\textbf{\underline{Comments}:} 
\begin{itemize}

\item The various solutions differ in the index of the Bessel
functions, as well as in the mass spectrum, which is obtained by
matching the b.c.'s according to the case.

\item Up to an overall phase, the gauge case coincides exactly with
the $(+,+)$ LH fermion case with $c = 1/2$ (or the $(-,-)$ RH fermion
case with $c = -1/2$). This limit would apply to gauginos in models
with unbroken $N=1$ SUSY.

\item The expression for possible 0-modes were given explicitly in
the previous section.

\end{itemize}

\newpage

%%%%%%%%%%%%%%%%%%%%%%%%%%%%%%%%%%%%%%%
\section{Lecture 3: Concrete Models}
\label{Models}
%%%%%%%%%%%%%%%%%%%%%%%%%%%%%%%%%%%%%%%

In this lecture, we will review a number of models that illustrate
various aspects of extra-dimensional phenomenology.  The results
summarized in the previous lecture constitute the basic language that
is required.  Fig.~\ref{fig:Guide} serves as a rough guide to the
organization of this lecture.

As already mentioned, a given model is defined by $i)$ specifying the
gauge and global symmetries (as in 4D models), and $ii)$ specifying
the field content, including which fields propagate in the bulk and
which are localized on the boundaries (physically, over a distance
smaller than the inverse cutoff, $\Lambda$).  In general, the
Lagrangian will contain a number of parameters without a 4D
counterpart.  An important example is provided by the Dirac masses
that control the localization of the fermion 0-modes.  The scale of
the new KK resonances is also typically a free parameter, one of
central importance for collider phenomenology.

The procedure to establish the connection to experiment is in
principle straightforward:
\begin{itemize}

\item Perform the KK decomposition for all the bulk fields. The
details depend on the metric background, which should be specified
consistently. Nevertheless, as we will see, there are a number of
qualitative features that hold in large (and interesting) classes of
backgrounds.

The data we obtain in this way are spectra (up to an overall scale)
and wavefunctions that determine the couplings among the KK modes
(including the 0-modes). In general, these will depend on several
microscopic model parameters.

\item Check for consistency with existing data, for instance

\begin{enumerate}

\item[$\star$] Electroweak precision constraints (EWPT).

\item[$\star$] Flavor/\cancel{CP} constraints.

\item[$\star$] Direct collider bounds.

\end{enumerate}

In other words, check that the properties of the 0-mode sector are
sufficiently similar to the SM, and that the new physics could have
escaped direct detection in high-energy (or other) processes.
Potential deviations from the SM at low energies are associated with
the heavy modes, and can be grouped in two classes:

\begin{enumerate}

\item \underline{Tree-level effects}: if present, these tend to imply
stringent constraints.

\item \underline{Loop effects}: these encode the quantum aspects of
the extra-dimensional model. Here one has to be careful due to the
possible sensitivity to physics at the cutoff scale. Sometimes, the
best one can do is estimate the expected size of such effects.
However, it may happen that the quantity in question, within a given
model, is dominated by the physics around the compactification scale,
not the cutoff scale. In such cases, a more detailed computation is
warranted. One should note that it may happen that the cutoff
sensitivity enters at a higher loop-order (for instance, at 2-loops
but not at 1-loop). Then we may regard the quantity as
\textit{effectively calculable}.

\end{enumerate}

The upshot of this analysis is typically a lower bound on the
compactification scale, as a function of other microscopic model
parameters.

\item At this point, one is ready to explore the collider
phenomenology and investigate how to test the predictions of the
model. This typically depends on a finite (not too large) number of
particles beyond the SM.

Nowadays, it is not too difficult to implement the relevant part of
the KK Lagrangian (just a 4D QFT) into programs such as
LanHEP~\cite{Semenov:2010qt} or FeynRules~\cite{Christensen:2008py}
that accept a Lagrangian written in a form closely related to the one
we are used to, compute the corresponding Feynman rules, and generate
code appropriate for a number of event generators. These can then be
further processed to reach detector-level objects. Needless to say,
it is important to have a solid understanding of the physics to
validate the output of such codes.

\item One might also be interested in other applications. Apart from
flavor or \cancel{CP} signals, a subject that is often relevant
relates to Dark Matter (DM) and the associated cosmology, as well as
expectations for direct and indirect DM detection. 

\end{itemize}

Our aim in this lecture is not to be exhaustive, but rather to
provide an educated \textit{feeling} for the possibilities opened up
to model building by the potential existence of compact extra
dimensions. We would like to emphasize the most important physics
aspects and highlight the features that are common/different in broad
classes of scenarios. We will start with the simplest models, and
slowly add structure that permits addressing issues, such as those
discussed in the first lecture.

%%%%%%%%%%%%%%%%%%%%%%%%%%%%%%%%%%%%%%%
\subsection{Universal Extra Dimensions (UED's)}
\label{UEDs}
%%%%%%%%%%%%%%%%%%%%%%%%%%%%%%%%%%%%%%%

Perhaps the most straightforward idea is to promote the SM (gauge
group and field content) to $1+(3+n)$ Minkowski spacetime.
Remarkably, the resulting scenarios can have interesting and
non-trivial features.

\medskip
\noindent
In a first approximation, UEDs are defined by:
\begin{itemize}

\item[1)]  The extra dimensions are assumed to be exactly flat.

\item[2)]  All SM fields are promoted to higher-dimensional (bulk)
fields.

\end{itemize}
As we will see, in practice, UED scenarios involve additional
assumptions to be specified shortly. We pointed out in the previous
lecture that the KK wavefunctions in this case are sines/cosines. We
reproduce them here for the case of 5D, to which we shall be mostly
confined, for easy reference:~\footnote{We do not show EWSB effects
here, although they can play an important role in certain aspects of
the phenomenology.}
\be
f^{(+,+)}_n &=& 
\left\{
\begin{array}{ccl}
\displaystyle 
 1                               & \textrm{for} & n = 0  \\ [0.5em]
\displaystyle 
\sqrt{2} \cos \frac{n y}{R}  & \textrm{for} & n \neq 0
\end{array}
\right.~,
\hspace{1cm}
\textrm{or}
\hspace{1cm}
f^{(-,-)}_n ~=~ \sqrt{2} \sin \frac{n y}{R}~,
\ee
obeying the orthonormality conditions
\be
\frac{1}{\pi R} \int_0^{\pi R} \! dy \, f^{(+,+)}_m f^{(+,+)}_n &=&
\delta_{mn}~,
\hspace{1cm}
\textrm{and}
\hspace{1cm}
\frac{1}{\pi R} \int_0^{\pi R} \! dy \, f^{(-,-)}_m f^{(-,-)}_n ~=~
\delta_{mn}~.
\ee
However, sines/cosines as above are very special in that they obey
additional ``selection rules", involving integrals of three or more
wavefunctions. For instance, we have
\be
\hspace{-5mm}
\frac{1}{\pi R} \int_0^{\pi R} \! dy \, f^{(+,+)}_{n_1}
f^{(+,+)}_{n_2} f^{(+,+)}_{n_3} &=& \frac{1}{\sqrt{2}\pi}
\left\{
\frac{\sin(n_1+n_2+n_3) \pi}{n_1+n_2+n_3} + \frac{\sin(n_1+n_2-n_3)
\pi}{n_1+n_2-n_3}
\right.
\nonumber \\ [0.5em]
& & \hspace{1cm} \left. \mbox{}
+ \frac{\sin(n_1-n_2+n_3) \pi}{n_1-n_2+n_3} + \frac{\sin(n_1-n_2-n_3)
\pi}{n_1-n_2-n_3}
\right\}~,
\ee
and, since $n_1$, $n_2$ and $n_3$ are integers, one can see that the
integral vanishes unless $n_1 \pm n_2 \pm n_3 = 0$ for some choice of
$\pm$ signs.

%%%%%%%%%%%%%%%%%%%%%
\begin{quote}

\textbf{\underline{Exercise}:}
Convince yourself that the above is true (within this example). It
may be useful to think in terms of the \textit{momentum components
contained in the trigonometric functions}. 
\end{quote}
%%%%%%%%%%%%%%%%%%%%%
%
In fact, the above selection rule can be generalized to
\be
n_1 \pm n_2 \pm \cdots \pm n_N = 0~,
\ee
for any $N$-point function, \textit{provided there is an even number
of $f^{(-,-)}_n$ insertions}. As it urns out this is the case for all
vertices associated with SM interactions; for instance
\begin{itemize}

\item Yukawa couplings:
\be
\begin{array}{ccc}
H \overline{Q} U  &  = & \underset{\rule{0mm}{3mm}(+,+)}{H}
\underset{(+,+)}{\overline{Q}_L} \underset{(+,+)}{U_R} +
\underset{\rule{0mm}{3mm}(+,+)}{H} \underset{(-,-)}{\overline{Q}_R}
\underset{\rule{0mm}{2.5mm}(-,-)}{U_L}
\end{array}~,
\nonumber
\ee

\item $A_5$ is $(-,-)$ but always comes with another $(-,-)$ field,
e.g.
\be
\begin{array}{ccc}
\overline{Q} \gamma_5 A_5 Q  &  = & \underset{(+,+)}{\overline{Q}_L}
\underset{\rule{0mm}{2.5mm}(-,-)}{A_5} \underset{(-,-)}{Q_R} -
\underset{(-,-)}{\overline{Q}_R}
\underset{\rule{0mm}{2.5mm}(-,-)}{A_5} \underset{(+,+)}{Q_L}
\end{array}~.
\nonumber
\ee

\end{itemize}
Note that, in UED models, the $A_5$'s for $SU(3)_C \times SU(2)_L
\times U(1)_Y$ and the ``wrong chirality fermions" are the only
$(-,-)$ fields.

The above selection rules have important consequences, e.g.
\be
\begin{array}{cccc}
\put(-20,-25){
\resizebox{3.5cm}{!}{\includegraphics{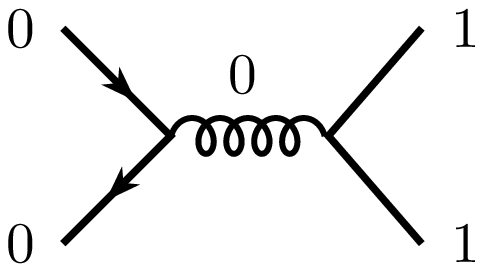}}
}
  &  \hspace{3.1cm}
   \textrm{is allowed, but}
  &   
\put(5,-25){
\resizebox{3.5cm}{!}{\includegraphics{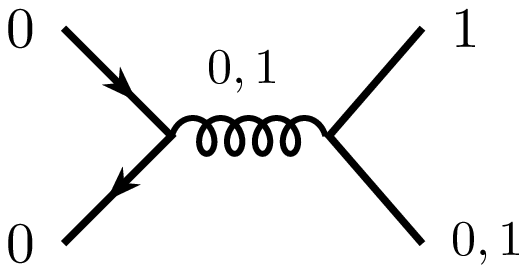}}
}
  &  \hspace{4cm}
  \textrm{is not allowed.}
\end{array}
\nonumber
\ee
Thus, \textit{at tree-level}, the new states can only be \textit{pair
produced}.
However, at loop level, the selection rules allow
\be
\put(-35,-24){
\resizebox{3.3cm}{!}{\includegraphics{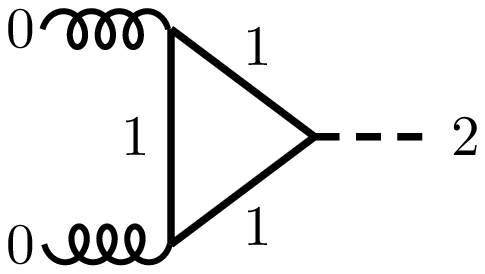}}
}
\hspace*{2.2cm}
& \propto& \frac{y g^2}{16\pi^2} \, \ln \Lambda~.
\nonumber
\ee
Thus, a second-level KK mode can be \textit{singly} produced. The
fact that the effect is (logarithmically) divergent means that there
must exist a local 5D operator associated to this process. On the
other hand, \textit{all bulk} operators lead to the tree-level
selection rule above, which is explicitly violated by a $0$--$0$--$2$
process. The resolution is that the effect corresponds to an operator
localized on the boundaries, with the same coefficient on both:
\be
{\cal L}_5 &\supset& \left[ \delta(y) + \delta(y - L) \right] {\cal
O}~.
\ee
%

%%%%%%%%%%%%%%%%%%%%%
\begin{quote}

\textbf{\underline{Exercise}:}
Show, by taking ${\cal O} = \Phi^3$, and replacing the UED KK
expansions, that $0$--$0$--$2$ processes exist, but not $0$--$0$--$3$
ones. Note the importance of the equality of the coefficients on both
boundaries.
\end{quote}
%%%%%%%%%%%%%%%%%%%%%

\begin{figure}[t]
\begin{center}
\includegraphics[width=0.45 \textwidth]{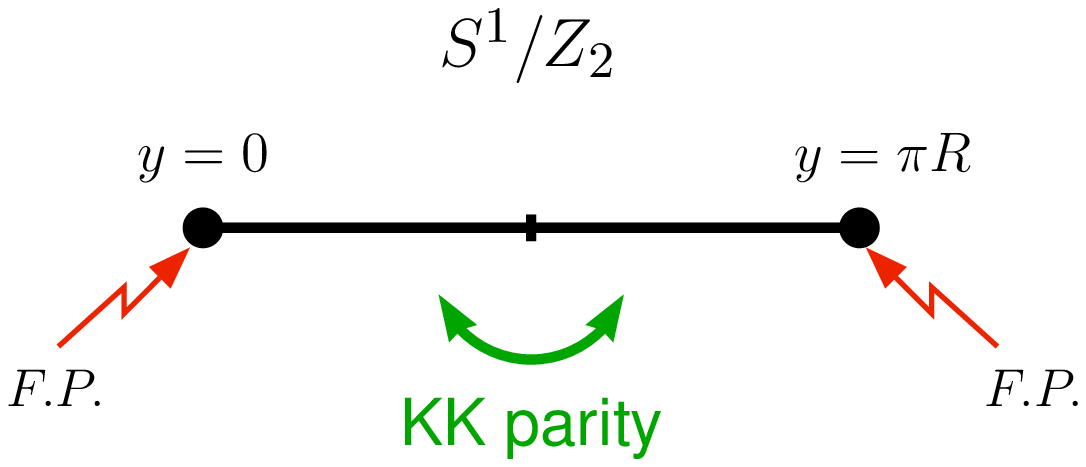}
\hspace{1cm}
\includegraphics[width=0.38 \textwidth]{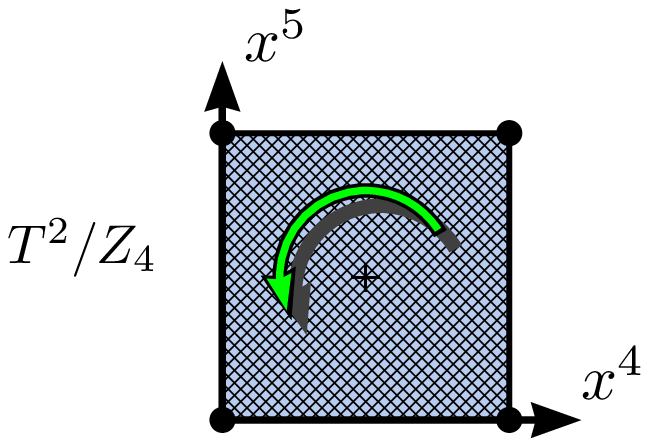}
\caption{The Kaluza-Klein parity in UED models arises from a
geometrical reflection symmetry. We show the 5D and 6D cases.
Sometimes these geometries are described as $S^1/Z_2$ and $T^2/Z_4$
(field theory) \textit{orbifolds}, respectively. Certain points,
marked here with large dots, are sometimes called ``fixed points".
The $S^1/Z_2$ orbifold corresponds to the interval compactification,
while the  $T^2/Z_4$ corresponds to the Chiral Square
compactification (see Fig.~\ref{fig:ChiralSquare}).}
\label{fig:KKParity}
\end{center}
\end{figure}

\medskip
\noindent
\textbf{\underline{KK-Parity}:} The bulk theory above has an exact
discrete symmetry under which
\be
\phi_n &\mapsto& (-1)^n \phi_n
\hspace{2cm}
\textrm{where } \phi ~=~ A_\mu, \psi, H~.
\ee
This symmetry, by treating different KK modes differently, must be a
spacetime symmetry. In fact, it follows from invariance under a
reflection about the middle of the interval, as shown in
Fig.~\ref{fig:KKParity}. This important property can be generalized
to more than 5D, as illustrated also in the figure.

Now we are ready to state the remaining assumptions that are part of
the definition of UED models (as understood in the literature).
\begin{itemize}

\item[3)] The reflection symmetry about $y=L/2$ is imposed,
i.e.~boundary operators are written symmetrically about the center.
In more detail, the theory is invariant under
\be
\begin{array}{lc|cl}
\displaystyle
A_\mu(x,y) \mapsto A_\mu(x,L-y)
  & & & 
\displaystyle
\Psi(x,y) \mapsto \pm \gamma_5 \Psi(x,L-y)
\\ [0.5em]
\displaystyle
A_5(x,y) \mapsto -A_5(x,L-y)
  & & & 
\displaystyle
H(x,y) \mapsto H(x,L-y)~.
\end{array}
\nonumber
\ee

\item[4)] It is assumed that the size of the coefficients of the
localized operators is of 1-loop order. Such an assumption is
technically natural, i.e.~they can be this small without tuned
cancellations (as estimated by NDA).

\end{itemize}
The above have important consequences
\begin{itemize}

\item KK-parity is an \textit{exact} symmetry. Although this is an
assumption, it is non-trivial that it can be imposed.

\item The single production of even KK-number states is 1-loop
suppressed (either because it corresponds to a bona fide loop effect
arising from bulk interactions, or because it arises at tree-level
from a localized operator, but with a small coefficient of 1-loop
order). \textit{Odd KK-number states cannot be singly produced}.

\item The corrections to EW observables arise at 1-loop order (or
have 1-loop size). Thus, the new physics could appear at the few
hundred GeV scale. For a light Higgs, such EW precision bounds can
push the KK scale to several hundred GeV.

\item The flavor structure of the bulk Lagrangian is as in the SM,
including the GIM mechanism. However, higher-dimension operators,
such as four-fermion interactions, can be dangerous due to the low
suppression scale. The UV completion above the cutoff $\Lambda$ must
have special flavor properties, if the compactification scale is as
low as allowed by EWPT. In any case, one should remember that UED
models do not pretend to provide a theory of flavor.

\item Localized operators (localized kinetic terms, in particular)
induce mass splittings within a given KK level. In the absence of
these, \textit{all} KK fields at the $n = 1$ level would have masses
given by $1/R$ (up to typically fairly small EWSB effects). 
The actual mass splittings are relatively small due to assumption~4).
This approximate degeneracy also plays an important role in the
phenomenology of UED scenarios.

\item The lightest $n=1$ state is exactly stable (i.e.~the lightest
state carrying the non-trivial KK-parity charge). This leads to a
(WIMP) dark matter candidate.

\end{itemize}

\medskip
\noindent
\textbf{\underline{Minimal UED's}:} (\textit{mUED})

Often one makes an additional assumption: that the localized
operators are generated \textit{entirely} by loops of bulk operators.
This allows to predict the spectrum, up to the overall scale. One
finds that 
\begin{itemize}

\item Strongly-interacting particles are the heaviest (receiving a
20-30\% upward mass-shift from 1-loop diagrams).

\item Weakly-interacting particles having $SU(2)_L$ interactions
receive a 5-10\% mass correction.

\item Particles with only $U(1)_Y$ quantum numbers are the lightest.

\item The lightest KK particle (the LKP) is $B^{(1)}_\mu$, the first
KK resonance of the hypercharge gauge boson. It receives a small and
negative 1-loop mass correction.

\end{itemize}

\noindent
The latter is a rather interesting DM candidate, and can provide the
observed DM relic density when its mass is in the several hundred GeV
range~\cite{Servant:2002aq,Kong:2005hn}. In addition, since
$B^{(1)}_\mu$ couples to hypercharge, it has a large annihilation
cross section into (RH) lepton pairs, leading to a characteristic
positron signal~\cite{Cheng:2002ej}. However, one should keep in mind
that mild departures from \textit{mUED} can lead to a significantly
different phenomenology (see e.g.~\cite{Burnell:2005hm}). 

%%%%%%%%%%%%%%%%%%%%%%%%%%%%%%%%%%%%%%%
\subsection{Warped Extra Dimensions}
\label{WarpedXDim}
%%%%%%%%%%%%%%%%%%%%%%%%%%%%%%%%%%%%%%%

We now turn our attention to cases where the spacetime curvature is
important (though still with vanishing 4D C.C.) Much of the
literature has focused on the AdS$_5$ background. However, the
central features are more general --as long as we are in the strong
warping regime-- and we will consider special backgrounds only when
necessary. We have in mind the line element of Eq.~(\ref{metric})
where $A(y)$ is a monotonically increasing function of
$y$.~\footnote{The function $A(y)$ is expected to be convex,
$A^{\prime\prime} \geq 0$. If a minimum is reached within  the region
of physical interest, one can treat the regions to the left and to
the right of this minimum separately, and then join them smoothly.}
This covers a relatively large class of backgrounds, including those
that arise from simple choices of $W$ in the \textit{superpotential
approach} described in Section~\ref{metricbackground}.

\begin{figure}[t]
\begin{center}
\includegraphics[width=0.45
\textwidth]{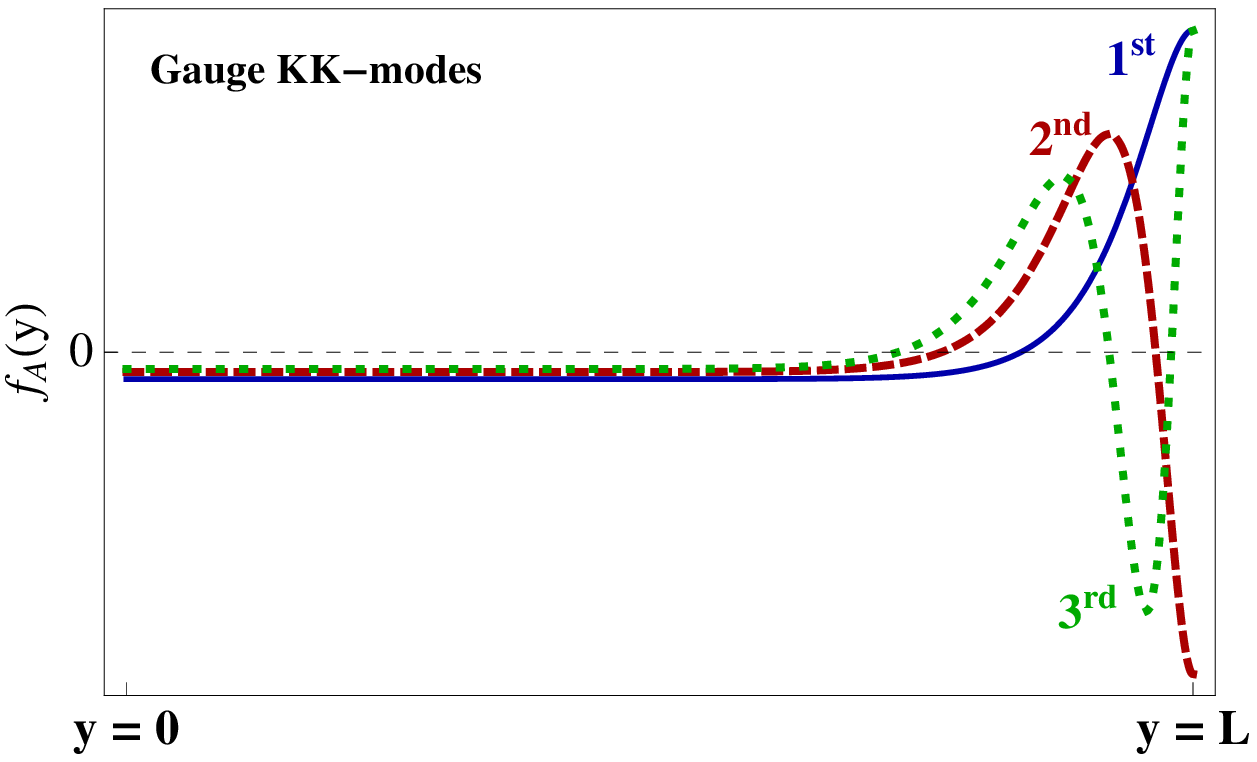}
\hspace{5mm}
\includegraphics[width=0.45
\textwidth]{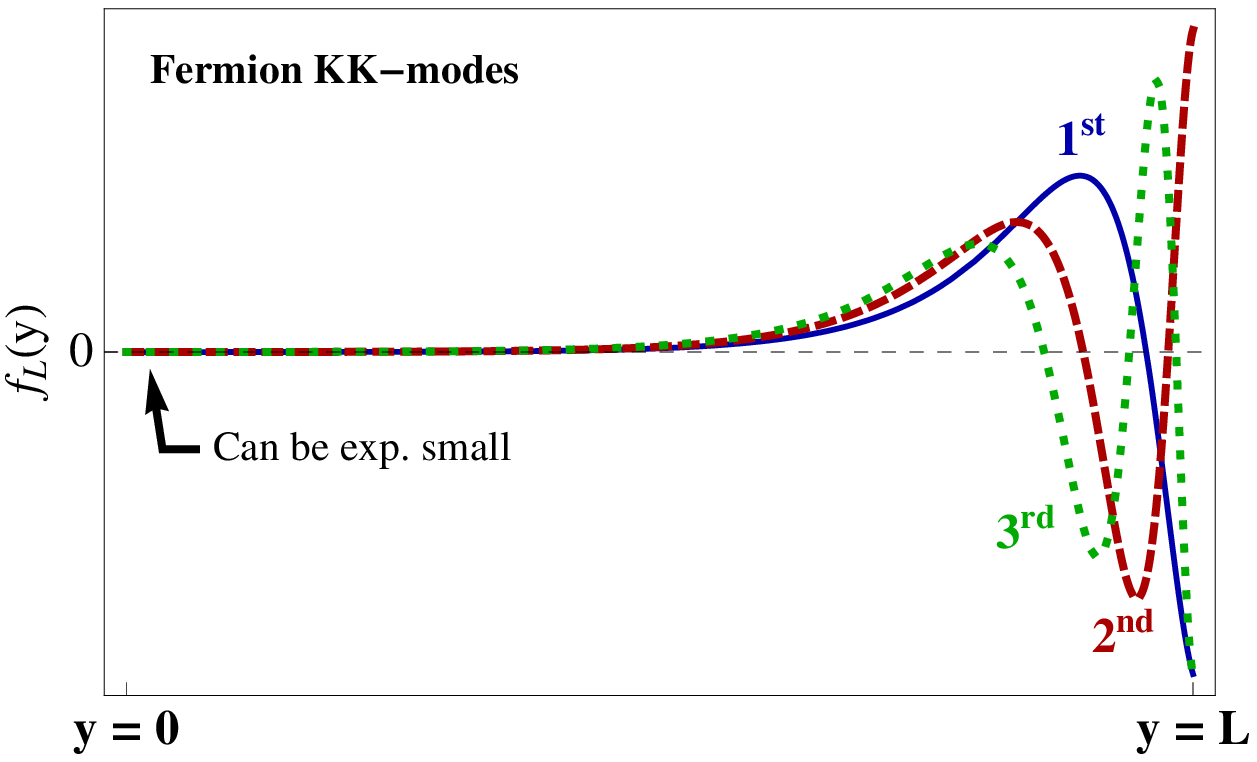}
\caption{Example of gauge (left panel) and fermion (right panel) KK
modes in \textit{strongly warped} backgrounds. The wavefunctions are
IR localized, but note the behavior near the UV brane.}
\label{fig:WarpedKKModes}
\end{center}
\end{figure}

What we mean by \textit{strong warping} corresponds to 
\be
A(L) \gg 1~,
\ee
where we assume, without loss of generality, that $A(0) = 0$. In this
case, the scale of the KK resonances is given by
\be
\tilde{k}_{\rm eff} \equiv A'(L) \, e^{-A(L)}~,
\label{WarpedKKScale}
\ee
that is, a measure of the \textit{warped-down curvature at $y=L$}.
One refers to $y=0$ as the UV boundary (brane), and to $y=L$ as the
IR boundary (brane). Under the assumption of strong warping, the
solutions to the various equations presented in the previous lecture
have the feature that the massive KK wavefunctions are IR localized,
as shown in Fig.~\ref{fig:WarpedKKModes} for the gauge and fermion
cases. Qualitatively, the generic \textit{strong warping} solutions
are similar to the explicit Bessel function solutions appropriate to
the AdS$_5$ background. The spectrum, in units of $\tilde{k}_{\rm
eff}$ is also very similar to the AdS$_5$ spectrum in units of
$\tilde{k} = k \, e^{-kL}$. Thus, the intuition gained from the
AdS$_5$ limit is of wider applicability.

The above feature of the KK wavefunctions has important consequences.
Consider a fairly generic interaction with the \textit{schematic}
structure
\be
{\cal L}_{\rm int} &\supset& \lambda_5 \Phi^N \Psi^M \left(
e^\mu_{\,\,\, \alpha} \gamma^\alpha \partial_\mu \right)^S
\left( g^{\beta \nu} A_\beta \partial_\nu \right)^K~,
\label{GenericInteraction}
\ee
involving bulk scalar, fermion and gauge fields, and with a number of
derivatives. Some of the latter may be contracted through the
f\"unfbein, while others are contracted via the metric. In
Eq.~(\ref{GenericInteraction}), the derivatives should be understood
to act on the fields in various ways. For simplicity, we focus on
$\partial_\mu$ derivatives, but our conclusion does not change when
$\partial_5$ derivatives are included. We could similarly include
\textit{brane-localized} fields. 

The 5D coupling has mass dimension
$[\lambda_5] = 5 - N \times \frac{3}{2} - M \times 2 - K \times
\frac{3}{2} - (S+K)$. Replacing our KK decompositions,
Eqs.~(\ref{KKDecompScalar}), (\ref{KKDecompFermion}) and
(\ref{KKDecompGauge}), we get interactions among KK fields
(schematically)
\be
\lambda_{n_1 \cdots n_N m_1 \cdots m_M k_1 \cdots k_K} \phi_{n_1}
\cdots \phi_{n_N} \psi_{m_1} \cdots \psi_{m_M} A_{k_1} \cdots A_{k_K}
\partial^{S+K}~,
\ee
where
\be
\lambda_{n_1 \cdots n_N m_1 \cdots m_M k_1 \cdots k_K} &=&
\frac{\lambda_5}{L^{(N+M+K)/2}} \, e^{-[\lambda_4] A(L)} \, \int_0^L
\! dy \, e^{-[\lambda_4] \left\{ A(y) - A(L) \right\}} f_{n_1}(y)
\cdots f_{k_K}(y)
\nonumber
\ee
has mass dimension
\be
[\lambda_4] = 4 - N \times 1 - M \times \frac{3}{2} - K \times 1 -
(S+K)~.
\ee
We now use the fact that the (massive) KK wavefunctions are localized
within a distance of order $k_{\rm eff}^{-1} \equiv [A'(L)]^{-1}$
from the IR brane. This means that the support of the integral is
dominated by the region $[L-k_{\rm eff}^{-1}, L]$, and that in this
region the exponential in the integrand above is of ``order one". In
addition, due to the normalization of the physical wavefunctions,
Eq.~(\ref{KKNorm}), we can estimate
\be
f_n(L) &\sim& \sqrt{k_{\rm eff} L}~,
\ee
and, therefore, the 4D coupling is of order
\be
\frac{\lambda_5}{L^{(N+M+K)/2}} \, e^{-[\lambda_4] A(L)} \,
\frac{1}{k_{\rm eff}} (k_{\rm eff} L)^{(N+M+K)/2}
&\sim&
\underset{\textrm{has mass dimension }
[\lambda_4]}{\underbrace{\lambda_5 \, k_{\rm eff}^{-1+(N+M+K)/2}}} \,
e^{-[\lambda_4] A(L)}~.
\nonumber
\ee
Thus, we learn that, quite generically, the scale of the effective 4D
coupling is \textit{red-shifted} precisely by its mass dimension.
\textit{This is how strongly warped scenarios generate exponentially
small scales starting from ``fundamental" ones}. The most famous
application is to the generation of the Planck/weak scale hierarchy,
through the redshift of the Higgs mass parameter.

\medskip
\noindent
\textbf{\underline{Comment}:} Above we simply estimated the size of
the overlap integral based on the generic localization of the KK
wavefunctions within a distance $k_{\rm eff}^{-1}$ from $y=L$. In
practice, the overlap integral may be further suppressed due to
destructive interference (from the oscillatory behavior). Such an
interference will be more pronounced when states with very different
masses are involved. Even more importantly, 0-modes need not be
IR-localized, as already illustrated in
Fig.~\ref{fig:FermionZeroModes}. Thus, integrals involving such modes
may be exponentially suppressed due to the wide separation of the
relevant wavefunctions. 
These additional suppressions from the details of the wavefunctions,
which appear on top of the redshift dictated by dimensional analysis,
are usefully described as the result of \textit{physical
localization} effects.. It is in this sense that our conventions for
the \textit{physical wavefunctions}, introduced in the previous
lecture, factor out the universal effects of the geometry
(background) from the \textit{geography of localization} that
characterizes a given model. It will be useful to remember that gauge
0-modes are flat, fermion 0-modes have profiles as in
Eqs.~(\ref{zeromodes}) or (\ref{zeromodescpar}), and scalar 0-modes
arising from 5D gauge fields obeying $(-,-)$ b.c.'s have $f^0_5(y)
\propto e^{2A(y)}$. These properties can have interesting
phenomenological applications, to be discussed below.

%%%%%%%%%%%%%%%%%%%%%%%%%%%%%%%%%%%%%%%
\subsection{Sample Scenarios}
\label{Scenarios}
%%%%%%%%%%%%%%%%%%%%%%%%%%%%%%%%%%%%%%%

We are now ready to present a few representative scenarios based on
warped extra dimensions.

%%%%%%%%%%%%%%%%%%%%%%%%%%%%%%%%%%%%%%%
\subsubsection{The Original Version: RS1}
\label{RS1}
%%%%%%%%%%%%%%%%%%%%%%%%%%%%%%%%%%%%%%%

All the SM fields, except for gravity, are assumed to be
$\delta$-function localized on the IR brane.  This model was proposed
as a solution to the Planck/weak scale hierarchy
problem~\cite{Randall:1999ee}, and as an alternative to supersymmetry:
\begin{itemize}

\item The background is taken as AdS$_5$ so that $A(y) = k y$ (see
Section~\ref{SpecialCases} in the previous lecture).

\item The warped-down curvature scale is taken as $\tilde{k}_{\rm
eff} \equiv k \sim$ TeV. The Higgs mass parameter, and hence EWSB, is
then naturally at the TeV scale.

\item The graviton 0-mode (i.e.~massless spin-2 fluctuations in the
KK spectrum of the 5D metric) is given by
\be
ds^2 &=& e^{-2ky} \left[ \eta_{\mu\nu} + h_{\mu\nu}(x) \right] dx^\mu
dx^\nu - dy^2~,
\label{metricfluctuations}
\ee
i.e.~it shares the same profile as the background. In our
KK-decomposition, we would factor out the $e^{-2ky}$ and define the
graviton \textit{physical} wavefunction as $f^0_G(y) = 1$. As for the
spin-1 case, the flatness of this wavefunction is closely connected
to the necessary universality of the gravitational interactions, here
following from general covariance. The couplings of the graviton are
always set by $M_{P}^2 = ( 1 - e^{-2kL}) \, M^{3}_{5}/k$,
independently of the localization properties of the gravitating
fields (see below).

\end{itemize}
In this model one expects massive spin-2 resonances that may be
narrow or wide, depending on the size of the curvature relative to
the 5D Planck scale. It should be emphasized that 4-fermion
operators, which are necessarily localized on the IR brane, are only
suppressed by ${\cal O}(\rm TeV)$. Thus, as in UED models, the UV
completion must have a special flavor structure to be consistent with
existing flavor (and \cancel{CP}) constraints. In general, one would
expect these models to suffer from severe FCNC and CP problems, which
has motivated the variants discussed next.

%%%%%%%%%%%%%%%%%%%%%%%%%%%%%%%%%%%%%%%
\subsubsection{The SM in a Warped Background}
\label{BulkRS}
%%%%%%%%%%%%%%%%%%%%%%%%%%%%%%%%%%%%%%%

It turns out that giving up the assumption that all the SM fields are
localized on the IR brane opens up a way to address the flavor
problem of RS1, without giving up the RS solution to the hierarchy
problem. The essential feature to be preserved is that the Higgs
(0-mode) be localized within a distance of order  $k_{\rm eff}^{-1}$
from $y=L$, as should be clear from our general discussion in
Section~\ref{WarpedXDim} regarding how the warped down scale is
generated. 
It is, however, not necessary that the SM fields be localized as in
the RS1 model, and in these scenarios one assumes that all fields can
propagate in the bulk.~\footnote{Sometimes one assumes that the Higgs
field is $\delta$-function localized, but even this is not necessary:
the localization need be only of order $k_{\rm eff}^{-1} \ll L$.}

\medskip
\noindent
\textbf{\underline{Gravitational Interactions}:} 

We will not review here the appropriate gauge fixing necessary to
perform the KK decomposition of the gravity sector. The KK expansion
for the spin-2 modes takes the form
\be
G_{\mu\nu}(x,y) &=& e^{-2A(y)} \sum_n g^n_{\mu\nu}(x) f^n_G(y)~,
\ee
where all quantities are \textit{dimensionless}. If we restrict to a
0-mode background, 
\be
ds^2 = e^{-2A(y)} g_{\mu\nu}(x) dx^\mu dx^\nu - dy^2~, 
\ee
we have that ${\cal R}_5[G] = e^{2A} {\cal R}_4[g]$, so that the 5D
Einstein-Hilbert action takes the form
\be
S_{\rm EH} &=& -\frac{1}{2} \int \! d^5 x \, \sqrt{G} \, M^3_5 \,
{\cal R}_5[G] 
\nonumber \\ [0.5em]
&=&
-\frac{1}{2} \int \! d^4 x \, \sqrt{g} \, M^3_5 \int_0^L \! dy \,
e^{-4A} \, e^{2A} \, {\cal R}_4[g]~,
\ee
which identifies the 4D Planck mass as
\be
M^2_P &=& M^3_5 \int_0^L \! dy \, e^{-2A(y)}~.
\ee
In the \textit{strong warping} limit the integrand has support over a
distance $1/k_{\rm UV} \ll L$ from $y = 0$, where $k_{\rm UV} =
A'(0)$ measures the curvature in the vicinity of the UV brane. We
therefore have
\be
M^2_P &\sim& \frac{M^3_5}{k_{\rm UV}}~,
\ee
which has no powers of the warp factor. Thus, we may take $M_5 \sim
k_{\rm UV} \sim M_P$, while still having at our disposal the much
smaller scale in Eq.~(\ref{WarpedKKScale}). 

%%%%%%%%%%%%%%%%%%%%%
\begin{quote}

\textbf{\underline{Exercise}:}
Use NDA to show that the cutoff associated with the 5D gravitational
interactions is given by $\Lambda_5^3 \sim M_5^3/l_5$, where $l_5 =
24\pi^3$ is the 5D loop factor. How does this cutoff compare to the
one associated with the $SU(3)_C$ interactions?
\end{quote}
%%%%%%%%%%%%%%%%%%%%%

\medskip
\noindent
\textbf{\underline{Flavor Anarchy}:} 

The freedom to localize the fermion 0-modes offers an appealing
understanding of the observed fermion mass hierarchies (and quark
mixing angles). Recall that the fermion masses span about six orders
of magnitude, from $y_e \sim 10^{-6}$ to $y_t \sim 1$. The neutrinos
may be associated with even smaller Yukawa couplings. 

\begin{figure}[t]
\begin{center}
\includegraphics[width=0.5 \textwidth]{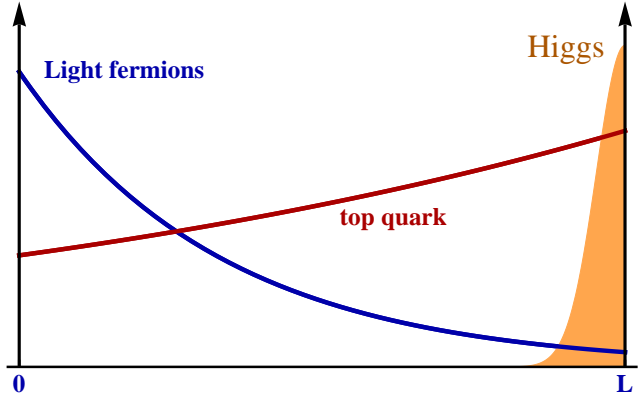}
\caption{Illustration of the ``flavor anarchy" setup. The light
fermions (including $b_R$) are localized towards the UV brane, while
the top is localized near the IR brane. The overlaps with the Higgs
wavefunction can naturally lead to exponential Yukawa hierarchies.}
\label{fig:Anarchy}
\end{center}
\end{figure}

In 5D, a bulk Yukawa operator leads to 
\be
y_5 H \overline{\Psi}_{1L} \Psi_{2L} + {\rm h.c.} &\mapsto&
y_4 H^0 \bar{\psi}^0_{1L} \psi^0_{2R} + {\rm h.c.} +
\textrm{``KK-modes"}~,
\ee
where
\be
y_4 &=& \left( \frac{y_5}{\sqrt{L}} \right) \, \frac{1}{L} \,
\int_0^L \! dy \, f^0_H f^0_{\psi_1} f^0_{\psi_2}
\nonumber \\ [0.5 em]
&\sim&
 \left( \frac{y_5}{\sqrt{L}} \right) \, \frac{1}{k_{\rm H} L} \,
\sqrt{k_{\rm H} L} \, f^0_{\psi_1}(L) f^0_{\psi_2}(L)~,
\ee
and in the second line we assumed that the Higgs 0-mode is localized
within a distance $k^{-1}_H$ of the IR brane, hence $f^0_H(L) \sim
\sqrt{k_{\rm H} L}$. Thus, the 4D Yukawa couplings depend not only on
the 5D Yukawa couplings, but also on the fermion localization. The
assumption of \textit{flavor anarchy} in this context\footnote{The
underlying idea of anarchy was first discussed in the 4D
context~\cite{Hall:1999sn,Nelson:2000sn}.} corresponds to taking the
5D Yukawa matrices to be generic matrices with all entries of the
same order, i.e.~no special flavor structure. The hierarchical
structure of the observed fermion masses (and mixing angles) would be
a consequence of the fermion geography, as illustrated in
Fig.~\ref{fig:Anarchy}. 
The first two generation quarks, plus the RH bottom, are localized
close to the UV brane, so that their overlap with the Higgs field is
small. The top quark, on the contrary, is localized near the IR
brane, and its overlap with the Higgs wavefunction is of order one so
that the top mass is naturally comparable to the electroweak vev. The
overlap of the Higgs profile with the gauge bosons ($Z$ and $W$) is
only mildly suppressed by a volume factor. Thus, these can also get
EWSB masses of order the electroweak vev. For clarity in the figure,
we have oversimplified by not distinguishing between chiralities and
the different light fermions. For instance, typically the RH top is
more localized towards the IR brane than the LH
top.\footnote{Although scenarios with the LH top closer to the IR
brane than the RH top can be motivated in certain scenarios.} The LH
bottom quark is then also somewhat localized towards the IR brane,
but the UV localization of the RH bottom can result in a suppressed
(4D) bottom Yukawa coupling. The upshot is that the observed
structure of fermion masses follows from underlying $c$-parameters in
the vicinity of $1/2$, in the language of
Fig.~\ref{fig:FermionZeroModes}. As a bonus, higher-dimension
operators, such as 5D four-fermion interactions involving the light
families, get also a suppression (beyond the standard redshift) that
amounts to an effective scale well-above the TeV. This dramatically
improves the tension with precision flavor constraints. 

One should point out that there are \textit{tree-level} FCNC effects
arising from the fact that the light families are not localized in
\textit{exactly} the same way. This means that there is some
flavor-dependence in the couplings of the fermions to KK gluons:
\be
g_c &=& \left( \frac{g_5}{\sqrt{L}} \right) \, \frac{1}{L} \,
\int_0^L \! dy \, \left[ f^0_c(y) \right]^2 f_{G'}(y)~,
\ee
where $f^0_c(y)$ are the fermion 0-mode wavefunctions given in
Eq.~(\ref{zeromodescpar}) and $f_{G'}(y)$ is the first KK-gluon
wavefunction (see the blue curve in the left panel of
Fig.~\ref{fig:WarpedKKModes}). Although these couplings are diagonal
in the fermion gauge eigenbasis, the fact that they are
fermion-dependent (through the $c$-dependence), means that in the
mass eigenbasis, off diagonal couplings are induced. For instance,
\be
\put(-35,-32){
\resizebox{4.5cm}{!}{\includegraphics{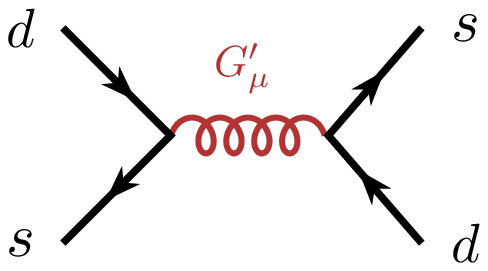}}
}
\hspace*{4cm}
& \Longrightarrow &K^0-\bar{K}^0~\textrm{oscillations :  } \Delta
m^2_K~.
\nonumber
\ee
However, as indicated in the left panel of
Fig.~\ref{fig:WarpedKKModes}, the KK-gluon wavefunction has the
property that it is almost flat, except very near the IR brane. Thus,
when $f^0_c$ is UV-localized, we have
\be
\frac{1}{L} \, \int_0^L \! dy \, \left[ f^0_c(y) \right]^2 f_{G'}(y)
&\approx&
\frac{f_{G'}(0)}{L} \, \int_0^L \! dy \, \left[ f^0_c(y) \right]^2
~=~f_{G'}(0)~,
\ee
i.e.~the c-dependence is very weak. This leads to suppressed
flavor-changing KK-gluon vertices, and is known as the RS-GIM
mechanism.

%%%%%%%%%%%%%%%%%%%%%
\begin{quote}

\textbf{\underline{Exercise}:}
\begin{itemize}

\item Check in the AdS$_5$ limit, using the explicit expressions for
the KK-gluon wavefunction, that the profile has the properties stated
above. Evaluate also the overlap integral to get a feeling for the
$c$-dependence.

\item Argue, from the gauge KK equation, that the previous property
holds more generally in strongly warped backgrounds.

\end{itemize}
\end{quote}
%%%%%%%%%%%%%%%%%%%%%

The above is remarkable: the anarchy assumption amounts to a ``theory
of flavor" with new physics at the TeV scale; and yet the scenario
can be roughly consistent with FCNC constraints. (Naively, such
flavor constraints would put bounds on the new physics scale of about
$10^3$ TeV or higher.) At the same time, the strongest constraints do
arise from the above KK-gluon exchange. When analyzed in detail, one
finds that some CP-violating phases have to be somewhat suppressed.
In spite of this, the above structure remains as one of the striking
properties of the bulk RS scenario.

\medskip
\noindent
\textbf{\underline{Electroweak Constraints}:} 

As already mentioned, one has to check that the SM properties are not
distorted too much by the new physics (i.e.~by the KK-modes). Given
that these are \textit{decoupling} effects, this can be used to set a
lower bound on the KK scale. The most sensitive and robust
constraints are derived from the EW precision measurements, mostly
done at LEP/SLD and the Tevatron. We will present a simple formalism
to perform such an analysis in the next lecture, but will discuss
here the physics in the context of several examples.

\begin{figure}[t]
\begin{center}
\includegraphics[width=0.4 \textwidth]{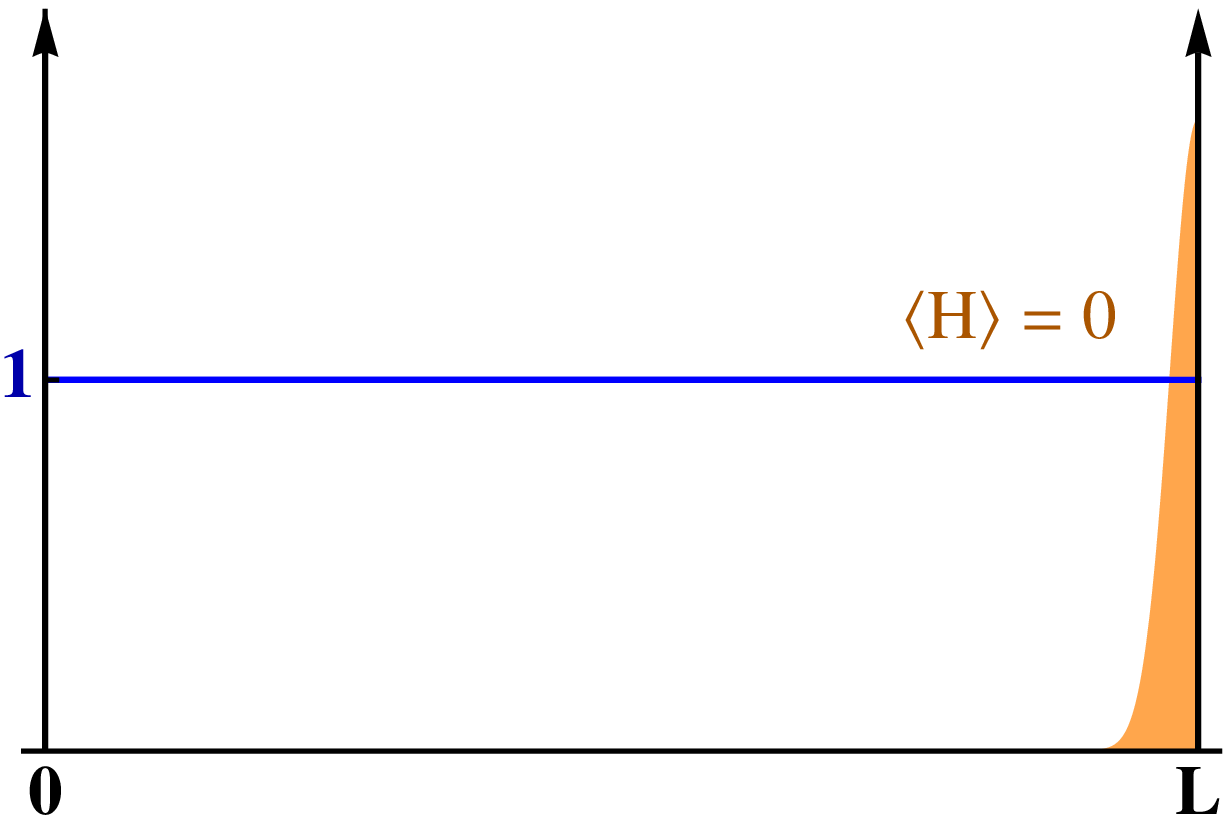}
\hspace{1cm}
\includegraphics[width=0.4 \textwidth]{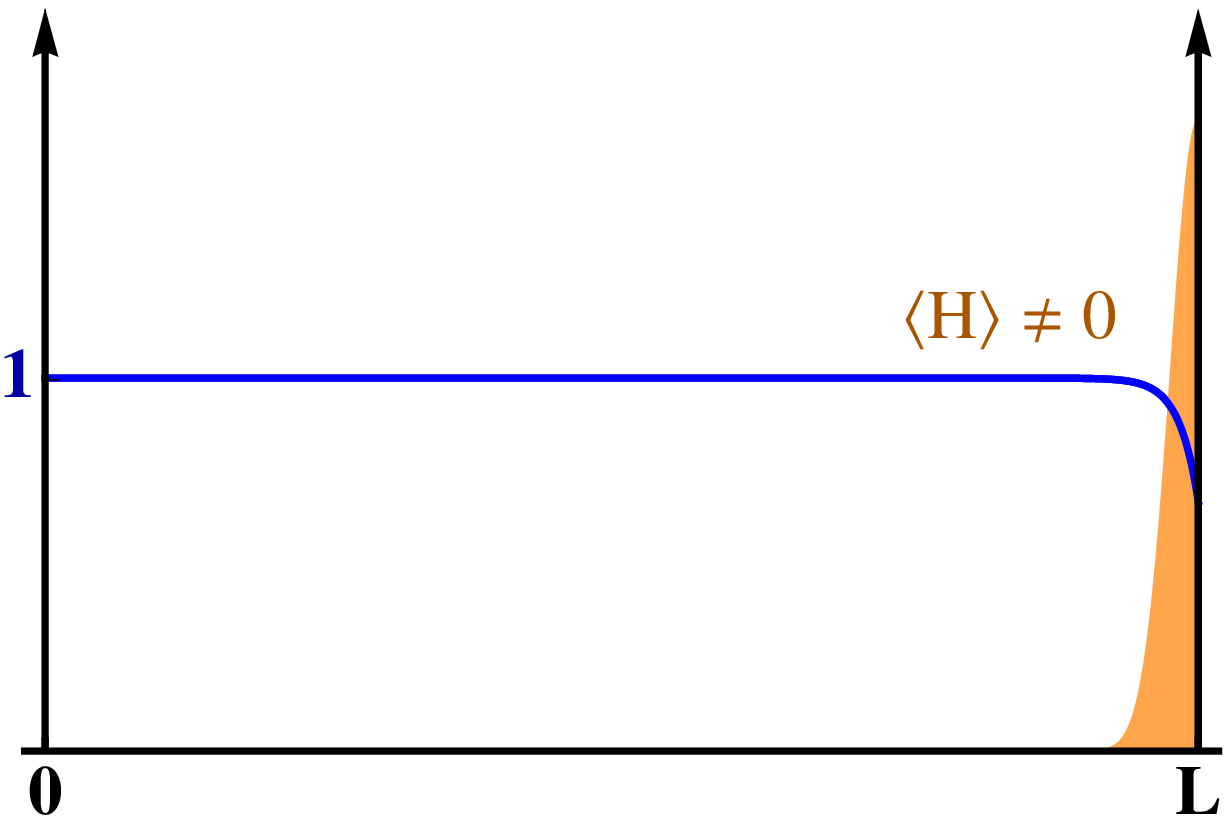}
\caption{Effect of a $y$-dependent Higgs vev on the gauge 0-mode. The
deformation of the wavefunction affects the $Z$ and $W$, but not the
gluons or the photon.}
\label{fig:EWDeformation}
\end{center}
\end{figure}

Generically, the largest deviations from SM properties arise from the
following observation: when the 0-mode Higgs, that by assumption has
an IR-brane localized profile (perhaps $\delta$-function), acquires a
vacuum expectation value, it adds a $y$-dependent mass to the gauge
EOM. The gauge spectrum is shifted (by a small amount if $v \ll
\tilde{k}_{\rm eff}$), and in particular no longer admits an exactly
massless solution.~\footnote{These solutions correspond to linear
superpositions of the ``unperturbed" KK modes described in the
previous sections (i.e.~with $v=0$). Thus, one can understand the
non-trivial effects on the EW precision observables as arising from
mixing of the 0-modes with their massive KK-towers.} The would-be
0-mode wavefunction is modified near the IR brane, as illustrated in
Fig.~\ref{fig:EWDeformation}, 
so that the $y$-gradient near this brane accounts for the non-zero
mass, which would be identified with $M_Z$ or $M_W$. The dominant
source of constraints arises from the requirement to fulfill the
relation $M^2_W \approx M^2_Z \cos^2\theta_w$, which is observed to
hold to excellent accuracy and suggests the existence of a
\textit{custodial symmetry}. More precisely, within the SM, if one
neglects the hypercharge and Yukawa interactions one can show that
the model has an exact global symmetry arising as
\be
SU(2)_L^{\rm gauge} \times SU(2)_R^{\rm global}
&\stackrel{\textrm{EWSB}}{\longrightarrow}&
SU(2)_{\rm Diag}^{\rm global} \equiv SU(2)_{\rm custodial}~.
\ee
Technically, the simplest way to see this is to \textit{rewrite} the
SM (with $g' = y_i = 0$) in terms of a Higgs $SU(2)_L \times SU(2)_R$
``bidoublet"
\be
\Phi &=& 
\left(
\begin{array}{ccc}
{H^{0}}^*  & H^+  \\
 -H^-     &  H^0  
\end{array}
\right)
\psset{unit=1pt}
\rput[lb](-40,-20){
\psline[linestyle=solid,linewidth=1,linecolor=
black,ArrowInside=-,ArrowInsidePos=0.6,arrowinset=0.3,arrowscale=1.5]{<->}(-20,0)(20,0)
\uput{4}[-90]{0}(0,0){\psscaleboxto(35,0){SU(2)_R}}
}
\rput[lb](10,3){
\psline[linestyle=solid,linewidth=1,linecolor=
black,ArrowInside=-,ArrowInsidePos=0.6,arrowinset=0.3,arrowscale=1.5]{<->}(0,-15)(0,15)
\uput{4}[0]{0}(0,0){\psscaleboxto(35,0){SU(2)_L}}
}
\nonumber
\ee
\\ [0.6em]
which implies that its vev, $\langle \Phi \rangle = v  \times
\mathds{1}_{2\times 2}$ preserves the diagonal subgroup of
$SU(2)_L^{\rm gauge} \times SU(2)_R^{\rm global}$. Since the three W
gauge bosons transform as a triplet under $SU(2)_L^{\rm gauge}$ (and
singlets under $SU(2)_R^{\rm global}$), hence like a triplet of
$SU(2)_{\rm custodial}$, the charged and neutral $W$ components are
forced to have the same mass, even after EWSB. Turning on the small
$g'$ induces mixing in the neutral gauge sector (between $W^3_\mu$
and the hypercharge gauge boson), and hence a relatively small mass
splitting between the $Z$ and $W^\pm$ masses, $M^2_W = M^2_Z
\cos^2\theta_w$, while the custodial-breaking effects from the Yukawa
couplings are even smaller, as they enter at loop level (dominantly
from the top Yukawa). 

What happens when we lift the SM to a warped extra dimension? Since
the Lagrangian has the same form as the SM, it is still the case that
for $g'= 0$ and in the absence of Yukawa couplings, there is a
custodial symmetry. Again, there are tree-level violations
proportional to $g'$, and loop-level corrections controlled by the
top Yukawa. The main difference is that the hypercharge KK-modes
couple with an \textit{enhanced} coupling to the Higgs
\be
g_{B^n} &=& \left( \frac{g'_5}{\sqrt{L}} \right) \, \frac{1}{L} \,
\int_0^L \! dy \, \left[ f^0_H(y) \right]^2 f_{B^n}(y)
~\sim~
g'\, \frac{1}{k_{\rm eff} L} \, \left( k_{\rm eff} L \right)^{3/2}~,
\ee
where we estimate the integral based on by now familiar arguments
(assuming, for simplicity, that all wavefunctions are localized
within a distance $1/k_{\rm eff}$ of $y=L$). Thus, $g_{B^n} \sim
\sqrt{k_{\rm eff} L} \, g' \gg g'$, since solving the hierarchy
problem is related to $k_{\rm eff} L \gg 1$. We then see that the
custodial violations due to the hypercharge KK-tower are enhanced
compared to that of the hypercharge 0-mode. The way to suppress them
is by increasing the KK-masses, i.e.~the overall KK scale
$\tilde{k}_{\rm eff}$. Indeed, in the limit that $\tilde{v} \ll
\tilde{k}_{\rm eff}$, where we use the notation $\tilde{v}$ to
emphasize that this is the \textit{warped-down} vev, to be identified
with the weak scale, the effect corresponds to
\be
\put(-35,-20){
\resizebox{5cm}{!}{\includegraphics{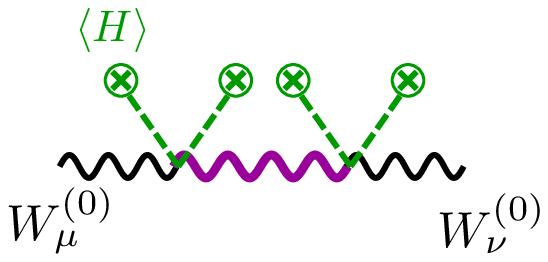}}
}
\hspace*{4cm}
& \propto& g^{\prime 4} \left( k_{\rm eff} L \right) \,
\frac{\tilde{v}^2}{\tilde{k}_{\rm eff}^2} \, \tilde{v}^2~.
\label{TreeLevelT}
\ee
where the internal (magenta) line denotes the exchange of massive KK
modes. Similarly, the couplings of the top KK-tower to the Higgs are
enhanced by $\sqrt{k_{\rm eff} L}$ compared to the top Yukawa, and
lead to
\be
\put(-35,-30){
\resizebox{6cm}{!}{\includegraphics{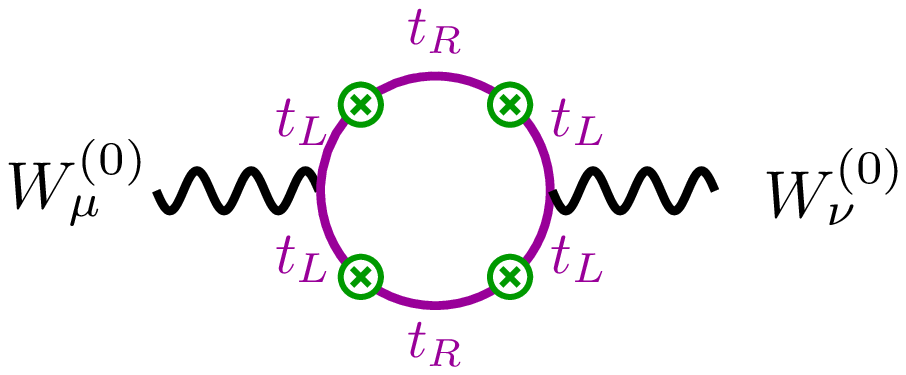}}
}
\hspace*{5cm}
& \propto& y^{4}_t \left( k_{\rm eff} L \right)^4 \,
\frac{\tilde{v}^2}{\tilde{k}_{\rm eff}^2} \, \tilde{v}^2 \times
\textrm{loop integral}~.
\nonumber
\ee
The integral is actually UV sensitive, and the best one can do at
this point is to estimate its size (e.g.~using the NDA philosophy).

The actual analysis shows that in the anarchic RS model (based on
AdS$_5$ and with a $\delta$-function localized Higgs), the first
gauge KK resonance should obey
\be
m_{\rm KK} &\gtrsim& 13~(8)~{\rm TeV}
\hspace{5mm}
\textrm{for}
\hspace{5mm}
m_h ~=~ 120~(500)~{\rm GeV}
\hspace{5mm}
\textrm{at}~
95 \%~\textrm{C.L.}
\nonumber
\ee
The bounds can be somewhat relaxed by allowing the Higgs to propagate
in the bulk (though still localized within $k^{-1}_{\rm eff}$ of
$y=L$), but they are still significant.  There are a number of
approaches one can take to alleviate these bounds and have the chance
of producing the new physics at the LHC. Some involve going beyond the
minimal 5D field content, but not all of them.

%%%%%%%%%%%%%%%%%%%%%%%%%%%%%%%%%%%%%%%
\subsubsection{Relaxing the Bounds from EW Constraints}
\label{EWPT}
%%%%%%%%%%%%%%%%%%%%%%%%%%%%%%%%%%%%%%%

As noted above, the severity of the bounds on the KK scale from EW
precision constraints is directly tied to the strength of the
couplings of the (localized) Higgs and the gauge/top KK towers. The
first class of proposals aim at reducing these couplings, i.e.~reduce
the relevant overlap integrals.

\medskip
\noindent
\textbf{$\bf 1)$ \underline{Brane-localized Kinetic Terms}:} 

We have mentioned several times the possibility of writing boundary
terms (in fact, we have argued that these must be written since they
are induced radiatively). The simplest example corresponds to brane
kinetic terms (BKT's):
\be
{\cal L}_5 &=& \delta(y) \left\{
-\frac{1}{4} r_{\rm UV} F_{\mu\nu} F^{\mu\nu} + \alpha_{\rm UV}
\overline{\Psi} i \gamma^\mu \partial_\mu \Psi
\right\} 
%\nonumber \\ [0.4em]
%& & \mbox{} 
+ \delta(y-L) \left\{
-\frac{1}{4} r_{\rm IR} F_{\mu\nu} F^{\mu\nu} + \alpha_{\rm IR}
\overline{\Psi} i \gamma^\mu \partial_\mu \Psi
\right\}~,
\nonumber
\ee
where the BKT coefficients, $r_i$ and $\alpha_i$, have mass dimension
$-1$, and we illustrate the idea with $\partial_\mu$ derivatives
($\partial_y$ derivatives require more care).  These terms can be
included in a straightforward manner when performing the KK
decomposition, as a modification of the b.c.'s (see previous lecture,
and also Section~\ref{Holography:AdS} in the next one).  They also
modify the orthonormality relations, for instance for the gauge KK
modes, Eq.~(\ref{GaugeKKNorm}), to:
\be
\frac{1}{L} 
\int_0^L \! f^m_A(y) f^n_A(y) \, dy +
\frac{r_{\rm UV}}{L} \, f^m_A(0) f^n_A(0) + \frac{r_{\rm IR}}{L} \,
f^m_A(L) f^n_A(L) 
&=& \delta_{mn}~,
\label{KKGaugeNormBKTs}
\ee
and a similar modification for the fermion wavefunctions with $r_i
\to \alpha_i$. 
\begin{figure}[t]
\begin{center}
\includegraphics[width=0.55 \textwidth]{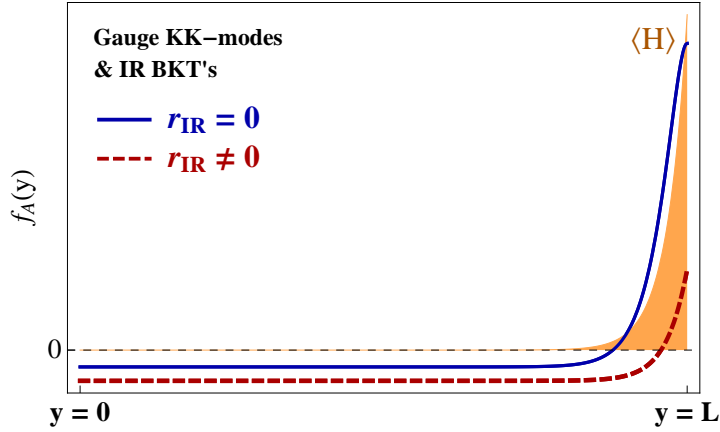}
\caption{Illustration of the modification of the $1^{\rm st}$-level
gauge KK mode when an IR brane-kinetic term (BKT) is turned on. The
KK mass is also lowered (for fixed $\tilde{k}_{\rm eff}$).}
\label{fig:BKTs}
\end{center}
\end{figure}
The effect of the BKT's is to \textit{repel} the wavefunctions from
the respective brane (for positive $r_i$, $\alpha_i$). In
Fig.~\ref{fig:BKTs}, we illustrate the effect of IR BKT's. Due to the
normalization condition, the (absolute) value of the wavefunction in
the UV region increases, and therefore also the couplings to UV
localized fields. We can see in the figure how the overlap with the
Higgs profile would be reduced compared to the case of vanishing
BKT's. In addition, for fixed $\tilde{k}_{\rm eff}$, the masses of
the KK modes are lowered. However, for these effects to be
phenomenologically significant, the size of the BKT coefficients has
to be larger than those induced radiatively. 

Let us also mention here that an important effect of the BKT's is to
modify the relation between the 5D and 4D gauge couplings,
Eq.~(\ref{TreeLevelMatching}):
\be
\frac{1}{g_4^{2}} &=& \frac{1}{g_5^2} \left[ L + r_{\rm UV} + r_{\rm
IR} \right]~.
\label{LoopLevelMatching}
\ee
Sometimes this is called \textit{loop-level matching}, especially
when the BKT's are assumed to arise dominantly from radiative
corrections (in which case $r_{\rm UV} \gg r_{\rm IR}$).

\newpage
\medskip
\noindent
\textbf{$\bf 2)$ \underline{Modifications of the Gravitational
Background}:} 

\begin{figure}[t]
\begin{center}
\includegraphics[width=0.45
\textwidth]{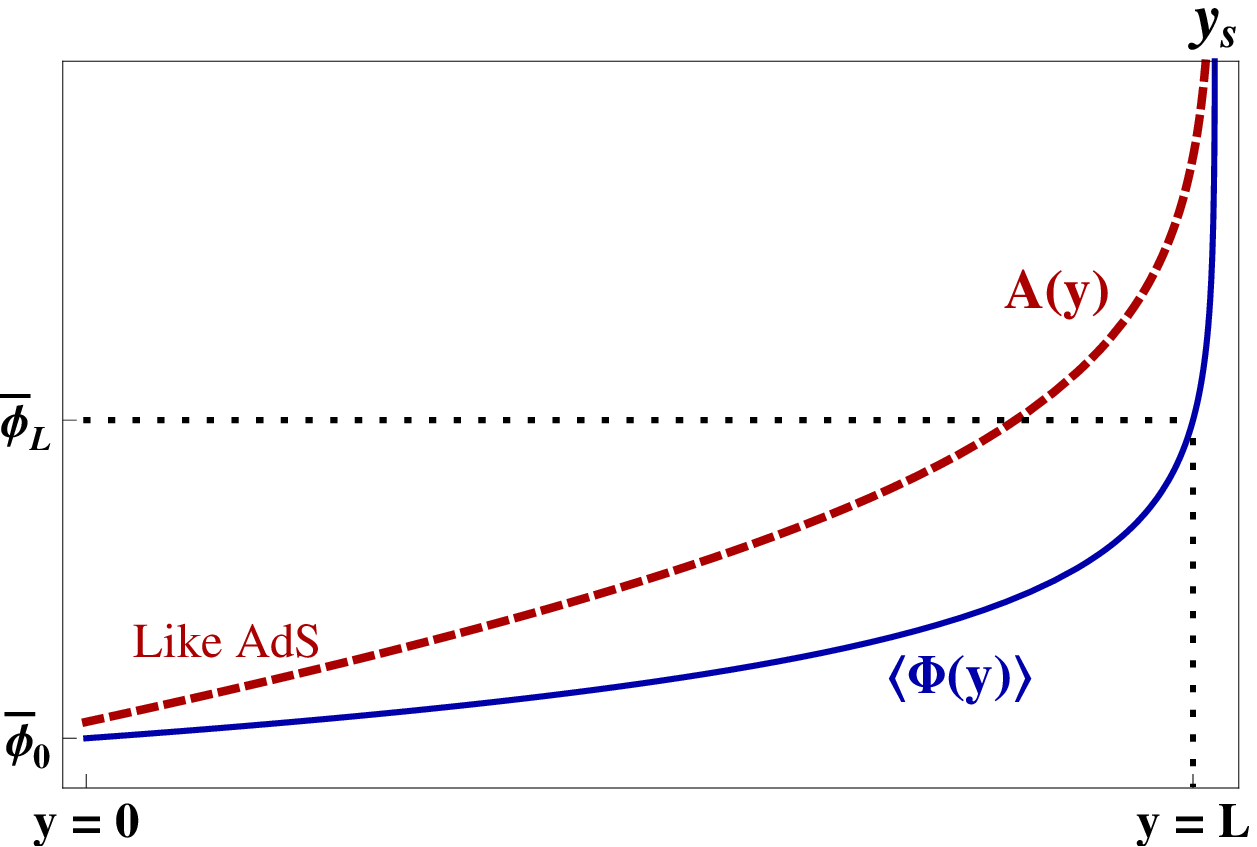}
\hspace{1cm}
\includegraphics[width=0.45 \textwidth]{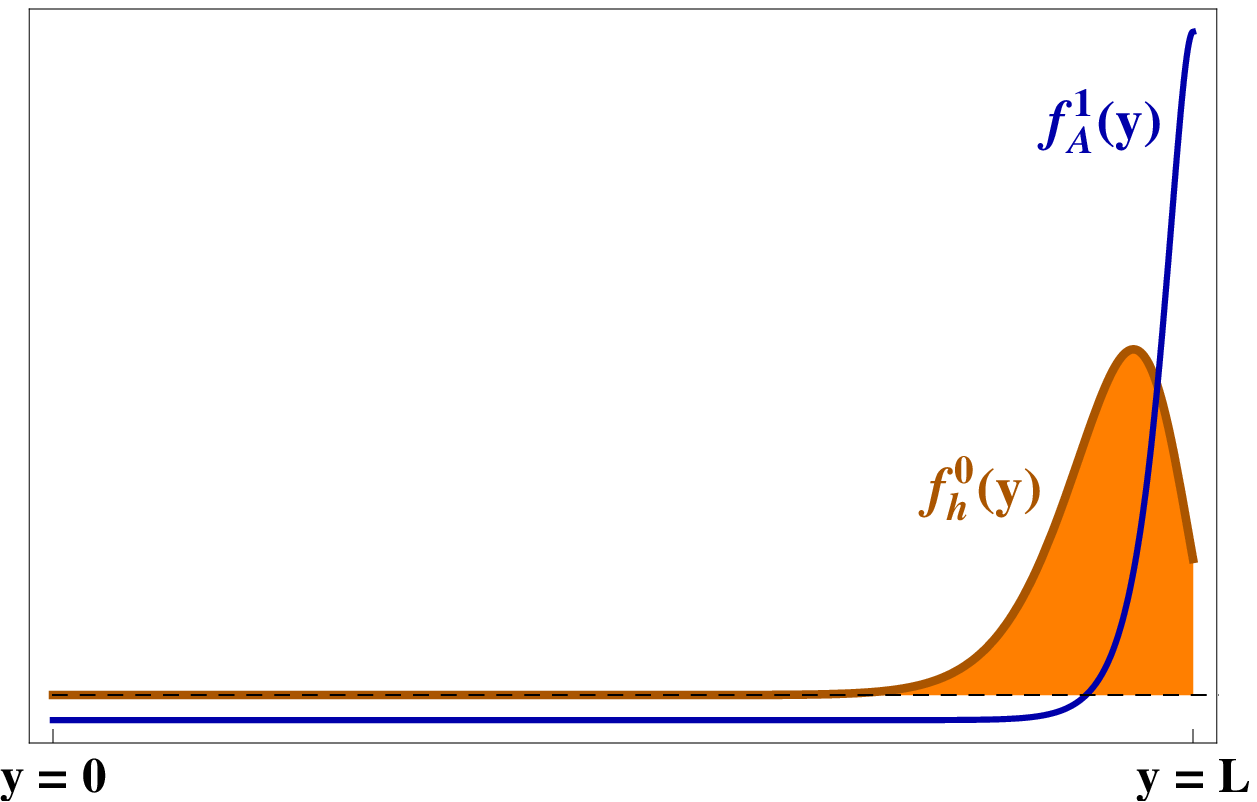}
\caption{Left panel: scalar and metric background profiles for the
superpotential of Eq.~(\ref{WSingular}). We use arbitrary units on
the vertical axis to show both profiles together. Right panel:
associated deformation of the Higgs profile, compared to the first
gauge KK mode.}
\label{fig:SingularBackgrounds}
\end{center}
\end{figure}

A different avenue to relax the bounds on the KK scale is to explore
non-trivial deviations from the AdS$_5$ background, as proposed
in~\cite{Cabrer:2010si}. For instance, including a bulk scalar with a
potential generated by the superpotential
\be
W(\Phi) &=& 6k \left( 1 + e^{\nu \Phi/\sqrt{6}} \right)~,
\label{WSingular}
\ee
the bulk solutions found from the superpotential method described in
the previous lecture give
\be
\langle \Phi \rangle &=& -\frac{\sqrt{6}}{\nu} \, \ln \left[ \nu^2 k
(y_s - y) \right]~,
\\ [0.5em]
A(y) &=& k y - \frac{1}{\nu^2} \, \ln \left( 1 - \frac{y}{y_s}
\right)~,
\ee
which are controlled by two parameters: $\nu$ and $y_s$. These bulk
solutions exhibit a singularity at $y=y_s$, which is chosen to lie
beyond $y=L$, but in its vicinity. In the left panel of
Fig.~\ref{fig:SingularBackgrounds}, we show both the scalar profile
and the metric background function $A(y)$. As explained in the
previous lecture, the relative position of the branes is determined
by the scalar boundary values (b.c.'s). Note that, except close to
the IR brane, $A(y)$ is approximately linear in $y$, which
corresponds to the AdS$_5$ limit. The nearby presence of the
singularity (even if outside the physical region) modifies the
background around the IR brane, which in turn affects the various
wavefunctions through their wave-equations. In order to discuss a
bulk Higgs, one chooses a coupling to the stabilizing scalar as
\be
{\cal L} &=& M(\Phi)^2 H^\dagger H~,
\ee
where $M(\Phi)^2 = a k [ a k - (2/3) W(\Phi)]$ for some dimensionless
$a > 2$. The \textit{physical} Higgs wavefunction (lightest mode) is
approximately given by
\be
f^0_h(y) &=& N_h \, e^{-A(y)} \, e^{a k y}~,
\hspace{2cm}
\frac{1}{L} \, \int_0^L \! dy \, \left[ f^0_h(y) \right]^2 ~=~ 1~.
\ee
The form of the above expression is similar to what one would find in
an AdS$_5$ background, except that here the function $A(y)$ is not
simply linear in $y$ near the IR brane, where the Higgs is localized.
We illustrate the result in the right panel of
Fig.~\ref{fig:SingularBackgrounds}, which illustrates how the overlap
integral between the Higgs profile and the gauge KK modes can be
suppressed by the ``repulsion" of $f^0_h(y)$ from the IR brane. The
gauge KK-wavefunctions are also affected by the deformed background,
being \textit{attracted} towards the IR brane. As a result, one can
have gauge KK resonances as light as $\sim 1~{\rm TeV}$, consistent
with EWPT.

The proposals $1)$ and $2)$ above do not require an extension of the
5D field content to lower the bound on the KK scale (apart from the
stabilizing scalar). The second class of proposals, to be discussed
next, focuses on protecting certain observables by the imposition of
symmetries, which require an extension of the field content. 

%%%%%%%%%%%%%%%%%%%%%%%%%%%%%%%%%%%%%%%
\subsubsection{Bulk Models with Custodial Symmetries}
\label{Custodial}
%%%%%%%%%%%%%%%%%%%%%%%%%%%%%%%%%%%%%%%

The idea here~\cite{Agashe:2003zs} is to force the custodial symmetry
to be as exact as possible by gauging it! Thus, the SM gauge group is
extended to
\be
SU(2)_L^{\rm gauge} \times \underset{\displaystyle \rule{0mm}{4mm}
\supset~U(1)_Y}{\underbrace{SU(2)_R^{\rm gauge} \times U(1)^{\rm
gauge}_X}}~,
\hspace{2cm}
(X = B-L)~.
\label{custodialsymmetry}
\ee
The hypercharge generator is identified with $Y = X + T^3_R$. This
gauge group is broken by boundary conditions, so that only the SM
subgroup has associated massless gauge bosons (before EWSB).
Specifically, one breaks the gauge symmetry down to $SU(2)_L \times
U(1)_Y$ on the UV brane:
\be
\begin{array}{rclcc}
\displaystyle 
W_R^{1,2} 
   & & & \hspace{5mm} &
\displaystyle 
(-,+)   \\ [0.5em]
\displaystyle
Z'  &
\displaystyle
=& 
\displaystyle
\frac{1}{\sqrt{g^2_R+g^{2}_X}} \{ g_R W^3_R - g_X X \}   & &
\displaystyle 
(-,+)
   \\ [0.5em]
\displaystyle 
W_L^{1,2,3}  &
   & & &
\displaystyle 
(+,+)   \\ [0.5em]
\displaystyle
B_\mu  &
\displaystyle
=& 
\displaystyle
\frac{1}{\sqrt{g^2_R+g^{2}_X}} \{ g_X W^3_R + g_R X \}   & &
\displaystyle 
(+,+)
\end{array}
\nonumber
\ee
Note that the LR symmetry remains unbroken on the IR brane, where the
strongly interacting states are localized. Thus, the custodial
symmetry is exact on this boundary, even for $g' \neq 0$.
Schematically,  the $T$-parameter receives contributions from the
massive $W_L$ and $W_R$ fields, leading to a good degree of
cancellation, since at the massive level the differences between L
and R are small, their wavefunctions being suppressed near the UV
brane:
\be
\put(-130,-20){
\resizebox{10cm}{!}{\includegraphics{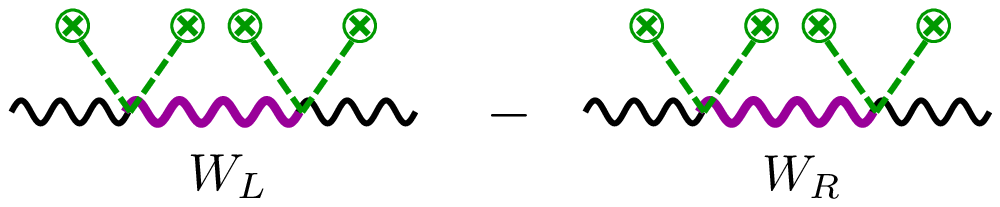}}
}
\nonumber
\ee
To be more precise, the leading $k_{\rm eff} L$ enhanced term  shown
in Eq.~(\ref{TreeLevelT}) cancels out. Of course, at the 0-mode
level, the custodial symmetry is broken, exactly as in the SM. There
are corresponding cancellations in loop diagrams: the bulk gauge
symmetry requires top partners in order to span full bulk gauge
multiplets, and these ensure that the loop diagrams are
\textit{finite}. The finite effect can still play a role in
constraining the model, and should be investigated on a case by case
basis, but the loop suppression factor brings it to the right level
to not be the overwhelming single source of constraints. Typically,
the dominant source of constraints in custodial models arises from
tree-level vertex diagrams, such as
\be
\put(-50,-20){
\resizebox{5cm}{!}{\includegraphics{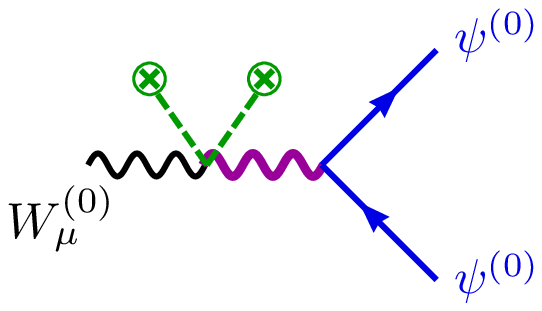}}
}
\nonumber
\ee
which, in anarchic scenarios, are nearly universal for the light
fermions and have no $k_{\rm eff} L$ enhancement.  In this case, they
can be thought as a contribution to the $S$-parameter.

A special case occurs for the well-measured $Zb_L\bar{b}_L$ coupling.
This is because the $(t_L,b_L)$ $SU(2)_L$ doublet cannot be localized
too far from the IR brane or else the top mass would be predicted to
be suppressed compared to the weak scale. This leads to a
\textit{non-universal vertex correction} in the $Zb_L\bar{b}_L$
coupling that can place significant bounds. Interestingly, this
vertex is protected when~\cite{Agashe:2006at}
\be
g_L &=& g_R
\hspace{1cm}
\textrm{and}
\hspace{1cm}
T^3_R(b_L) ~=~ T^3_L(b_L) ~=~ -\frac{1}{2}~.
\ee
Due to the connection to the underlying custodial structure,
Eq.~(\ref{custodialsymmetry}), this is known as a \textit{custodial
protection of $Zb_L\bar{b}_L$}. Realizing the above quantum number
assignments requires the embedding of the $(t_L,b_L)$ $SU(2)_L$
doublet into a $SU(2)_L \times SU(2)_R$ \textit{bidoublet} with
$X=2/3$:
\be
Q_L &=& 
\left(
\begin{array}{ccc}
t_L~(+,+) & \chi_L^{5/3}~(-,+)  \\
b_L~(+,+) & \chi_L^{2/3}~(-,+)
\end{array}
\right)
\psset{unit=1pt}
\rput[lb](-68,-20){
\psline[linestyle=solid,linewidth=1,linecolor=
black,ArrowInside=-,ArrowInsidePos=0.6,arrowinset=0.3,arrowscale=1.5]{<->}(-40,0)(40,0)
\uput{4}[-90]{0}(0,0){\psscaleboxto(35,0){SU(2)_R}}
}
\rput[lb](10,3){
\psline[linestyle=solid,linewidth=1,linecolor=
black,ArrowInside=-,ArrowInsidePos=0.6,arrowinset=0.3,arrowscale=1.5]{<->}(0,-15)(0,15)
\uput{4}[0]{0}(0,0){\psscaleboxto(35,0){SU(2)_L}}
}
\nonumber
\ee
\\ [0.6em]
Here we also exhibit the boundary conditions for the LH components of
the fermion bi-doublet, $Q$, which are chosen to preserve the full
$SU(2)_L \times SU(2)_R$ on the IR brane, but to lift the unwanted
fermion 0-modes by the choice on the UV. The b.c.'s on the RH
chiralities can be read from the above, following the general rules
explained in the previous lecture. The superscripts on the new 5D
fermions, $\chi_L^{5/3}$ and $\chi_L^{2/3}$, indicate their electric
charges, where $Q = T^3_L + T^3_R + X$. The new charge-$2/3$ tower,
with $T^3_L(\chi^{2/3}_L) = - T^3_L(t_L)$, is responsible for the
required cancellations in the fermion loop contributions to the
$T$-parameter. Interestingly, a prediction of the setup is the
presence of  fermion resonances with an exotic electric charge, $\pm
5/3$. A global fit to the EW precision data shows
that~\cite{Delaunay:2010dw}, under the assumption of flavor anarchy,
\be
m_{\rm KK} &\gtrsim& 4~{\rm TeV}
\hspace{5mm}
\textrm{at}~
95 \%~\textrm{C.L.}
\nonumber
\ee
However, if one relaxes the anarchy assumption, one can have
\be
m_{\rm KK} &\gtrsim& 2~{\rm TeV}
\hspace{5mm}
\textrm{at}~
95 \%~\textrm{C.L.}
\nonumber
\ee
The original RS flavor problem can still be addressed through a
version of the \textit{General Minimal Flavor Violation} ansatz: in
this extra-dimensional realization, the SM flavor symmetries are
gauged, but broken by b.c.'s on the UV brane, in analogy to the
breaking of the LR symmetry described above.

%%%%%%%%%%%%%%%%%%%%%%%%%%%%%%%%%%%%%%%
\subsubsection{Models of Dynamical EWSB}
\label{DynESWB}
%%%%%%%%%%%%%%%%%%%%%%%%%%%%%%%%%%%%%%%

So far, we have treated the Higgs as a ``fundamental" field, while
EWSB is described --in analogy with the SM-- by writing an
appropriate phenomenological Higgs potential (with the important
difference that the warping provides the TeV scale). However, it is
possible to go further, and use the extra-dimensional structure to
construct models where the EW symmetry is broken
\textit{dynamically}. Here we mention two examples:
\begin{itemize}

\item \underline{Condensation}: no Higgs scalar is introduced. The
strong interactions associated with the KK gluons may trigger the
condensation of fermion 0-mode pairs, if these are localized
sufficiently close to the IR brane. If the fermion bilinear has
non-trivial EW quantum numbers, EWSB
ensues~\cite{Dobrescu:1998dg,Burdman:2007sx,Bai:2008gm}. A relatively
light scalar bound state of these fermions unitarizes $W_L W_L \to
W_L W_L$ scattering.

\item \underline{Gauge-Higgs Unification (GHU)}: in these models
there are no 5D scalars, but 4D scalars arise from the
extra-dimensional polarizations of gauge fields obeying $(-,-)$
b.c.'s. The 5D gauge symmetry forbids a tree-level (5D) potential for
these modes, but the compactification leads to a calculable (i.e.~UV
insensitive) potential~\cite{Hosotani:1983xw}. The interplay between
the top Yukawa and weak gauge interactions generates a non-vanishing
expectation value for these scalars, which if appropriately charged
under the EW group induce EWSB.

\end{itemize}
Let us only mention a few aspects of one of the most studied GHU
scenarios~\cite{Agashe:2004rs}. One needs a gauge group large enough
to contain the SM group plus additional gauge fields that can give
rise to the 4D scalar 0-modes with the quantum numbers of the Higgs.
One may also want to embed the custodial symmetry in the model. The
smallest group that satisfies these requirements is
\be
SO(5) \times U(1)_{B-L} 
&
\psset{unit=1pt}
\rput[lb](0,4){
\psline[linestyle=solid,linewidth=0.8,linecolor=black,ArrowInside=-,ArrowInsidePos=0.6,arrowinset=0.2,arrowscale=1.3]{->}(0,0)(25,14)
\psline[linestyle=solid,linewidth=0.8,linecolor=black,ArrowInside=-,ArrowInsidePos=0.6,arrowinset=0.2,arrowscale=1.3]{->}(0,0)(25,-14)
}
\hspace{1cm}
&
\begin{array}{lcl}
\displaystyle
SU(2)_L \times U(1)_Y  & &
\textrm{on the UV brane}    \\ [1em]
\displaystyle
SU(2)_L \times SU(2)_R \times U(1)_{B-L}  & &
\textrm{on the IR brane} 
\end{array}~,
\nonumber
\ee
where the breaking occurs by b.c.'s. The Higgs scalars are embedded as
\be
A_5^{\hat{a}} T^{\hat{a}} &=&
\frac{1}{\sqrt{2}}
\left(
\begin{array}{cccc:c}
  &   &  &  & H^+   \\ [0.3em]
  &   &\multirow{3}{*}{$0_{4 \times 4}$}  &  & H^0   \\ [0.3em]
  &   &  &  & -H^-   \\  [0.3em]
  &   &  &  & -{H^0}^*   \\ [0.3em]
\hdashline
\rule{0mm}{5mm}
H^- & {H^0}^* & -H^+ & -H^0 & 0   \\ 
\end{array}
\right)~,
\label{A5Higgs}
\ee
where $\hat{a}$ runs over the $SO(5)/SO(4)$ generators.
The embedding of the fermion sector is somewhat complicated and there
are several possible realizations. Thus, we will be content with
describing the results in general terms. The computation of the Higgs
potential was illustrated, in flat space, in the TASI 2005 Lecture on
``To the $5^{th}$ Dimension and Back", by
R.~Sundrum~\cite{Sundrum:2005jf}. An elegant computation of the
computation for arbitrary warp factor was given
in~\cite{Falkowski:2006vi}. The object of interest is the
Coleman-Weinberg potential, which for a real bulk scalar with
non-trivial $SO(5) \times U(1)_{B-L}$ quantum numbers takes the form
\be
V &=& \frac{1}{2} \, \sum_n \, \int \! \frac{d^4 p}{(2\pi)^4} \, \ln
\left( p^2 + m^2_n \right)~.
\ee
Here $m_n$ is the spectrum obtained by the scalar KK-decomposition
\textit{in the presence of the Higgs background},
Eq.~(\ref{A5Higgs}), and contains the information about the metric
background as well as the imposed boundary conditions. The
contributions from fields of other spins are obtained by using the
corresponding spectrum and multiplying by the number of 4D massive
degrees of freedom (2 for complex scalars, 4 for fermions, 3 for real
gauge bosons, 5 for the graviton), and including a $(-1)$ for the
fermionic contributions.

\begin{figure}[t]
\begin{center}
\includegraphics[width=0.5 \textwidth]{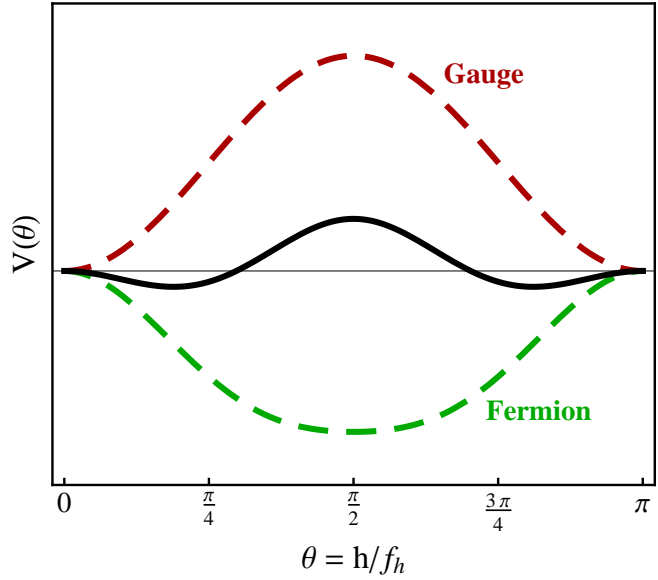}
\caption{Radiatively induced Higgs potential in GHU models,
illustrating the interplay between the gauge (stabilizing) and
fermion (destabilizing) effects. The black curve displays a
non-trivial minimum in the \textit{angular} variable $h/f_h$.}
\label{fig:GHUPotential}
\end{center}
\end{figure}

The above Coleman-Weinberg potential can be manipulated into a
convergent integral
\be
V &=& \sum_{r = {\rm fields}} \frac{N_r}{16\pi^2} \, \int_0^\infty \!
dp \, p^3 \ln \left[ 1 + F_r(-p^2) \,\sin^2\left( \frac{\lambda_r
h}{f_h} \right) \right]~,
\label{GHUPotential}
\ee
where $\lambda_r$ is a representation-dependent, order one constant,
\be
f^2_h &=& \frac{1}{g^2_5 \, \int_0^L \! dy \, e^{-2A(y)}}
\ee
is the \textit{Higgs decay constant}, and $h$ is the Higgs vev. The
dependence on the latter enters only through the explicit sine in
Eq.~(\ref{GHUPotential}). The \textit{form factor} $F_r(-p^2)$
contains the model information. Falkowski found simple approximations
that capture the physics quite well~\cite{Falkowski:2006vi}
\be
\begin{array}{lcrcl}
\textrm{Gauge:}  & \hspace{5mm} &
\displaystyle
F_{(1)}(-p^2)  &\approx& 
\displaystyle
\frac{g^2 f^2_h}{M^2_{\rm KK} \sinh^2\left( p/M_{\rm KK} \right)}~,
\\ [1em]
\textrm{Fermion:}  & &
\displaystyle
F_{(1/2)}(-p^2)  &\approx& 
\displaystyle
\frac{y^2 f^2_h}{M^2_{\rm KK} \sinh^2\left( p/M_{\rm KK} \right)}~,
\end{array}
\ee
where $g$ and $y$ are the 0-mode gauge and Yukawa couplings,
respectively, and
\be
M_{\rm KK} &\equiv& \frac{1}{\int_0^L \! dy \, e^{-A(y)}}~.
\ee
In Fig.~\ref{fig:GHUPotential} we illustrate how the gauge and Yukawa
contributions can generate a non-vanishing minimum for $h$. This
minimum can lie at $h \ll f_h$, which corresponds to a limit where
the Higgs has SM-like properties.
We refer the reader to the literature for further details of GHU
scenarios, and close here our (non-exhaustive) survey of
extra-dimensional models with phenomenological applications at the
weak scale. 

\newpage
%%%%%%%%%%%%%%%%%%%%%%%%%%%%%%%%%%%%%%%
\section{Lecture 4: Introduction to Holography}
\label{Holography}
%%%%%%%%%%%%%%%%%%%%%%%%%%%%%%%%%%%%%%%

In the previous lectures we have analyzed compact extra-dimensional
theories by \textit{rewriting} them in terms of Kaluza-Klein modes.
This language is of direct interest for the collider phenomenology,
since it is these (mass eigenstates) that would be produced in
high-energy collisions.

However, the KK decomposition is not the only rewriting that is
useful. In this lecture, we review the \textit{holographic approach},
which in certain cases provides the most direct way of deriving
properties of the theory. One such application is to the analysis of
EW constraints. In addition, holography gives rise to a \textit{dual
4D description of extra-dimensional models} that is quite interesting
in its own right.

%%%%%%%%%%%%%%%%%%%%%%%%%%%%%%%%%%%%%%%
\subsection{The Holographic Approach: Scalar Case}
\label{Holography:Scalars}
%%%%%%%%%%%%%%%%%%%%%%%%%%%%%%%%%%%%%%%

It will be useful to illustrate the idea, and its connection to the
KK picture, in the simplest case of a real bulk scalar, propagating
in our generic background, Eq.~(\ref{metric}). First, let us
generalize slightly our previous free scalar action to
\be
S &=& \int \! d^5x \sqrt{g} \, \left\{ \frac{1}{2} \partial_M \Phi
\partial^M \Phi - \frac{1}{2} \, M(y)^2 \Phi^2 \right\}
+ \int_{\rm UV} \!\!\!\!  d^4 x \sqrt{\bar{g}} \, {\cal L}_0(\Phi) + 
\int_{\rm IR} \!\!\!  d^4 x \sqrt{\bar{g}} \, {\cal L}_L(\Phi)~,
\label{ScalarActionHolography}
\ee
by allowing a potentially $y$-dependent mass, as well as (quadratic)
boundary terms.\footnote{The philosophy here is the same as in the KK
approach: use the free theory to define a useful set of variables,
and then treat possible interactions perturbatively, after rewriting
them in terms of the ``free variables".} The latter will typically
translate into a certain form of the boundary conditions, as dictated
by the variational principle discussed in Lecture~2. For instance,
consider
\be
{\cal L}_i(\Phi) &=& \frac{1}{2} \, r_i \, \bar{g}^{\mu\nu}
\partial_\mu \Phi \partial_\nu \Phi - \frac{1}{2} \, M_i \Phi^2~,
\label{BoundaryL}
\ee
where $\bar{g}_{\mu\nu}$ denotes the induced metric, and $r_i$ and
$M_i$ have mass dimension $-1$ and $1$, respectively (recall that
$[\Phi] = 3/2$).  If we write, for convenience~\footnote{This is
particularly useful when $k_{\rm UV} \sim k_{\rm IR} \equiv k_{\rm
eff}$, and $M^2 = (\alpha^2 - 4) \, k^2_{\rm eff}$.}
\be
M_0 &=& - (\alpha - 2) \, k_{\rm UV} + m_{\rm UV}~,
\\ [0.5em]
M_L &=& + (\alpha - 2) \, k_{\rm IR} + m_{\rm IR}~,
\ee
where $k_{\rm UV} \equiv A'(0)$ and $k_{\rm IR} \equiv A'(L)$, the
associated b.c.'s read
\be
\left. \partial_y \Phi \rule{0mm}{3.5mm} \right|_0 &=& \left.  -
\left[ \left( \alpha - 2 \right) + b_{\rm UV} \!\! \left( \frac{i
\partial_\mu}{k_{\rm UV}} \right) \right] k_{\rm UV} \Phi  \right|_0~,
\label{PhiUVBC}
\\ [0.5em]
\left. \partial_y \Phi \rule{0mm}{3.5mm} \right|_L &=& \left.  -
\left[ \left( \alpha - 2 \right) - b_{\rm IR} \!\! \left(
\frac{e^{A(L)} \, i \partial_\mu}{k_{\rm IR}} \right) \right] k_{\rm
IR} \Phi  \right|_L~,
\label{PhiIRBC}
\ee
where we defined, for $i= {\rm UV}, {\rm IR}$,
\be
b_i(z) &=& \hat{r}_i z^2 - \hat{m}_i~,
\hspace{2cm}
\hat{r}_i ~\equiv~r_i k_i~,~~
\hat{m}_i ~\equiv~ \frac{m_i}{k_i}~.
\label{bi}
\ee
It is understood here that when $z \propto \partial_\mu$, then $z^2
\propto \partial_\mu \partial^\mu \equiv \eta^{\mu\nu} \partial_\mu
\partial_\nu$, i.e.~the Lorentz contraction is done with the Minkowski
metric.  Our sign conventions are such that \textit{positive} $r_i$
($m_i$) \textit{decrease} (\textit{increase}) the KK masses, as could
be expected from physical considerations.

%%%%%%%%%%%%%%%%%%%%%
\begin{quote}

\textbf{\underline{Exercise}:}
Convince yourself that by sending $\hat{m}_i \to \infty$ one recovers
the Dirichlet b.c.'s on the respective boundary. 
\end{quote}
%%%%%%%%%%%%%%%%%%%%%

In the holographic approach, the UV and IR boundaries are treated
asymmetrically. In particular, special importance is given to the UV
brane (where $A(0) = 0$). The scalar theory above is subjected to a
two-step process. In the first step, while the b.c.~on the IR brane
is imposed as in the KK analysis, one requires that on the UV
brane the field attains a prescribed value, $\phi_0(x)$:
\be
\Phi(x,y=0) &=& \phi_0(x)
\hspace{1cm}
\overset{\textrm{4D mom.~space}}{\rule{0mm}{3mm}\Longrightarrow}
\hspace{1cm}
\tilde{\Phi}(p,y=0) ~=~\tilde{\phi}_0(p)~,
\ee
where the tildes on the fields indicate a 4D Fourier transform.  To
keep track of the connection to our KK analysis, we can factor out one
power of the warp factor [as we did to define the \textit{physical
wavefunctions} in Eq.~(\ref{KKDecompScalar})]
\be
\tilde{\Phi}(p,y) &\equiv& e^{A(y)} \, \tilde{\phi}(p,y)~,
\label{PhysNorm}
\ee
so that the EOM and IR b.c.~read [c.f.~Eq.~(\ref{ScalarfnEOM})]:
\be
0 &=& \partial^2_y \tilde{\phi} - 2 A' \partial_y \tilde{\phi} + 
\left[ A^{\prime\prime} - 3 (A')^2 - M^2 + e^{2A} p^2 \right]
\tilde{\phi}~,
\label{HolEOM}
\\ [0.5 em]
\left. \partial_y \tilde{\phi} \rule{0mm}{3mm} \right|_L &=&  -
\left[ \left( \alpha - 1 \right) - b_{\rm IR} \!\! \left( \frac{p
\, e^{A(L)}}{k_{\rm IR}}\right)  \right] k_{\rm IR} \left.
\tilde{\phi} \rule{0mm}{3mm} \right|_L~.
\label{HolBC}
\ee
Note that compared to the KK equations we have $m_n \to p$.
If we replace such a solution back into the action and integrate by
parts, we find that
\begin{enumerate}

\item The bulk part is proportional to the EOM, Eq.~(\ref{HolEOM}).

\item The surface term generated at $y=L$ from the integration by
parts combines with ${\cal L}_L(\Phi)$ to reproduce the IR b.c.,
Eq.~(\ref{HolBC}).

\end{enumerate}
It follows that these two contributions vanish, and the action
evaluated on the above solution is determined by ${\cal L}_0(\Phi)$
and the generated surface term at $y=0$:
\be
S[\phi_0] &=& \int \! \frac{d^4 p}{(2\pi)^4} \, \left. \left\{
\frac{1}{2} \, \tilde{\Phi} \, \partial_y \tilde{\Phi} + {\cal
L}_0(\tilde{\Phi})
\right\} \right|_0
\nonumber \\[0.5em]
&=& \int \! \frac{d^4 p}{(2\pi)^4} \, \left. \left\{
\frac{1}{2} \, \tilde{\phi} \, \partial_y \tilde{\phi} + \frac{1}{2}
k_{\rm UV} \tilde{\phi}^2 + {\cal L}_0(\tilde{\phi})
\right\} \right|_0~.
\label{Stmp}
\ee
\noindent
Note that this action depends only on $\tilde{\phi}_0$, the UV
boundary field, that will be called the \textit{holographic field}.
In particular, $\left.  \partial_y \tilde{\phi} \right|_0$ is fixed
by the holographic field and the IR boundary condition. 

%%%%%%%%%%%%%%%%%%%%%
\begin{quote}

\textbf{\underline{Exercise}:}
Derive the previous result, Eq.~(\ref{Stmp}).

\end{quote}
%%%%%%%%%%%%%%%%%%%%%

At this point it is convenient to define a \textit{holographic
profile}, $K(p,y)$, with $\tilde{\phi}_0$ factored out:
\be
\tilde{\phi}(p,y) &=& \tilde{\phi}_0(p) \cdot K(p,y)~,
\ee
where $K(p,y)$ obeys the same EOM and IR b.c.~as $\tilde{\phi}$,
Eqs.~(\ref{HolEOM}) and (\ref{HolBC}), and in addition
\be
K(p,y=0) &=& 1~.
\ee
In terms of this profile, we get the \textit{holographic action}
\be
S[\phi_0] &=& \int \! \frac{d^4 p}{(2\pi)^4} \, \left\{
\frac{1}{2} \, \tilde{\phi}^2_0 \left. \partial_y K(p,y)
\rule{0mm}{3.5mm} \right|_0 + \frac{1}{2} k_{\rm UV} \tilde{\phi}^2_0
+ {\cal L}_0(\tilde{\phi}_0)
\right\}~,
\label{HolographicAction}
\ee
where $\tilde{\phi}_0^2$ is understood as $\tilde{\phi}_0(p)
\tilde{\phi}_0(-p)$.  In the second step of the holographic approach,
one requires that $S[\phi_0] $ be stationary against variations in
$\phi_0$:
\be
\delta S[\phi_0] &=& \int \! \frac{d^4 p}{(2\pi)^4} \, \left\{
\tilde{\phi}_0 \left. \partial_y K(p,y) \rule{0mm}{3.5mm} \right|_0 +
k_{\rm UV} \tilde{\phi}_0 + \frac{\partial {\cal L}_0}{\partial
\tilde{\phi}_0}
\right\} \delta \tilde{\phi}_0
\nonumber \\ [0.5em]
&=& 0 
\hspace{2cm}
\textrm{for any}~\delta \tilde{\phi}_0~.
\ee
It follows that
\be
\left. \partial_y \tilde{\phi} \rule{0mm}{3.5mm} \right|_0 &=& \left.
- k_{\rm UV} \tilde{\phi} - \frac{\partial {\cal L}_0}{\partial
\tilde{\phi}} \right|_0~.
\ee
For ${\cal L}_0$ as given in Eq.~(\ref{BoundaryL}), we get
\be
\left. \partial_y \tilde{\phi} \rule{0mm}{3mm} \right|_0 &=&  -
\left[ \left( \alpha - 1 \right) k_{\rm UV} + r_{\rm UV} \, p^2 -
m_{\rm UV} \right] \left. \tilde{\phi} \rule{0mm}{3mm} \right|_0~,
\ee
which matches the UV b.c.~Eq.~(\ref{PhiUVBC}) after recalling
Eq.~(\ref{PhysNorm}).

\medskip
\noindent
\textbf{\underline{The lesson is that}:} 

\begin{enumerate}

\item Considering classical solutions that take a \textit{fixed}
field value at $y=0$, and 

\item Then allowing the fixed value to vary arbitrarily
(i.e.~integrating over it in the path-integral sense),

\end{enumerate}
is completely equivalent to performing a KK analysis. This
observation is, in a sense, remarkable. We have that the 4D
holographic theory, Eq.~(\ref{HolographicAction}), treated with the
usual tools (i.e.~deriving the EOM by the principle of least action
w.r.t.~$\phi_0$), contains all the information of the 5D theory. The
latter describes an infinite number of 4D degrees of freedom (the KK
modes), yet we seem to have a single $\phi_0(x)$ in the holographic
action.

On closer inspection, we see that $S[\phi_0]$ is highly non-local,
i.e.~the momentum dependence of $\left. \partial_y K(p,y)
\rule{0mm}{3.5mm} \right|_0$ can be non-analytic in $p$, as dictated
by the 5D EOM. What happens is that
\be
\Pi \,\tilde{\phi}_0 &\equiv& \left\{
\left. \partial_y K(p,y) \rule{0mm}{3.5mm} \right|_0 + k_{\rm UV}
\right\} \tilde{\phi}_0 + \frac{\partial {\cal L}_0}{\partial
\tilde{\phi}_0}
\label{Pi}
\ee
vanishes for an infinite number of values of $p$. When ${\cal
L}_0$ is quadratic, these solutions coincide with the KK masses: $p^2
= m^2_n$.

To exhibit this in more detail, consider the \textit{Neumann
propagator}, $G^N(p; y,y')$, satisfying
\be
\left\{
\partial^2_y - 2 A' \partial_y + 
\left[ A^{\prime\prime} - 3 (A')^2 - M^2 + e^{2A} p^2 \right]
\right\}G^N(p; y,y') &=& e^{2A} \delta(y-y')~,
\label{NeumannPropEOM}
\ee
as well as the boundary conditions
\be
\left. \left\{ \partial_y + \left[ \left( \alpha - 1 \right) +
b_{\rm UV} \!\! \left( \frac{p}{k_{\rm UV}}\right)  \right] k_{\rm
UV} \right\} G^N(p; y,y') \rule{0mm}{3mm} \right|_{y=0} &=& 0~,
\label{PropUVBC}
\\ [0.5em]
\left. \left\{ \partial_y + \left[ \left( \alpha - 1 \right) -
b_{\rm IR} \!\! \left( \frac{p \, e^{A(L)}}{k_{\rm IR}}\right)
\right] k_{\rm IR} \right\} G^N(p; y,y') \rule{0mm}{3mm}
\right|_{y=L} &=& 0~.
\label{PropIRBC}
\ee
To find an expression for $G^N(p; y,y')$ it is convenient to use the
\textit{holographic profile} $K(p,y)$, together with a second
function that also obeys the scalar EOM, Eq.~(\ref{HolEOM}),  but
satisfies the b.c.'s~\cite{Falkowski:2008yr}
\be
S(p,y=0) &=& 0~,
\\ [0.5em]
\left. \partial_y S(p,y)  \rule{0mm}{3mm} \right|_{y=0} &=& 1~.
\ee
The Neumann propagator can then be expressed as 
\be
G^N(p; y,y') &=& \frac{1}{\Pi(p)} \, K(p,y) K(p,y') - S(p,y_<) K(p,y_>)~,
\label{GN}
\ee
where $\Pi$ is defined by Eq.~(\ref{Pi}) and $y_> = \textrm{max}\{
y,y' \}$, $y_< = \textrm{min}\{ y,y' \}$. 

%%%%%%%%%%%%%%%%%%%%%
\begin{quote}

\textbf{\underline{Exercise}:}
\begin{enumerate}

\item[a)] Check that the above expression for $G^N(p; y,y')$ indeed
satisfies Eqs.~(\ref{NeumannPropEOM})--(\ref{PropIRBC}).

\item[b)] Show that $G^D(p;y,y') \equiv -S(p,y_<) K(p,y_>)$ is the
\textit{Dirichlet propagator} satisfying Eq.~(\ref{NeumannPropEOM})
and the IR b.c., Eq.~(\ref{PropIRBC}), but obeying the UV
b.c.~$G^D(p;y=0,y') = 0$. 

\end{enumerate}
\end{quote}
%%%%%%%%%%%%%%%%%%%%%
%
It follows that
\be
G^N(p; 0,0) &=& \frac{1}{\Pi(p)}~,
\label{Pizeroes}
\ee
and therefore that the holographic action,
Eq.~(\ref{HolographicAction}), can be written as (recall we are
assuming here that ${\cal L}_0$ is quadratic in the field)
\be
S[\phi_0] &=& \int \! \frac{d^4 p}{(2\pi)^4} \, \frac{1}{2} \,
\tilde{\phi}_0 \, \Pi \, \tilde{\phi}_0
\nonumber \\ [0.4em]
&=&  \int \! \frac{d^4 p}{(2\pi)^4} \, \frac{1}{2} \, \tilde{\phi}_0
\left[ G^N(p; 0,0) \right]^{-1} \tilde{\phi}_0~.
\ee

On the other hand, if we define the KK wavefunctions by applying the
variational procedure reviewed in Lecture 2 to the scalar action,
Eq.~(\ref{ScalarActionHolography}), namely
\be
\left\{
\frac{d^2}{dy^2} - 2 A' \frac{d}{dy} + 
\left[ A^{\prime\prime} - 3 (A')^2 - M^2 + e^{2A} m^2_n \right]
\right\} f_n(y) &=& 0~,
\label{fnEOM}
\\ [0.5em]
\left. \left\{ \frac{d}{dy} + \left[ \left( \alpha - 1 \right) +
b_{\rm UV} \!\! \left( \frac{m_n}{k_{\rm UV}}\right)  \right] k_{\rm
UV} \right\} f_n(y) \rule{0mm}{3mm} \right|_{y=0} &=& 0~,
\label{fnBCUV}
\\ [0.5em]
\left. \left\{ \frac{d}{dy} + \left[ \left( \alpha - 1 \right) -
b_{\rm IR} \!\! \left( \frac{m_n \, e^{A(L)}}{k_{\rm IR}}\right)
\right] k_{\rm IR} \right\} f_n(y) \rule{0mm}{3mm} \right|_{y=L} &=& 0~,
\label{fnBCIR}
\ee
where the $f_n$ are normalized according to [see also
Eq.~(\ref{KKGaugeNormBKTs})]
\be
\frac{1}{L} 
\int_0^L \! f_n(y) f_m(y) \, dy +
\frac{r_{\rm UV}}{L} \, f_n(0) f_m(0) + \frac{r_{\rm IR}}{L} \,
f_n(L) f_m(L) 
&=& \delta_{mn}~,
\ee
it is known on general grounds that the (Neumann) Green function can
be represented as
\be
G^N(p; y,y') &=& \frac{1}{L} \, \sum_n \, \frac{f_n(y) f_n(y')}{p^2 -
m^2_n}~.
\ee
It is then clear from Eq.~(\ref{Pizeroes}) that the zeros of $\Pi$
coincide with the poles of $G^N(p; 0,0)$, i.e.~the KK masses $p^2 =
m^2_n$, as we wanted to show.

\medskip
\noindent
\textbf{\underline{Comments}:} The \textit{holographic profile},
$K(p,y)$, can be written as
\be
K(p,y) &=& \Pi(p) \, G^N(p; 0,y) ~=~ \frac{G^N(p; 0,y)}{G^N(p; 0,0)}~,
\ee
which we will call the \textit{amputated boundary-to-bulk
propagator}, where \text{amputation} here means dividing by the
boundary-to-boundary propagator, $G^N(p; 0,0)$.

Note also that, for $p\sim m_n$, the holographic Lagrangian reads
\be
{\cal L}_{\rm hol} &\sim& \frac{L}{2} \, \frac{1}{\left[ f_n(0)
\right]^2} \,
\tilde{\phi}_0 \left( p^2 - m^2_n \right) \tilde{\phi}_0 +
\textrm{suppressed}~,
\ee
and, therefore, the wavefunctions evaluated at $y=0$ can be read from
the coefficient of the ``kinetic term" in the region dominated by the
corresponding KK mode. Normalizing canonically, these wavefunctions
will control the interaction terms in the more general interacting
theory.

%%%%%%%%%%%%%%%%%%%%%%%%%%%%%%%%%%%%%%%
\subsection{Explicit Expressions for Wavefunctions, Propagators, etc.}
\label{Holography:Explicit}
%%%%%%%%%%%%%%%%%%%%%%%%%%%%%%%%%%%%%%%

It is possible to write completely general and fairly explicit
formulas for all the central objects in our discussion.  In this
section we provide such expressions, and also give the formulas in the
AdS$_5$ limit, for easy reference.

%%%%%%%%%%%%%%%%%%%%%%%%%%%%%%%%%%%%%%%
\subsubsection{General Case}
\label{Holography:General}
%%%%%%%%%%%%%%%%%%%%%%%%%%%%%%%%%%%%%%%

Assume that $h_1(p,y)$ and $h_2(p,y)$ are two \textit{independent}
solutions to the bulk EOM, Eq.~(\ref{HolEOM}).  In particular, their
Wronskian is non-vanishing (see also footnote~\ref{Wronskian} in
Section~\ref{KKDecomp} of Lecture 2):
\be
W(y) &=& h_1 \partial_y h_2 - h_2 \partial_y h_1 ~\neq~ 0~.
\ee
The brane-localized terms, which determine the b.c.'s, enter only
through
\be
\tilde{h}^{\rm UV}_i(p) &\equiv& \left. \left\{ \partial_y h_i(p,y) + \left[ \left( \alpha - 1 \right) +
b_{\rm UV} \!\! \left( \frac{p}{k_{\rm UV}}\right)  \right] k_{\rm UV} \, h_i(p,y) \right\} \right|_{y=0}~,
\label{h1tildeUV}
\\ [0.5em]
\tilde{h}^{\rm IR}_i(p) &\equiv& \left. \left\{ \partial_y h_i(p,y) + \left[ \left( \alpha - 1 \right) -
b_{\rm IR} \!\! \left( \frac{p \, e^{A(L)}}{k_{\rm IR}} \right)  \right] k_{\rm IR} \, h_i(p,y) \right\} \right|_{y=L}~,
\label{h1tildeIR}
\ee
for $i=1,2$, where $b_{\rm UV}(z)$ and $b_{\rm IR}(z)$ are as in
Eq.~(\ref{bi}).  We also define the auxiliary functions
\be
f^{\rm UV}(p,y) &=& \tilde{h}^{\rm UV}_2(p) \, h_1(p,y) - \tilde{h}^{\rm UV}_1(p) \, h_2(p,y)~,
\\ [0.5em]
f^{\rm IR}(p,y) &=& \tilde{h}^{\rm IR}_2(p) \, h_1(p,y) - \tilde{h}^{\rm IR}_1(p) \, h_2(p,y)~,
\ee
which are solutions to the EOM, obeying UV / IR $(+)$ b.c.'s,
respectively.  Imposing the $(+)$~type b.c.~on the other brane for
either of these functions results, up to normalization, in the KK
wave-functions, $f_n(y) \propto f^{\rm UV}(m_n,y) \propto f^{\rm
IR}(m_n,y)$, where the KK masses are solutions, $p=m_n$, of
\be
D_N(p) &\equiv& \tilde{h}^{\rm UV}_1(p) \, \tilde{h}^{\rm IR}_2(p) 
- \tilde{h}^{\rm UV}_2(p) \, \tilde{h}^{\rm IR}_1(p) ~=~ 0~.
\ee
If the KK problem required Dirichlet b.c.'s on the UV brane (but $(+)$
on the IR brane), the KK wavefunctions would be given by $f_n(y)
\propto f^{\rm IR}(m_n,y)$ with the spectrum now determined by
\be
D_D(p) / k_{\rm UV} &\equiv& f^{\rm IR}(p,0) ~=~ h_1(p,0) \, 
\tilde{h}^{\rm IR}_2(p) - h_2(p,0) \, \tilde{h}^{\rm IR}_1(p) ~=~ 0~.
\ee
Note also that $W(0) = f^{\rm UV}(p,0)$ and $W(L) = f^{\rm IR}(p,L)$.
In terms of the above we have that
\be
K(p,y) &=& \frac{f^{\rm IR}(p,y)}{f^{\rm IR}(p,0)}~,
\label{KGeneral}
\\ [0.3em]
S(p,y) &=& \frac{1}{W(0)} \, \left[ h_1(p,0) \, h_2(p,y) - h_2(p,0) \, h_1(p,y) \right]~,
\label{SGeneral}
\\ [0.3em]
G^N(p;y,y') &=& \frac{f^{\rm UV}(p,y_<) \, f^{\rm IR}(p,y_>)}{W(0) \, D_N(p)}~.
\label{GNGeneral}
\ee
In particular, $G^N(p;0,0) = D_D(p) / [k_{\rm UV} D_N(p)]$, a relation
of central importance in the holographic interpretation to be
presented in Section~\ref{Holography:Dual}.  The student should have
no problem checking the above formulas.

%%%%%%%%%%%%%%%%%%%%%%%%%%%%%%%%%%%%%%%
\subsubsection{The AdS$_5$ Limit}
\label{Holography:AdS}
%%%%%%%%%%%%%%%%%%%%%%%%%%%%%%%%%%%%%%%

Here we collect the expressions for the wavefunctions and propagators
in AdS$_5$.  The formulas below are also useful for fields of other
spins.  For fermions with a LH (RH) 0-mode, one should identify
$\alpha = c+1/2$ ($\alpha = c-1/2$), and set $m_{\rm UV} = m_{\rm IR}
= 0$.  For the spin-1 case, one should take $\alpha=1$.  Up to
respective factors of $k$ and $k \, e^{kL}$, the functions of
Eqs.~(\ref{h1tildeUV}) and (\ref{h1tildeIR}) read:
\be
\tilde{J}^{\rm UV}_{\alpha}(z) &=& z J_{\alpha-1}(z) + b_{\rm UV}(z)J_\alpha(z)~,
\\ [0.5em]
\tilde{J}^{\rm IR}_{\alpha}(z) &=& z J_{\alpha-1}(z) - b_{\rm IR}(z) J_\alpha(z)~,
\ee
with analogous definitions for $\tilde{Y}^{\rm UV}_{\alpha}(z)$ and
$\tilde{Y}^{\rm IR}_{\alpha}(z)$.  In terms of these, the KK
wavefunctions, obeying Eqs.~(\ref{fnEOM}) and (\ref{fnBCIR}) are
\be
f_n(y) &=& N_n \, e^{k y} \left[ J_\alpha \! \left( \frac{m_n}{k} \, 
e^{ky} \right) + B^{\rm IR}_\alpha(m_n) \, Y_\alpha \! 
\left( \frac{m_n}{k} \, e^{ky} \right) \right]~,
\ee
where $N_n$ is determined by the normalization condition, Eq.~(\ref{KKNorm}), and
\be B^{\rm IR}_\alpha(p) &=& -\frac{\tilde{J}^{\rm IR}_{\alpha}\!
\left( \frac{p}{k} e^{kL} \right)}{\tilde{Y}^{\rm IR}_{\alpha}\!
\left( \frac{p}{k} e^{kL} \right)}~.  \ee
The UV b.c., Eq.~(\ref{fnBCUV}), determines the KK masses from
\be
D_N(m_n) &\equiv& 
\tilde{J}^{\rm UV}_{\alpha}\! \left( \frac{m_n}{k} \right) \tilde{Y}^{\rm IR}_{\alpha}\! \left( \frac{m_n}{k} e^{kL} \right)
- \tilde{Y}^{\rm UV}_{\alpha}\! \left( \frac{m_n}{k} \right) \tilde{J}^{\rm IR}_{\alpha}\! \left( \frac{m_n}{k} e^{kL} \right)
~=~ 0~.
\label{DN}
\ee
If instead we were to impose Dirichlet b.c.'s on the UV brane, the
spectrum would be given by
\be
D_D(m_n) &\equiv& 
J_{\alpha}\! \left( \frac{m_n}{k} \right) \tilde{Y}^{\rm IR}_{\alpha}\! \left( \frac{m_n}{k} \, e^{kL} \right)
- Y_{\alpha}\! \left( \frac{m_n}{k} \right) \tilde{J}^{\rm IR}_{\alpha}\! \left( \frac{m_n}{k} \, e^{kL} \right)
~=~ 0~.
\label{DD}
\ee
The holographic profile $K(p,y)$ and the function $S(p,y)$ take the
form:
\be
K(p,y) &=& 
e^{ky} \, \frac{\left[ J_\alpha \! \left( \frac{p}{k} \, e^{ky} \right) + B^{\rm IR}_\alpha(p) \, Y_\alpha \! \left( \frac{p}{k} \, e^{ky} \right) \right]}
{\left[ J_\alpha \! \left( \frac{p}{k} \right) + B^{\rm IR}_\alpha(p) \, Y_\alpha \! \left( \frac{p}{k} \right) \right]}~,
\\ [0.6em]
S(p,y) &=& 
- \frac{\pi}{2k} \, e^{ky} \left[ Y_\alpha \! \left( \frac{p}{k} \right) J_\alpha \! \left( \frac{p}{k} \, e^{ky} \right) -  
J_\alpha \! \left( \frac{p}{k} \right) Y_\alpha \! \left( \frac{p}{k} \, e^{ky} \right) \right]~.
\ee
Then the Neumann propagator is given by
\be
G^N(p;y,y') &=& \frac{\pi \, e^{k(y + y')}}{2kD_N(p)} \,
\left[ 
\tilde{Y}^{\rm UV}_{\alpha}\! \left( \frac{p}{k} \right) J_{\alpha}\! \left( \frac{p}{k} \, e^{ky} \right)
- \tilde{J}^{\rm UV}_{\alpha}\! \left( \frac{p}{k} \right) Y_{\alpha}\! \left( \frac{p}{k} \, e^{ky} \right)
\right]
\nonumber \\ [0.5em]
&& \hspace{2cm}
\times
\left[ 
\tilde{Y}^{\rm IR}_{\alpha}\! \left( \frac{p}{k} \, e^{kL} \right) J_{\alpha}\! \left( \frac{p}{k} \, e^{ky'} \right)
- \tilde{J}^{\rm IR}_{\alpha}\! \left( \frac{p}{k} \, e^{kL} \right) Y_{\alpha}\! \left( \frac{p}{k} \, e^{ky'} \right)
\right]~,
\label{GNAdS}
\ee
while the Dirichlet propagator reads:
\be
G^D(p;y,y') &=& \frac{\pi \, e^{k(y + y')}}{2kD_D(p)} \,
\left[ 
Y_{\alpha}\! \left( \frac{p}{k} \right) J_{\alpha}\! \left( \frac{p}{k} \, e^{ky} \right)
- J_{\alpha}\! \left( \frac{p}{k} \right) Y_{\alpha}\! \left( \frac{p}{k} \, e^{ky} \right)
\right]
\nonumber \\ [0.5em]
&& \hspace{2cm}
\times
\left[ 
\tilde{Y}^{\rm IR}_{\alpha}\! \left( \frac{p}{k} \, e^{kL} \right) J_{\alpha}\! \left( \frac{p}{k} \, e^{ky'} \right)
- \tilde{J}^{\rm IR}_{\alpha}\! \left( \frac{p}{k} \, e^{kL} \right) Y_{\alpha}\! \left( \frac{p}{k} \, e^{ky'} \right)
\right]~.
\label{GDAdS}
\ee
One can easily check that the above expressions obey the EOM and all
the relevant b.c.'s.

%%%%%%%%%%%%%%%%%%%%%%%%%%%%%%%%%%%%%%%
\subsection{Application: EW Constraints in Anarchic Models}
\label{Holography:EWPT}
%%%%%%%%%%%%%%%%%%%%%%%%%%%%%%%%%%%%%%%

The holographic rewriting of the 5D theory allows for a simple
derivation of the expressions for the \textit{oblique} parameters
that control some of the most important corrections to the EW
observables in models with flavor anarchy (i.e.~with the light
families localized near the UV brane). Non-universal corrections,
such as $\delta g_{Zb_L\bar{b}_L}$ require more work.

Consider a bulk $SU(2)_L \times SU(2)_R \times U(1)_X$ custodial
model.  It will be easy to specialize the results to models with only
the SM gauge group by setting to zero the extra gauge
bosons/propagators.  Our starting action is then
\be
S^{\rm Cust.}_{\mathrm{gauge}}&=& \int \! \frac{d^4p}{(2\pi)^4}
\int_0^L \! dy \, \sqrt{g} \left\{
-\frac{1}{4g_{5L}^2} (L^a_{MN})^2
-\frac{1}{4g_{5R}^2} (R^a_{MN})^2
-\frac{1}{4g_{5X}^{2}} (X_{MN})^2
\right.
\nonumber \\
&& \hspace{3.5cm} \left. \mbox{}
+ \frac{1}{2} \, \textrm{v}(y)^2 
(L^a_M-R^a_M)^2
\right\}~,
\ee
where $L^a_{MN}$, $R^a_{MN}$ and $X_{MN}$ are the field strengths for
$SU(2)_L$,  $SU(2)_R$, and $U(1)_X$, respectively, while $g_{5L}$,
$g_{5R}$ and $g_{5X}$ are the corresponding gauge couplings. Note
also that we allow for a $y$-dependent vev, as would arise from a
bulk Higgs with a non-trivial profile.

%%%%%%%%%%%%%%%%%%%%%
\begin{quote}

\textbf{\underline{Exercise}:}
Show that the $(L-R)^2$ structure for the gauge mass term arises when
the Higgs is a bidoublet of $SU(2)_L \times SU(2)_R$. 
\end{quote}
%%%%%%%%%%%%%%%%%%%%%
%
It is useful to rewrite the above action in the ``$V-A$" basis:
\be
V_M^a=\frac{g_{5R}}{g_{5L}}L_M^a + \frac{g_{5L}}{g_{5R}} R_M^a~,
\qquad \qquad
A_M^a=L_M^a-R_M^a~, 
\ee
so that
\be
S^{\rm Cust.}_{\mathrm{gauge}}&=& \int \! \frac{d^4p}{(2\pi)^4}
\int_0^L \! dy \, \sqrt{g} \left\{
-\frac{1}{4g_{5Z}^2} \left[ (V^a_{MN})^2 + (A^a_{MN})^2 \right]
-\frac{1}{4g_{5X}^{2}} (X_{MN})^2
+\frac{1}{2} \textrm{v}(z)^2 A_M^2
\right\}~,
\nonumber \\
\label{VAAction}
\ee
where $g_{5Z}^2 \equiv g_{5L}^2 + g_{5R}^2$, and $V^a_{MN}$ and
$A^a_{MN}$ are the field strengths associated with $V_M$ and $A_M$,
respectively. We can now rewrite this action holographically by
defining the holographic profiles $K_A$, $K_V$ and $K_X$, satisfying
[see also Eq.~(\ref{fAEOM})]
\be
\partial_y \left[ e^{-2 A} \partial_y K_A \right] - \left[ M_A(y)^2 -
p^2 \right] K_A &=& 0~,
\label{KAEOM}
\\ [0.5 em]
\partial_y \left[ e^{-2 A} \partial_y K_{V,X} \right] + p^2 K_{V,X}
&=& 0~,
\label{KVEOM}
\ee
where $M_A(y)^2 = g^2_{5Z} \textrm{v}(y)^2$. For simplicity, we
assume that the possible quadratic IR-localized terms vanish, so that
the $K_i$ simply obey 
\be
K_i(p,y=0) &=& 1~,
\\ [0.5em]
\left. \partial_y K_i(p,y) \rule{0mm}{3.5mm} \right|_{y=L} &=& 0~.
\ee
for $i = A, V, X$. In these models all fields obey Neumann ($+$)
b.c.'s at $y=L$ so that the custodial symmetry is well-preserved in
the IR region. On the UV brane, we imagine there is physics that
``lifts" the unwanted gauge 0-modes:
\be
R^b &\equiv& W^b_R~,
\hspace{2cm} \textrm{for}~b = 1,2
\nonumber \\ [0.5em]
Z' &=& \frac{1}{\sqrt{g_{5R}^2 + g_{5X}^2}} \, \left( R^3 - X
\right)~,
\ee
by adding a UV-localized Lagrangian
\be
{\cal L}_0 &=& \frac{1}{2} M_R^2 \left[ \left( W_R^1 \right)^2 +
\left( W_R^2 \right)^2 \right] + \frac{1}{2} M_{Z'}^2 \left( Z'
\right)^2~,
\ee
and sending $M_R^2, M_{Z'}^2 \to \infty$. This imposes effectively
Dirichlet ($-$) b.c.'s on the UV brane, so that there are no
corresponding holographic fields (the boundary values vanish). 

We are therefore left with the SM fields, related to $A$, $V$ and $X$
by
\be
A^a=W_L^a -\delta^{a3}B~, \qquad
V^a=\frac{g_{5R}}{g_{5L}} \, W_L^a
+ \delta^{a3}\frac{g_{5L}}{g_{5R}} \, B~, \qquad
X=B~,
\label{AVXWB}
\ee
where we switched to the more familiar notation $W^a_L \equiv L^a$.
We then have
\be
S[\bar{W}_L, \bar{B}] &\stackrel{\rm Hol}{=}&
\int \! \frac{d^4p}{(2\pi)^4} \left( - \frac{1}{2} P^{\mu\nu} \right)
\left\{
 \bar{A}^a_\mu \Pi_A(p^2) \bar{A}^a_\nu
+ \bar{V}^a_\mu \Pi_V(p^2) \bar{V}^a_\nu
+ \bar{X}_\mu \Pi_X(p^2) \bar{X}_\nu
\right\}
\nonumber \\ [0.5em]
&=&
\int \! \frac{d^4p}{(2\pi)^4} \left( - P^{\mu\nu} \right) \left\{
\bar{W}^+_\mu \Pi_{+-} \bar{W}^-_\nu
+ \frac{1}{2} \, \bar{W}^3_\mu \Pi_{33} \bar{W}^3_\nu
+ \frac{1}{2} \, \bar{B}_\mu \Pi_{BB} \bar{B}_\nu
+ \bar{W}^3_\mu \Pi_{3B} \bar{B}_\nu
\right\}~,
\nonumber \\
\ee
where $P^{\mu\nu} = \eta^{\mu\nu} - p^\mu p^\nu / p^2$ is the
transverse projector, and we use here a bar to denote the boundary
values (i.e.~the holographic fields).  We also defined $\bar{W}^{\pm}
\equiv (\bar{W}^1 \mp i \bar{W}^2)/\sqrt{2}$.  To obtain the second
line, we used Eq.~(\ref{AVXWB}) applied to the holographic fields.  In
doing so, we identify
\be
\Pi_{aa} &=& \Pi_A + \frac{g^2_{5R}}{g^2_{5L}} \, \Pi_V~,
\hspace{2cm} a=1,2,3
\\ %[0.4em]
\Pi_{BB} &=& \Pi_A + \frac{g^2_{5L}}{g^2_{5R}} \, \Pi_V + \Pi_X~,
\\ [0.5em]
\Pi_{3B} &=& -\Pi_A + \Pi_V~,
\ee
where
\be
\Pi_{A,V} &=& \left. \frac{1}{g^2_{5Z}} \, \partial_y K_{A,V}
\right|_{y=0}~,
\hspace{2cm}
\Pi_{X} ~=~ \left. \frac{1}{g^2_{5X}} \, \partial_y K_{X}
\right|_{y=0}~,
\label{PisAVX}
\ee
are the boundary propagators, which take into account the gauge
couplings in front of the gauge kinetic terms in Eq.~(\ref{VAAction}).

\medskip
\noindent
\textbf{\underline{UV-localized Fermions}}
\medskip

Up to this point we have not mentioned anything about fermions. A
holographic approach for fermions has been discussed
in~\cite{Contino:2004vy}. If these are UV-localized their couplings
to gauge bosons are nearly independent of the precise fermion
localization profiles. In this case, the couplings of the fermionic
holographic fields take the \textit{universal} form
\be
\mathcal{L}_{A\bar{f}f}^{\mathrm{Univ.}}
= \bar{W}^a_\mu J^{a\,\mu}_L + \bar{B}_\mu J^{\mu}_Y~,
\label{UniversalFermionGauge}
\ee
where
\be
J^{a\,\mu}_{L} \equiv  
\sum_{\psi} \bar{\psi} \gamma^\mu T^a_{L} \psi~,
\hspace{2cm}
J^{\mu}_{Y} \equiv 
\sum_{\psi} \bar{\psi} \gamma^\mu Y \psi~.
\label{J:f}
\ee
The holographic procedure then automatically gives an effective 4D
theory in \textit{oblique form}.

When the theory has a mass gap, so that 
\be
\Pi_i(p^2) &=& \Pi_i(0) + p^2 \, \Pi'_i(0) + \frac{1}{2} \, p^4 \,
\Pi^{\prime \prime}_i(0) + \cdots
\ee
have an analytic (Taylor) low-energy expansion, the effects of the
new physics (contained in the $\Pi$'s) can be parameterized in terms
of the following~\cite{Barbieri:2004qk}:
\be
\hat{T}&=& 1-\frac{\Pi_{33}(0)}{\Pi_{+-}(0)}~, 
\hspace{1cm}
W ~=~\frac{1}{2} \, g^2 M_W^2 \, \Pi_{33}^{\prime\prime}(0)~, 
\label{TW} \\ [0.5em]
\hat{S}&=& g^2 \Pi_{3B}^\prime(0)~, 
\hspace{1.62cm}
Y~=~W~,
\label{SY}
\ee
where
\be
\frac{1}{g^2} ~=~ \Pi^\prime_{11}(0)~, 
\hspace{1cm}
\frac{1}{g^{\prime\,2}} ~=~ \Pi^\prime_{BB}(0)~, 
\hspace{1cm}
-\frac{M^2_{W}}{g^2} ~=~ \Pi_{+-}(0)~.
\label{Oblique:normalization}
\ee
The first two parameters are related to the Peskin-Takeuchi $S$ and
$T$~\cite{Peskin:1991sw}
\be
\alpha S &=& 4s_W^2\hat{S}~,
\hspace{1.5cm}
\alpha T ~=~ \hat{T}~.
\ee
Given a specific background, one can solve for the holographic
profiles to find the $\Pi_i(p^2)$, and use the previous general
formulas to compare to the EW data.

As an example, for the AdS$_5$ background, and assuming that the
Higgs is $\delta$-function IR-localized, one finds in the
non-custodial case (just getting rid of the $(-,+)$ fields above
completely)
\be
\hat{T}_{UV} &=& \frac{g^{\prime\,2}}{2g^2} \left(
\frac{M_W}{\tilde{k}} \right)^2 \, kL~, 
\hspace{2cm}
W_{UV} ~=~ \frac{1}{4 kL} \, \left( \frac{M_W}{\tilde{k}} \right)^2~,
\label{UVTW}
\\ [0.4em]
\hat{S}_{UV} &=& \frac{1}{2} \, \left( \frac{M_W}{\tilde{k}}
\right)^2~, 
\hspace{3.03cm}
Y_{UV}~=~ W_{UV}~,
\label{UVST}
\ee
where the UV subscripts remind us that the fermions are assumed to be
UV localized. One should note the pattern of $kL$ enhancements and
suppressions. In the presence of the custodial structure, one finds
$\hat{T}_{UV} = 0$, while the other remain unchanged.

\medskip
\noindent
\textbf{\underline{Arbitrary Fermion Localization}}
\medskip

It is possible to extend the holographic approach to cases where the
fermion 0-modes are arbitrarily localized. We illustrate the idea in
a toy model with a single gauge field coupled to a (single) fermion
current. The new effects can be expressed in the following
diagrammatic form:

\medskip
\noindent
\underline{Vertex corrections}:
\be
\put(5,-60){
\resizebox{3cm}{!}{\includegraphics{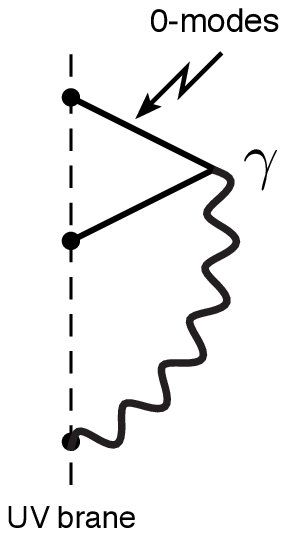}}
}
\hspace*{3cm}
&=& \underset{\displaystyle \rule{0mm}{5mm} g^2_5 /
K'(p,0)}{\frac{1}{\underbrace{G^N(p;0,0)}}} \, \int_0^L \! dy \left[
f^0_\psi(y) \right]^2 
\overset{\displaystyle g^2_5 \,
\frac{K(p;y)}{K'(p,0)}}{\overbrace{\rule{0mm}{4mm} G^N(p;0,y)}}
~=~ 
\overset{\displaystyle \equiv \bar{g}(p)}{\overbrace{\rule{0mm}{8mm} 
\int_0^L \! dy \left[ f^0_\psi(y) \right]^2 K(p,y)}}~,
\psset{unit=1pt}
\rput[lb]{0}(-268,-31){
\psline[linestyle=solid,linewidth=1,linecolor=black,arrowinset=0.2,arrowscale=1]{<-}(-10,0)(10,0)
\uput{4}[0]{0}(10,0){\psscaleboxto(110,0){\textrm{Amputate
w.r.t.~gauge}}}
\uput{4}[0]{0}(10,-15){\psscaleboxto(140,0){\textrm{boundary field
(for matching)}}}
}
\nonumber
\ee
where, for simplicity, we omit the spinor indices provided by the
Dirac $\gamma^\gamma$.  We see how we obtain the overlap integral of
the gauge and fermion wavefunctions, which we call $\bar{g}(p)$.  Note
that it depends on the 4D momentum, and also on the fermion type
through its wavefunction profile [recall the $c$-parameters introduced
in Eqs.~(\ref{MproptoA}) and (\ref{zeromodescpar})].

\medskip
\noindent
\textbf{\underline{Comment}:} at zero momentum, $K(p=0,y)$ is
precisely the profile for the gauge 0-mode, so the above integral
matches exactly our KK notion of \textit{couplings as overlap
integrals}. Also, if the gauge field is massless, one has
$\bar{g}(p=0) = 1$. For a full comparison with our KK results, recall
that here the gauge field is not canonically normalized, but rather
has a factor $L/g^2_5$ in front of the kinetic term.

\medskip
\noindent
\underline{4-fermion interactions}:
\be
\put(0,-60){
\resizebox{3cm}{!}{\includegraphics{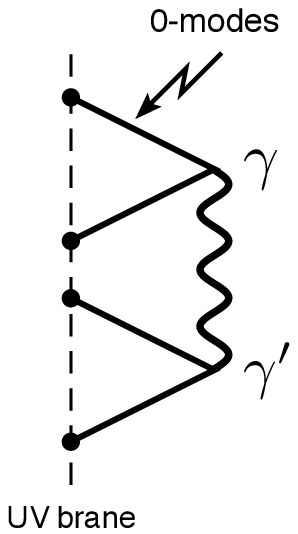}}
}
\hspace*{3cm}
&=& \int_0^L \! dy \, dy' \left[ f^0_\psi(y) \right]^2 G^N(p;y,y')
\left[ f^0_\psi(y') \right]^2
\nonumber \\ [0.5em]
\psset{unit=1pt}
\rput[lb]{0}(10,25){
\displaystyle
=~ G^N(p;0,0) \, \bar{g}(p)^2 + \frac{g_5^2}{L} \,
\gamma_{\psi\psi'}~,
}
\nonumber
\ee
where again we omitted the spinor structure and a Lorentz
$\eta_{\gamma\gamma'}$.  In the second line we used
Eqs.~(\ref{GN})-(\ref{Pizeroes}):
\be
G^N(p;y,y') &=& G^N(p;0,0) K(p,y) K(p,y') + G^D(p;y,y')~,
\ee
where $G^D(p;y,y')$ is the \textit{Dirichlet propagator} [see the
exercise after Eq.~(\ref{GN}), but note that we include a factor
$g^2_5$ in the definition of both $G^D$ and $G^N$, due to the
non-canonical gauge normalization of this section.  See also
Eq.~(\ref{PisAVX})].  In addition, we defined
\be
\gamma_{\psi\psi'} &\equiv& \frac{1}{g^2_5 L} \int_0^L \! dy \, dy'  \left[
f^0_\psi(y) \right]^2 G^D(p;y,y') \left[ f^0_\psi(y') \right]^2~,
\ee
which has mass dimension $-2$.  In this definition, the explicit
factor of $1/g^2_5$ simply cancels the corresponding factor in
$G^D(p;y,y')$.  Note also that our wavefunctions, which are
dimensionless, differ by a factor of $\sqrt{L}$ (and powers of the warp 
factor) from those used in Ref.~\cite{Davoudiasl:2009cd}.

The above effects can be encoded in the 4D effective Lagrangian:
\be
{\cal L}_{\rm eff} &=& \frac{1}{2} \left( - P^{\mu\nu} \right)
\bar{A}_\mu \left[ G^N(p;0,0) \right]^{-1} \bar{A}_\nu
%\nonumber \\ [0.5em]
%& & \mbox{}
+ \bar{g}(p) \bar{A}_\mu J^\mu + \frac{1}{2} \, \frac{g_5^2}{L}
\gamma_{\psi\psi'} J_\mu J^\mu~.
\put(-150,-50){
\resizebox{1.5cm}{!}{\includegraphics{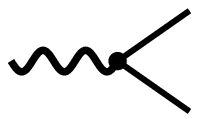}}
\put(-37,40){$\underbrace{\rule{1.6cm}{0mm}}$}
\put(0,8){$\displaystyle = \bar{g}$}
}
\put(-70,-55){
\resizebox{1cm}{!}{\includegraphics{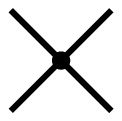}}
\put(-40,45){$\underbrace{\rule{2.5cm}{0mm}}$}
\put(0,10){$\displaystyle = \frac{g_5^2}{L} \gamma_{\psi\psi'}$}
}
\ee
so that the net 4-fermion operator in the effective theory is
\be
\put(0,-12){
\resizebox{4cm}{!}{\includegraphics{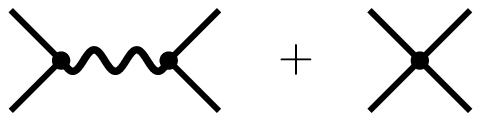}}
}
\hspace*{4.5cm}
=~ \bar{g}(p) \, G^N(p;0,0) \, \bar{g}(p) + \frac{g_5^2}{L} \,
\gamma_{\psi\psi'}~,
\nonumber
\ee
thus matching the full theory result above.  It is then
straightforward to put these ingredients together in the context of a
full $SU(2)_L \times SU(2)_R \times U(1)_X$ model to include the cases
where some fermions are not UV localized.  The most notable example
arises from the third generation $(t_L,b_L)$ $SU(2)_L$ doublet, which
can lead to important non-universal corrections to the coupling of the
LH bottom to the $Z$ gauge boson.

\medskip
\noindent
\textbf{\underline{Comments}:} 
\begin{itemize}

\item The propagators used in this section have mass dimension $-2$,
unlike those introduced in the generic discussion based on a bulk
scalar field of Section~\ref{Holography:Scalars}, which had 5D
normalization.  The difference in dimensionality is provided by the
factors of $g^2_5$ already mentioned above.

\item In our diagrammatic arguments, for simplicity we omitted the
factors of $i$ in the Feynman rules, as well as the spinor/Lorentz
structure, but they can be easily restored.

\item The above 4D effective Lagrangian is not in the oblique form
even when the fermion profiles are universal, so that $J^\mu =
\sum_\psi \bar{\psi} \gamma^\mu T \psi$ and $\bar{g}(p)$ is
independent of the fermion type. This is because the factor in front
of $\bar{A}_\mu J^\mu$ is not~1, and also because there are 4-fermion
interactions. However, when the fermion profiles are universal, one
can redefine the holographic gauge fields to put the effective
Lagrangian in the oblique form. For an explicit example, in the
$SU(2)_L \times SU(2)_R \times U(1)_X$ case, the student is referred
to~\cite{Davoudiasl:2009cd}.

\item An alternative way of obtaining the same result is to integrate
out the heavy KK modes, and match onto the SM + dimension-6
operators. The above reference also reviews the method in detail.
Here we rather quote, for reference, the elegant expressions obtained
in~\cite{Cabrer:2011fb,Cabrer:2011qb}
\be
\alpha T &=& s_W^2 M_Z^2 \, L \, \int_0^L \! dy \, e^{2A(y)} \left[
\Omega_\psi - \Omega_h \right]^2~,
\\ [0.5em]
\alpha S &=& 8 s_W^2 c_W^2 M_Z^2 \, L \, \int_0^L \! dy \, e^{2A(y)}
\left[ \Omega_\psi - \frac{y}{L} \right] \left[ \Omega_\psi -
\Omega_h \right]~,
\\ [0.5em]
Y &=& W ~=~ c_W^2 M_Z^2 \, L \, \int_0^L \! dy \, e^{2A(y)} \left[
\Omega_\psi - \frac{y}{L} \right]^2~,
\ee
where
\be
\Omega_{\psi,h}(y) &\equiv& \frac{1}{L} \int_0^y \! dy' \left[
f^0_{\psi,h}(y') \right]^2~,
\ee
which obeys $\Omega_{\psi,h}(L) = 1$ due to the normalization of the
wavefunctions. These expressions hold for any background and any
Higgs vev profile. They only assume that the $f^0_\psi$'s lead to
universal shifts (so that the oblique parameter analysis applies),
and that $v \ll \tilde{k}_{\rm eff}$, the scale of new physics.

\item We also provide general expressions that allow to compute the
most important non-oblique effect in anarchic scenarios with
custodial symmetry:
\be
\delta g_{Z\bar{b}_Lb_L} &=& \frac{g^2 v^2}{2c_W^2} \left\{
\big[g_L^2 T^3_L -g_R^2 T^3_R \big] \beta^D_Q
+\big[g_L^2 T^3_L - g^{\prime\,2} Y \big] 
\left( \beta^N_Q - \beta^N_{\mathrm{UV}}-\beta^D_Q
\right)
\right\},
\label{deltagZbb:general}
\ee
where $T^3_L = \frac{1}{2}$ and $T^3_R$ are the $SU(2)_L \times
SU(2)_R$ quantum numbers for the LH bottom (also $Y = \frac{1}{6}$),
and we defined the mass-dimension $-2$ quantities
\be
\beta^{N,D}_\psi &=& \frac{1}{g^2_5 L} \, \int_0^L \! dy \, dy' \, \left[
f^0_\psi(y) \right]^2 \tilde{G}^{N,D}(p=0; y, y') \left[ f^0_h(y')
\right]^2~,
\label{beta}
\ee
with $\beta^{N,D}_Q = \beta^{N,D}_{b_L}$,
$\beta^{N}_{\rm UV} = \beta^{N}_{\textrm{light
fermions}}$. In Eq.~(\ref{beta}), the tilde on the propagator
indicates that the 0-mode needs to be subtracted (for the Neumann
b.c.'s).

When $g_R = g_L$ and $T^3_R = T^3_L$ (custodial protection of $\delta
g_{Z\bar{b}_L b_L}$), the above simplifies to
\be
\delta g_{Z\bar{b}_Lb_L} &=&
- \frac{g^2 v^2}{2c_W^2} \,
\Big[ \frac{1}{2} \, g^2 + \frac{1}{6} \, g^{\prime\,2} Y\Big] \left(
\beta^N_Q - \beta_{\mathrm{UV}}^N  -\beta^D_Q
\right)~.
\ee
It is also worth noting the alternate forms for the $\beta$ coefficients:
\be
\beta^N_\psi &=& L \, \int_0^L \! dy \, e^{2A(y)}  \left[
\Omega_\psi - \frac{y}{L} \right] \left[ \Omega_h - \frac{y}{L}
\right]~,
\\ [0.5em]
\beta^D_\psi &=& L \, \int_0^L \! dy \, e^{2A(y)}  \Omega_\psi
\Omega_h~.
\ee

\end{itemize}
%

%%%%%%%%%%%%%%%%%%%%%%%%%%%%%%%%%%%%%%%
\subsection{The 4D Dual Interpretation}
\label{Holography:Dual}
%%%%%%%%%%%%%%%%%%%%%%%%%%%%%%%%%%%%%%%

After this excursion into the depths of EW precision constraints, let
us go back to the holographic rewriting of the 5D theory. We will
offer some elementary remarks on how the 4D dual interpretation
arises. For further details, I refer the student to T.~Gherghetta's
2010 TASI Lectures on ``A Holographic view and BSM physics".

We will illustrate the main points in the context of the scalar
example (in sufficient generality to make it easily applicable to
other spins).  We found that, as a functional of the 5D scalar
boundary value, the holographic action reads~\footnote{Recall that, in
coordinate space, $[\phi_0] = \frac{3}{2}$, not $[\phi_0] =1$.}
\be
S[\phi_0] &=& \int \! \frac{d^4 p}{(2\pi)^4} \, \frac{1}{2} \,
\tilde{\phi}_0 \, \Pi \, \tilde{\phi}_0~,
\label{HolographicActionAgain}
\ee
where $\Pi(p) = \left[ G^N(p;0,0) \right]^{-1}$ is the inverse
boundary-to-boundary propagator. This propagator obeys b.c.'s on the
UV brane that take into account possible UV localized terms. For
instance, if we write a UV localized mass term, and then take it to
infinity, we get $G^N \to G^D$, the propagator obeying Dirichlet
b.c.'s on the UV brane.

The holographic procedure of replacing the classical bulk EOM, for
prescribed $\phi_0$, back into the 5D action corresponds to the
leading-order in the evaluation of the (Euclidean) path integral 
\be
Z_h[\phi_0] &=& C \, \int_{\Phi(y=0)=\phi_0} \! {\cal D} \Phi \,
e^{-S_{\rm Bulk}[\Phi]} ~\approx~ e^{-S[\phi_0]}~,
\label{Zh}
\ee
where $C$ is a normalization constant chosen so that $Z_h[0] = 1$,
and the second equality follows from the evaluation of the integral
with the method of steepest descent that leads to the semi-classical
expansion. Since $S[\phi_0]$ is a 4D scalar action, we may regard
$Z_h[\phi_0]$ as the generating functional for correlators in some
(not explicitly specified) 4D theory, with $\phi_0$ acting as an
external source. More precisely, $\phi_0$ is interpreted as the
source for some operator ${\cal O}$ in the \textit{dual theory}:
\be
Z[\phi_0] &=& C' \, \int \! {\cal D} \chi \, e^{-S_{\rm Dual}[\chi] +
\int \! d^4 x \, \phi_0 {\cal O}}~,
\label{ZDual}
\ee
where we denote the dynamical variables in the dual theory by $\chi$,
and the operator ${\cal O}$ is built out of the $\chi$'s. The ${\cal
O}$-correlators are given by
\be
\langle {\cal O}_1 \ldots {\cal O}_n \rangle &=& \left.
\frac{\delta}{\delta \phi_0(x_1)} \cdots \frac{\delta}{\delta
\phi_0(x_n)} \, Z[\phi_0] \right|_{\phi_0 = 0}~.
\ee
In general, for a given 5D theory, we do not explicitly know its
dual, $S_{\rm Dual}[\chi]$, but by identifying $Z[\phi_0] =
Z_h[\phi_0]$ we can find the correlators for ${\cal O}$ from the
holographic action arising from the 5D theory. The
\underline{connected} Green functions are then
\be
\langle {\cal O}_1 \ldots {\cal O}_n \rangle^c &=& - \left.
\frac{\delta}{\delta \phi_0(x_1)} \cdots \frac{\delta}{\delta
\phi_0(x_n)} \, S[\phi_0] \right|_{\phi_0 = 0}~,
\ee
where we used the semi-classical leading order (or tree-level) term.

In particular, the two-point function (which is the only
non-vanishing one if we use the free bulk scalar action) is given by
\be
\langle {\cal O}(p) {\cal O}(-p) \rangle &=& - \Pi(p) ~=~ - \left[
G^N(p;0,0) \right]^{-1}~.
\ee
Thus the \textit{zeros} of $G^N(p;0,0)$ give the masses of the
states in the dual theory that can be created by ${\cal O}$.

In general, loops involving the $\chi$ fields will induce terms that
depend on $\phi_0$, giving rise to a contribution to the effective
dual action of the form (we use here $L$ to match dimensions)
\be
\Delta S_{\rm Dual} &=& \int \! d^4x \, \left\{
\frac{L}{2} \, {\cal Z} \, \partial_\mu \phi_0 \partial^\mu \phi_0 +
\frac{L}{2} \Delta m^2 \phi^2_0 + \cdots
\right\}~.
\ee
On the holographic side, these are matched by the terms we already
wrote:
\be
S_0 &=& \int \! d^4 x \, {\cal L}_0(\Phi) ~=~ \int \! d^4 x \left\{
\frac{1}{2} \, r_0 \, \partial_\mu \phi_0 \partial^\mu \phi_0 +
\frac{1}{2} \, M_0 \phi_0^2
\right\}~.
\label{BoundaryS}
\ee
We again see that such boundary terms are not really optional: they
are generated by quantum effects.  If we integrate, in the
path-integral sense, over $\phi_0$ (as we know we have to do in order
to recover the correct b.c.'s on the UV brane), the above terms imply
that $\phi_0$ is a \textit{dynamical} field.  Then the term $\int \!
d^4 x \, \phi_0 {\cal O}$ corresponds to a coupling of the dual
d.o.f.~to $\phi_0$, which should be thought as a field external to the
dual theory.  The physical states of the theory should then correspond
to admixtures between the dual fields (also called \textit{composite
states}) and the \textit{elementary} field $\phi_0$.  We can,
nevertheless, consider a limit where such a mixing is absent.  If the
localized mass $M_0$ is very large, the fluctuations in $\phi_0(x)$
will be very suppressed.  In the limit $M_0 \to \infty$, the EOM will
force $\phi_0$ to vanish, so that the dual theory contains no $\phi_0
{\cal O}$ term when on-shell.  Since in this limit the holographic
theory is equivalent to a KK theory obeying Dirichlet b.c.'s on the UV
brane, we can expect that the Dirichlet KK states map exactly into the
\textit{pure} dual states, without mixing with $\phi_0$.

We can make the above more precise by recalling our general
expressions for the Neumann propagator, given in
Section~\ref{Holography:General}, and in particular, the result
mentioned after Eq.~(\ref{GNGeneral}):
\be
G^N(p;0,0) &=& \frac{D_D(p)}{k_{\rm UV} D_N(p)}~,
\ee
where $D_D(m_n) = 0$ and $D_N(m_n) = 0$ determine the KK spectrum for
$(-)$ and $(+)$ b.c.'s on the UV brane, respectively.  Thus, we see
that the poles of $\langle {\cal O}(p) {\cal O}(-p) \rangle = -\Pi(p)$
indeed coincide with the Dirichlet KK masses.  In summary, we learn
that
\be
\textrm{Dirichlet KK states} & \quad \longleftrightarrow \quad&
\textrm{pure dual states}
\nonumber
\ee
There is also a sense in which the associated KK wavefunctions,
obeying Dirichlet b.c.'s on the UV brane, characterize the properties
of pure dual states.  The student is referred to
Ref.~\cite{Gherghetta:2010cj} for full details.

\begin{figure}[t]
\begin{center}
\includegraphics[width=0.5 \textwidth]{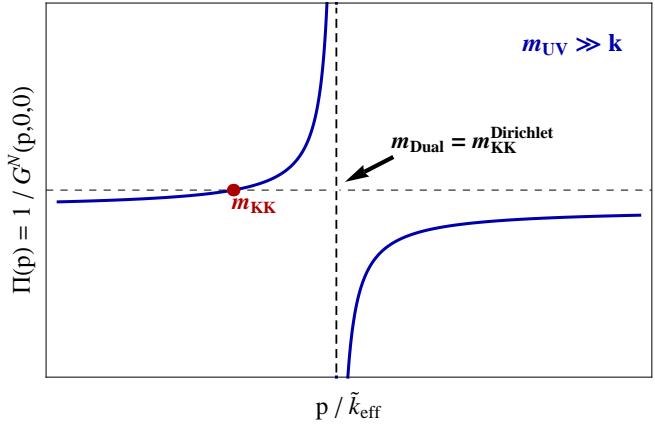}
\caption{Detail of the inverse brane-to-brane Neumann propagator (for
concreteness, in the AdS$_5$ case).  The poles occur at the Dirichlet
KK masses, which can be interpreted as the masses of the \textit{dual
states}.  The zeros, i.e.~the poles of the Neumann propagator, are the
generic KK masses (red dot).  For large UV localized mass $M_0$
(equivalently $m_{\rm UV}$) the two coincide.  The mass splitting is
interpreted in the dual theory as arising from mixing between the
elementary $\phi_0$ and the dual states $\chi$.  In the AdS$_5$
background, one can further interpret the dual states as bound states
in a CFT where the conformal symmetry has been \textit{spontaneously}
broken (by the IR boundary).  The radion/dilaton is the (pseudo)
Nambu-Goldstone mode corresponding to the conformal breaking.  A
stabilizing scalar field would correspond to \textit{explicit}
breaking of the conformal symmetry, as is the Planck mass associated
to the UV brane.}
\label{fig:HolographicStates}
\end{center}
\end{figure}
For finite $M_0$, the KK spectrum [i.e.~the poles of $G^N(p;0,0)$]
differs from the Dirichlet spectrum [i.e.~the poles of the Dirichlet
propagator, $G^D(p;0,0)$, which coincide with the zeros of
$G^N(p;0,0)$].  These ``mass shifts" in the KK theory (illustrated in
Fig.~\ref{fig:HolographicStates}) can be interpreted as due to kinetic
and mass mixing between $\phi_0$ and the $\chi$ states in the dual
theory.  In Ref.~\cite{Batell:2007jv}, such a picture has been worked
out in full detail in the AdS$_5$ limit, and assuming that $r_0 = M_0
= 0$ (so that $\Phi$ obeys standard Neumann b.c.'s, $\left.
\partial_y \Phi \right|_{0} = 0$ on the UV brane) by
\textit{diagonalizing the Dirichlet basis to the Neumann basis in the
presence of mixing}.

Therefore, one obtains a generic interpretation of the KK states as
mixed dual/elementary states:
\be
| \phi_n \rangle &=& \sin\theta_n |\phi_0 \rangle + \cos\theta_n
|\chi \rangle~,
\ee
where the $| \phi_n \rangle$ are the generic KK states, which are mass
eigenstates, and are the ones we would see in experiments.  Here we
used a schematic notation, as if this was a two-state system, which it
is not, but should be sufficient to illustrate the general idea (from
the comments above, we can think of the $\chi$ as the Dirichlet KK
states).  The point of physics is that the massive KK states of the
theory ($n\neq 0$) have $\sin\theta_n \ll 1$, $\cos\theta_n \approx
1$, so that they are mostly \textit{composite}.  For the 0-modes the
elementary/composite content depends on the localization:
\be
\begin{array}{lclcl}
\textrm{UV localization}          & \quad \longrightarrow \quad & 
\textrm{mostly elementary }   & &
\displaystyle 
(\sin\theta_0 \approx 1, \cos\theta_0 \ll 1)~,   \\ [0.8em]
\textrm{IR localization}           & \quad \longrightarrow \quad & 
\textrm{mostly composite }   & &
\displaystyle 
(\sin\theta_0 \ll 1, \cos\theta_0 \approx 1)~,   \\ [0.8em]
\textrm{Approximately flat}           & \quad \longrightarrow \quad & 
\hspace{-1.8mm}
\begin{array}{l}
\textrm{largely elementary,} \\
\textrm{but some compositeness} 
\end{array}  & &
\displaystyle 
(\cos\theta_0 \sim 1/\sqrt{k_{\rm eff} L})~.
\end{array}
\nonumber
\ee
One should emphasize that this holographic language simply provides a
way to look at the physics obtained from the more simple-minded
Kaluza-Klein approach (after all, the KK states \textit{are} the
physical mass eigenstates).  In fact, the way we have described it
here (which is well-defined operationally) the whole re-interpretation
amounts to a \textit{definition}, since we do not have independent
access to the dual theory $S_{\rm Dual}[\chi]$.

Nevertheless, in concrete examples like AdS$_5$, the detailed
properties of the various propagators allow for non-trivial checks
that the interpretation makes physical sense.  Of course, the AdS$_5$
limit is closely tied to the setting on which the holographic duality
was first proposed~\cite{Maldacena:1997re}, and on which the method is
inspired~\cite{Witten:1998qj}.  In this case, both theories are known
and justify the name \textit{duality}, which indicates a strong/weak
coupling mapping:
\be
\begin{array}{c}
\textrm{Type IIB String Theory} \\
\textrm{on AdS$_5 \times S^5$} 
\end{array}
& \quad \stackrel{\textrm{Dual}}{\rule{0mm}{3mm}\Longleftrightarrow}
\quad& 
\begin{array}{c}
\textrm{4D } {\cal N} = 4~SU(N) \\
\textrm{super Yang-Mills}
\end{array}
\nonumber
\ee
Many non-trivial checks (computations on both sides) strongly suggest
that the conjectured duality is indeed true. 

One simple map is between the isometries of AdS$_5$ and the (super)
conformality of ${\cal N} = 4$ SYM.
Thus, one expects, more generally, that AdS$_5$ backgrounds map onto
\textit{conformal} theories (where the conformal symmetry is
explicitly broken by the UV brane, and spontaneously broken by the IR
brane). This is why in many applications one thinks of a \textit{CFT
dual}. For further aspects of the dictionary
see~\cite{ArkaniHamed:2000ds,Rattazzi:2000hs}.

Let us conclude this lecture by using the \textit{holographic
interpretation} to understand some of the results we have derived in
the KK language:
\begin{itemize}

\item The Higgs field, being IR localized, is almost pure \textit{CFT
composite}. Since the compositeness scale is of order $\tilde{k}_{\rm
eff} \ll M_P$, we can understand why $v_{\rm EW} \ll M_P$ (up to
little hierarchies). 

\item Gauge fields, being flat, have a non-negligible CFT admixture.
Thus their properties can receive sizable corrections from the
strongly coupled CFT sector (the $k_{\rm eff} L$ enhancements we have
found).

\item When light fields are UV localized, they are mostly
\textit{elementary}. The $S$-parameter involves both these light
fermions and the Higgs: it can be important, but not as much as the
$T$-parameter which involves only the Higgs (and gauge bosons). In
custodial models, the custodial symmetry of the CFT sector ensures
that its contribution to the $T$ parameter vanishes.

\item The top quark, as well as the LH bottom quark, have a sizable
CFT content, and their properties can receive sizable corrections
(e.g.~$\delta g_{Z\bar{b}_L b_L}$). Since both chiralities of the top
quark are mostly composite, its mixing with the CFT sector leads to a
large top mass, comparable to the EWSB vev, which is also connected
to the CFT strong dynamics. The bottom quark can be lighter, since
only the LH component has a composite nature, but the RH bottom is
mostly elementary.

\item 
\textbf{\underline{Custodial Models}:} 

\noindent
Consider the boundary value of a gauge field $\bar{A}^a_\mu(x)$. If
it is to be interpreted as a dynamical field coupled to a dual 4D
theory
\be
{\cal L}_{\rm Dual} &\supset& \bar{A}^a_\mu J^{\mu \, a}~,
\ee
the \textit{CFT currents} $J^{\mu \, a}$ better be
\underline{conserved}. Thus, the dual theory must possess a
\underline{global} symmetry that is weakly gauged by
$\bar{A}^a_\mu(x)$. If the $\bar{A}^a_\mu(x)$ obey Dirichlet boundary
conditions on the UV brane (add a mass, then send it to infinity)
then the CFT is left just with the global symmetry. This leads to the
important mapping
\be
\begin{array}{c}
\textrm{5D gauge symmetries}
\end{array}
& \quad \stackrel{\textrm{Dual}}{\rule{0mm}{3mm}\Longleftrightarrow}
\quad& 
\begin{array}{l}
\textrm{(weakly gauged) CFT global symmetries}  \\
\textrm{\ldots depending on whether the gauge field} \\
\textrm{obeys Dirichlet b.c.'s on the UV, or not.}
\end{array}
\nonumber
\ee
In the custodial models we have a bulk
\be
SU(2)_L \times SU(2)_R \times U(1)_X
\ee
gauge symmetry, but all non-SM gauge fields obey Dirichlet b.c.'s on
the UV brane. Thus, they correspond to \textit{global} symmetries of
the CFT:
\be
\left[ SU(2)_L \times SU(2)_R \times U(1)_X \right]^{\rm global}
&\supset&
\textrm{weakly gauged}~SU(2)_L \times U(1)_Y~.
\ee
Similarly to the SM, it is this global symmetry that protects the
relation $M^2_W \approx M^2_Z \cos^2\theta_w$.

Recall that the SM gauge fields (0-modes) are mostly elementary,
hence not quite part of the CFT. In particular, $g'$ breaks the
custodial symmetry, much as in the SM.

%%%%%%%%%%%%%%%%%%%%%
\begin{quote}

\textbf{\underline{Exercise}:}
What about the loop effects associated with the top tower?
\end{quote}
%%%%%%%%%%%%%%%%%%%%%

\end{itemize}

\centerline{
Have a productive extra-dimensional journey!}

%%%%%%%%%%%%%%%%%%%%%%%%%%%%%%%%%%%%%%%%%
\section*{Acknowledgments}

I would like to thank K.~Matchev and T.~Tait for the invitation to
lecture, and also for organizing an exciting program.  My thanks go
also to Professors T.~DeGrand and K.T.~Mahanthappa, as well as to the
students of TASI 2011, for helping to create a stimulating atmosphere.
The author is supported by DOE grant DE-FG02-92ER40699.

%%%%%%%%%%%%%%%%%%%%%%%%%%%%%%%%%%%%%%%

\end{document}